\documentclass[a4paper,12pt]{article}
\usepackage[table]{xcolor}				
\usepackage{booktabs}					

\usepackage{amsmath,amssymb}			
\usepackage{dsfont}						

\usepackage{graphicx}					
\usepackage{subfig}						

\usepackage[pdftex,pdfborder={0 0 0}]{hyperref}	
\usepackage{tikz}	
\usepackage{ifthen}						

\newlength{\xtrawidth}
\newlength{\xtraheight}
\numberwithin{equation}{section}		
\numberwithin{table}{section}
\numberwithin{figure}{section}

\graphicspath{{figs/}}

\def\bC{\mathbb{C}}

\def\bZ{\mathbb{Z}}
\def\cC{\mathcal{C}}
\def\cD{\mathcal{D}}
\def\cF{\mathcal{F}}

\def\cH{\mathcal{H}}
\def\cJ{\mathcal{J}}

\def\cN{\mathcal{N}}
\def\cO{\mathcal{O}}
\def\cR{\mathcal{R}}

\def\IF{{\mathbb F}}

\def\tmu{\mu_{\rm eff}}
\def\tE{E_{\rm eff}}
\def\tJ{\widetilde{J}}
\def\tu{\tilde{u}}
\def\tF{\tilde{F}}

\def\um{\underline{m}}

\def\Ftop{F^{\rm inst}_{\rm top}}
\def\Fns{F^{\rm inst}_{\rm NS}}

\def\bc{\mathbf{c}}
\def\bd{\mathbf{d}}
\def\bt{\mathbf{t}}
\def\bB{\mathbf{B}}
\def\bT{\mathbf{T}}
\def\BPSN{N^{\mathbf{d}}_{j_L,j_R}}

\DeclareMathOperator{\vol}{vol}

\def\sx{\mathsf{x}}
\def\sy{\mathsf{y}}

\def\sO{\mathsf{O}}
\def\sO0{\mathsf{O}_{\mathbb{F}_0}}

\def\hm{\hat{m}}
\def\hht{\hat{t}}
\def\hu{\hat{u}}
\def\hv{\hat{v}}
\def\hmu{\hat{\mu}}
\def\hF{\hat{F}}
\def\hJ{\hat{J}}
\def\hTheta{\hat{\Theta}}

\def\hQm{\widehat{Q}_m}

\def\({\left(}
\def\){\right)}

\newcommand{\fad}{\operatorname{\Phi}_{\mathsf{b}}}

\newcommand{\mypsi}[2]{\operatorname{\Psi}_{#1,#2}}

\newcommand{\myh}{\mathsf{h}_{\mathsf{b}}}

\newcommand{\mX}{{\mathsf{X}}}
\newcommand{\mY}{{\mathsf{Y}}}

\newcommand{\mO}{{\mathsf{O}}}
\newcommand{\mb}{{\mathsf{b}}}
\newcommand{\map}{{\mathsf{p}}}
\newcommand{\mq}{{\mathsf{q}}}
\newcommand{\tr}{{\rm Tr}}

\def\Mgn[#1]#2{{\overline{\cal M}_{#1,#2}}}
\def\pqs[#1,#2]{{\footnotesize{$\left[\begin{array}{c} #1\\#2  \end{array}\right]$}}} 
\def\pqsu[#1,#2]{\left[\begin{array}{c} #1\\#2  \end{array}\right]} 
\def\pqssu[#1,#2]{{\footnotesize{\left[\begin{array}{c} #1\\#2  \end{array}\right]}}} 
\def\pqh[#1,#2]{{\footnotesize{$\left[\begin{array}{c} #1\\#2  \end{array}\right]$}}} 
\def\pqhu[#1,#2]{\left[\begin{array}{c} #1\\#2  \end{array}\right]} 
\newcommand{\ban}{\begin{eqnarray}}
\newcommand{\ean}{\end{eqnarray}}
\newcommand{\be}{\begin{equation}}
\newcommand{\ee}{\end{equation}}
\numberwithin{equation}{section}

\newcommand{\IC}{\mathbb{C}}
\newcommand{\IP}{\mathbb{P}}

\newcommand{\IR}{\mathbb{R}}

\newcommand{\im}{{\rm Im \,}}

\newcommand{\ba}{\begin{aligned}}
\newcommand{\ea}{\end{aligned}}
\newcommand{\ben}{\begin{eqnarray}\displaystyle}
\newcommand{\een}{\end{eqnarray}}

\newcommand{\CB}{{\cal B}}

\newcommand{\CF}{{\cal F}}

\newcommand{\Tr}{{\rm Tr}}
\newcommand{\re}{{\rm e}}
\newcommand{\ri}{{\rm i}}
\newcommand{\rd}{{\rm d}}
\newcommand{\mx}{{\mathsf{x}}}
\newcommand{\my}{{\mathsf{y}}}

\def\CURVEDEGREE{c}		

\ifthenelse{\equal{\CURVEDEGREE}{c}}
	{\def\tca{\tilde{c}_\alpha}}
	{\def\tca{\tilde{d}_\alpha}}
\ifthenelse{\equal{\CURVEDEGREE}{c}}
	{\def\dca{c_\alpha}}
	{\def\dca{d_\alpha}}
\ifthenelse{\equal{\CURVEDEGREE}{c}}
	{\def\dc{c}}
	{\def\dc{d}}

\begin{document}

\begin{titlepage}
\begin{center}
\hfill BONN-TH-2015-09\\
\vskip 0.55in

{\Large \bf Exact solutions to quantum spectral curves by topological string theory}

\vskip 0.3in

{ Jie Gu${}^{a}$, Albrecht Klemm${}^{a}$, Marcos Mari\~{n}o${}^{b}$, Jonas Reuter${}^{a}$}\\

\vskip 0.2in
{\footnotesize
\begin{tabular}{c}
${}^{\, a}${\em Bethe Center for Theoretical Physics, Physikalisches Institut, }\\
$\phantom{{}^{\, a}}${\em Universit\"at Bonn, 53115 Bonn, Germany} \\
$\phantom{{}^{\, a}}${\em } \\
${}^{\, b}${\em D\'{e}partement de Physique Th\'{e}orique et Section de Math\'{e}matiques, }\\
$\phantom{{}^{\, b}}${\em Universit\'{e} de Gen\`{e}ve, Gen\`{e}ve, CH-1211 Switzerland} \\
$\phantom{{}^{\, b}}${\em }
\end{tabular}
}
\end{center}
\vskip1ex

\begin{center} {\bf Abstract} \end{center}

We generalize the conjectured connection between quantum spectral problems and topological strings to many local almost del Pezzo surfaces with arbitrary mass parameters. The conjecture uses perturbative information of the topological string in the unrefined and the Nekrasov--Shatashvili limit to solve non-perturbatively the quantum spectral problem. We consider the quantum spectral curves for the local almost del Pezzo surfaces of $\IF_2$, $\IF_1$, $\CB_2$ and a mass deformation of the $E_8$ del Pezzo corresponding to different deformations of the three-term operators $\mO_{1,1}$, $\mO_{1,2}$ and $\mO_{2,3}$. To check the conjecture, we compare the predictions for the spectrum of these operators with numerical results for the eigenvalues. We also compute the first few fermionic spectral traces from the conjectural spectral determinant, and we compare them to analytic and numerical results in spectral theory. In all these comparisons, 
we find that the conjecture is fully validated with high numerical precision. For local $\IF_2$ we expand the spectral determinant around the orbifold point and find intriguing relations for Jacobi 
theta functions. We also give an explicit map between the geometries of $\IF_0$ and $\IF_2$ as well as a systematic way to derive the operators $\mO_{m,n}$ from toric geometries.

\vfill

\noindent June, 2015

{
\let\thefootnote\relax
\footnotetext{\tt jiegu@th.physik.uni-bonn.de, aklemm@th.physik.uni-bonn.de, Marcos.Marino@unige.ch, jreuter@th.physik.uni-bonn.de}
}

\end{titlepage}

\tableofcontents
\newpage

\section{Introduction}

 Topological string theory on Calabi--Yau (CY) threefolds can be regarded as a simplified model for string theory 
 with many applications in both mathematics and physics. Topological 
 strings come in two variants, usually called the A-- and the B--models, related by mirror symmetry. 
 When the CY is toric, the theory can be solved at all orders in perturbation theory with different techniques. 
 The A--model can be solved via localization 
 \cite{kon,kz} or the topological 
 vertex \cite{akmv}, while the B--model can be solved with the holomorphic anomaly equations \cite{bcov,Haghighat:2008gw} or with topological recursion 
 \cite{Marino:2006hs,Bouchard:2007ys}. 
 Part of the richness and mathematical beauty of the theory in the toric case stems from 
 the interplay between these different approaches, which involve deep relations to knot theory, matrix models, and integrable systems.
 
 In spite of these developments, there are still many open questions. Motivated by instanton counting in gauge theory \cite{n}, it was noted \cite{ikv} that 
 the topological string on toric CYs can be ``refined," and an additional coupling constant 
 can be introduced. Although many of the standard techniques in topological string theory can be extended to the refined case \cite{hk,kw,ckk,no}, 
 this extension is not as well understood as it should 
 (for example, it does not have a clear worldsheet interpretation). 
 Another realm where there is much room for improvement is the question of the non-perturbative completion of the theory. Topological string theory, as any other string theory, is in principle 
 only defined perturbatively, by a genus expansion. An important question is whether this perturbative series can be regarded as the asymptotic expansion of a well-defined quantity. In the 
 case of superstring theories in AdS, such a non-perturbative completion is provided conjecturally by a CFT on the boundary. In the case of topological string theory on CY threefolds, 
 there is a similar large $N$ duality with Chern--Simons theory on three-manifolds, but this duality only applies to very special CY backgrounds \cite{gv,akmv-cs}\footnote{It is sometimes believed that the Gopakumar--Vafa/topological vertex 
 reorganization of the topological string free energy provides a non-perturbative completion, but this is not the case. This reorganization is not a well-defined function, since for real string coupling it has 
 a dense set of poles on the real axis, and for complex string coupling it does not seem to lead to a convergent expansion \cite{hmmo}.}.
 
 One attractive possibility is that the topological string emerges from a simple quantum system in low dimensions, as it happens with non-critical (super)strings. 
 Since the classical or genus zero limit of topological 
 string theory on a toric CY is encoded in a simple algebraic mirror curve, it has been hoped that the relevant quantum system 
 can be obtained by a suitable ``quantization" of the mirror curve \cite{Aganagic:2003qj}. In \cite{acdkv}, it was shown that a formal 
 WKB quantization of the mirror curve makes it possible to recover the refined topological string, but in a special limit --the Nekrasov--Shatashvili (NS) limit-- first discussed in the context of gauge theory in \cite{ns}. 
 The quantization scheme in \cite{acdkv} is purely perturbative, and the Planck constant associated to the quantum curve is the coupling constant appearing in the NS limit. 
 
 Parallel developments \cite{Marino:2011eh,Hatsuda:2012dt,cm,hmo2,hmmo,Kallen:2013qla} in the study of the matrix model for ABJM theory \cite{kwy} 
 shed additional light on the quantization problem. It was noted in \cite{Kallen:2013qla} that the quantization of the mirror curve leads to a quantum-mechanical operator with a 
 computable, discrete spectrum. The solution to this spectral problem involves, in addition to the NS limit of the refined topological string, a non-perturbative 
 sector, beyond the perturbative WKB sector studied in \cite{acdkv}. Surprisingly, this sector involves the {\it standard} topological string. 
 The insights obtained in \cite{Kallen:2013qla} 
 thanks to the ABJM matrix model apply in principle only to 
 one particular CY geometry, but they were extended to other CYs in \cite{Huang:2014eha}, which generalized 
 the method of \cite{Kallen:2013qla} for solving the spectral problem. A complete picture 
 was developed in \cite{ghm}, which made 
 two general conjectures valid in principle for arbitrary toric CYs based on del Pezzo surfaces: first, the quantization of the mirror curve to a local del 
 Pezzo leads to a positive-definite, trace class operator on $L^2(\IR)$. 
 Second, the spectral or Fredholm determinant of this operator can be computed in closed form from the standard and NS topological string free energies. 
 The vanishing locus of this spectral determinant gives an exact quantization condition which determines the spectrum of the corresponding operator. 
 The first conjecture was proved, to a large extent, in \cite{km}, where it was also shown that the integral 
 kernel of the corresponding operators can be expressed in many cases in terms of the quantum dilogarithm. The second conjecture has been tested in \cite{ghm} in various examples. 
 
 The conjecture of \cite{ghm} establishes a precise link between the spectral theory of trace class operators and the enumerative geometry of CY threefolds. 
 From the point of view of spectral theory, it leads to a new family of trace class operators whose spectral determinant can be written in closed form --- a relatively rare commodity. From the point 
 of view of topological string theory, the spectral problem provides a non-perturbative definition of topological string theory. For example, one can show that, as a consequence of \cite{ghm}, the 
 genus expansion of the topological string free energy emerges as the asymptotic expansion of a 't Hooft-like limit of the spectral traces of the operators \cite{mz}. 
 
The conjecture of \cite{ghm} concerning the spectral determinant has not been proved, but some evidence was given for some simple CY geometries in \cite{ghm}. Since the conjecture 
holds in principle for {\it any} local del Pezzo CY, it is important to test this expectation in some detail. In addition, working out the consequences of the conjecture in particular geometries 
leads to many new, concrete results for both, spectral theory and topological string theory. The goal of this paper is to test the conjecture in detail for many different 
del Pezzo geometries, in particular for general values of the mass parameters, 
and to explore its consequences. In order to do this, we use information on the refined topological string amplitudes to high genus, which lead for example to precision 
tests of the formulae for the spectral traces of the corresponding operators. 

In more detail, the content of this paper is organized as follows. In Section 2 we explain in detail how to obtain the geometries appropriate for operator analysis from 
mirror symmetry of global orbifolds. As an example, we work out the mass-deformed $E_8$ del Pezzo, which realizes a perturbation of the three-term operator 
$\mO_{2,3}$ considered in \cite{km}. In Section 3, we review and expand the conjecture of \cite{ghm}, as well as some of the results on the spectral theory of quantum curves 
obtained in \cite{km, Kashaev:2015wia}. In Section 4, we apply these general ideas and techniques to four different geometries: local $\IF_2$, local $\IF_1$, local $\CB_2$ and 
the mass deformed $E_8$ del Pezzo surface. In all these cases we compute the spectrum as it follows from the conjectural correspondence, and we compare it 
to the numerical results obtained by direct diagonalization of the operators. We also compute the first few fermionic spectral traces, as they follow from the conjectural 
expressions for the spectral determinants, and we compare them with both analytic and numerical results. In the case of local $\IF_2$, we work out the explicit 
expansion at the orbifold point. This leads to analytic expressions for the spectral traces, in terms of Jacobi theta functions and their derivatives. In the case of the 
$\mO_{2,3}$ operator, we also compare the large $N$ limit of its fermionic spectral traces, obtained in \cite{mz}, to topological string theory at the conifold point. The conjecture turns out 
to pass all these tests with flying colors. In the Appendices, we collect information on the Weierstrass and Fricke data of local CY manifolds, and we explain 
the geometric equivalence between local $\IF_2$ and local $\IF_0$.

\section{Orbifolds, spectral curves and operators}
As we mentioned in the introduction, the conjecture of \cite{ghm} associates a trace class operator to mirror spectral curves. Let us denote the variables 
appearing in the mirror curve by $x$, $y$. The corresponding Heisenberg operators, which we will denote by $\mx$, $\my$, satisfy the canonical 
commutation relation 
\begin{equation}
\label{heis-cr}
	[\sx, \sy] = \ri \hbar. 
\end{equation}
Since the spectral curves involve the exponentiated variables $\re^x, \re^y \in \IC^*$, after quantization one finds the Weyl operators
 \be
 \mX=\re^\mx, \qquad \mY= \re^\my.  
 \ee
As shown in \cite{km}, the simplest trace class operator built out of exponentiated Heisenberg operators is 
\be
\rho_{m,n}=\mO^{-1}_{m,n}
\ee
where
\be
\label{omn}
\mathsf{O}_{m,n}=\re^{\mathsf{x}}+\re^{\mathsf{y}} +\re^{-m\mathsf{x}-n\mathsf{y}},\quad m,n\in\mathbb{R}_{>0}.
\end{equation}
For example, the operator $\mO_{1,1}$ arises in the quantization of the mirror curve to the local $\IP^2$ geometry. Since these operators 
can be regarded as building blocks for the spectral theory/topological string correspondence studied in this paper, it is natural to 
ask how to construct local toric geometries which lead to $\mO_{m,n}$ operators after quantization.

It turns out that, to do this, one has to consider an $\mathbb{C}^3/G$ orbifold with a crepant 
resolution. This means that the resolution space $\widehat {\mathbb{C}^3/G}$ is a non-compact 
Calabi-Yau manifold, i.e. it has to have trivial canonical bundle. The section of 
the latter $\Omega={\rm d }z_1\wedge{\rm d }z_2\wedge{\rm d }z_3$ on $\mathbb{C}^3$ 
has to be invariant and it is not hard to see that this condition is also sufficient. 
For abelian groups, $G=\mathbb{Z}_{N_1}\times \mathbb{Z}_{N_2}$ is the most general choice in 
the geometrical context\footnote{Non-abelian groups $G$, which leave $\Omega$ invariant are also 
classified \cite{YauYu}, however they lead in general to non-toric A--model 
geometries. It is  still a challenge to figure out the general B--model 
geometry using non-abelian gauge linear $\sigma$ models.}, and 
$\widehat {\mathbb{C}^3/G}$ has a toric description. In fact all 
local toric Calabi-Yau spaces $X$ can be obtained by elementary 
transformations, i.e. blow ups and blow downs in codimension two, 
from  $\widehat {\mathbb{C}^3/G}$.       

\subsection{Toric description of the resolution of abelian orbifolds}  
\label{abelianorbifolds}
Let $N$ be the order of $G$. Invariance of $\Omega$ implies that the 
exponents $n^p_k\in \mathbb{N}_0$ of the orbifold action of the  
$\mathbb{Z}_{N_p}$ group factor on the $\mathbb{C}^3$ coordinates defined by 
\begin{equation}
z_k\mapsto  \exp \left(\frac{2 \pi  n^p_k}{N}\right) z_k, \quad  k=1,2, 3, \quad p=1,2 
\label{action}
\end{equation}
add up to $\sum_{k=1}^3 n^p_k=0\ {\rm mod}\ N$ for $p=1,2$.  
The  resolution leading to the A--model geometry $\widehat {\mathbb{C}^3/G} $ with $G$ abelian  
is described by standard toric techniques~\cite{MOP}, while the procedure that leads to the 
B--model curve is an adaptation of Batyrev's construction  to the local toric 
geometries~\cite{Katz:1996fh}\cite{Hori:2000kt}. The toric description of the 
resolution, see~\cite{MOP}, is given by a non-complete three dimensional fan $\Sigma_X$ in  
$\mathbb{Z}_{\mathbb{R}}^3$, whose trace at distance one from the origin is given by an 
integral simplicial two dimensional lattice polyhedron $\Delta$. 
Let $n^{(j)}_k$, $j=1,\ldots, |G|$, $k=1,2,3$, 
be the set of exponents  of all elements of $G$, then the two dimensional polyhedron $\Delta$ 
is simplicial and is the convex hull of 
\begin{equation}
\Delta=\{(m_1,m_2,m_3) \in \mathbb{N}_{\ge 0}^3| \sum_{k=1}^3 m_i=N, \exists j 
\ {\rm with}\ m_k-n^{(j)}_k=0 \ {\rm mod } \ N, \forall k \} \ ,  
\label{Delta} 
\end{equation}
in the smallest lattice $\Gamma$ generated by the points $(m_1,m_2,m_3)$. Let us give the two fundamental  
types of examples of this construction. 

Consider  as type (a) $G=\mathbb{Z}_N$ generated by (\ref{action})  ${\underline n}=(1,m,n)$, with $1+m+n=N$ and 
$m>0,n>0$, $\Delta$ is the convex hull of $\{\hat \nu_1=(0,N,0), \hat \nu_2=(0,0,N), (N,0,0)\}$. 
The point  $\nu_{\cal O}=(1,m,n)$ is by (\ref{Delta}) an inner point of $\Delta$, which we choose to 
be the origin of $\Gamma$, while $\Gamma$ is spanned by $\hat e_i=\nu_i-\nu_{\cal O}$, $i=1,2$.
Choosing the canonical basis $e_1=(0,1)$ and $e_2=(1,0)$ for $\Gamma=\mathbb{Z}^2$ and 
dropping the redundant first entry in the coordinates of $\Delta$, we find that 
\begin{equation} 
\Delta={\rm conv}(\{(1,0),(0,1),(-m,-n)\}) \subset \mathbb{Z}^2_{\mathbb{R}} \ .
\label{eq:DeltaZn}  
\end{equation}
We will argue below that the mirror curve seen as the Hamiltonian always contains an 
operator of type $\mO_{m,n}$.  

Consider as type (b) $G=\mathbb{Z}_{N_1}\times \mathbb{Z}_{N_2}$ with $|G|=N=N_1\times N_2$ 
generated by (\ref{action}), where ${\underline n}^{(1)}=N_2 (1,m,0)$ with $1+m=N_1$, and  
${\underline n}^{(2)}=N_1 (0,n,1)$ with $1+n=N_2$. We require $m>0$ and $n>0$ and either\footnote{The $\mathbb{Z}_{2}\times \mathbb{Z}_{2}$  
orbifold with $n=m=1$ has no inner point, as ${\underline n}^{(1)}\circ {\underline n}^{(2)}=(2,0,2)$ is as any point with one zero entry on the edge.}
$m>1$ or $n>1$. The point  $\nu_{\cal O}={\underline n}^{(1)}\circ {\underline n}^{(2)}=(N_2,N_1 N_2-N_1-N_2,N_1)$ 
is by (\ref{Delta}) an inner point of $\Delta$, which we choose to be the origin of 
$\Gamma$, while we can span $\Gamma$ by $\hat \nu_1=(N_1,N-N_1,0)-\nu_{\cal O}$ and $\hat \nu_2=(0,N-N_2,N_2)-\nu_{\cal O}$. 
Choosing  the canonical basis $e_1=(0,1)$ and $e_2=(1,0)$ for $\Gamma=\mathbb{Z}^2$ we find similarly as before

\begin{equation} 
\Delta={\rm conv}(\{(-m,1),(1,-n),(1,1)\}) \subset \mathbb{Z}^2_{\mathbb{R}} \ .
\label{eq:DeltaZnZn}  
\end{equation} 

Let $I_n(\Delta)$ be the number of all lattice points of $\Delta$ 
that lie only inside faces of dimension $n$ and not inside faces of 
dimension $k<n$, and $\bar I_n(\Delta)$ all points on dimension $n$ faces.   
$I_2(\Delta)$, i.e. the number of lattice points inside $\Delta$, counts  
compact (exceptional) divisors of the smooth non-compact Calabi-Yau 3-fold  
$X=\widehat {\mathbb{C}^3/G}$, while $I_1(\Delta)$, i.e. the number of lattice points 
inside  edges, counts  non-compact (exceptional) divisors of $X$, which are 
line bundles over exceptional $\mathbb{P}^1$'s. Their structure 
can be understood as follows. If $\mathbb{Z}_d\subset \mathbb{Z}_N$ with 
$d|N$ is a subgroup of $G$ that leaves a coordinate in $\mathbb{C}^3$ 
invariant, then it acts as $\mathbb{C}^2/\mathbb{Z}_d$ on the remaining 
$\mathbb{C}^2$ and its local resolution contains an $A_{d-1}$ type Hirzebruch 
sphere tree of $\mathbb{P}^1$'s whose intersection in  $\widehat { \mathbb{C}^2/\mathbb{Z}_d}$ 
is the negative Cartan matrix of the Lie algebra $A_{d-1}$. These 
$\mathbb{P}^1$'s are represented in the toric diagram as lattice 
points on the edge of $\Delta$ that is dual to the invariant 
coordinate.

In the mirror geometry described below, $I_2(\Delta)$ is identified with the genus and the 
number of complex structure parameters deformations $\tu_i$, $i=1,\ldots, 
I_2(\Delta)$,  of the family of mirror curves ${\cal C}$, while 
$I_1(\Delta)$ counts independent residua $m_k$, $k=1,\ldots, I_1(\Delta)$, of the 
meromorphic differential $\lambda$ on that curve. In the field theory, $\tu_i$ 
correspond to vevs of \emph{dynamical fields} while the $m_k$ are \emph{mass 
parameters}\footnote{These statements remain true for local Calabi-Yau 
3-folds $X$ described by non-compact fans whose traces $\Delta$ are 
non-simplicial, only that $I_1(\Delta)$ has 
to be replaced with  $\bar I_1(\Delta)-3$.}. In the 
resolution $X=\widehat {\mathbb{C}^3/G}$, the $\tu_i$ 
parameters are associated by the mirror map to the volumes of the
curves determining the volume of the compact (exceptional) 
divisors, while the $m_i$ parameters are associated by the mirror 
map to the volumes of the $\mathbb{P}^1$ of the sphere trees in the 
resolution of  the ${\widehat {\mathbb{C}^2/\mathbb{Z}_d}}$  singularities. 
The curve classes that bound the K\"ahler cone are linear combinations 
of these curves classes. The precise curve classes $[C_\alpha]$ with that 
property are encoded in the generators $l^{(\alpha)}$ of the Mori cone.

For orbifolds $\Delta$ is simplicial. Thus it is elementary to count
\begin{equation} 
I_2(\Delta)=\left\lfloor \frac{N}{2} 
\right\rfloor-\left\lceil \frac{I_1(\Delta)}{2}\right\rceil\ ,
\label{inner}
\end{equation}
where 
\begin{equation}
I_1(\Delta)=\left\{\begin{array}{rl} {\rm gcd}(m+1,n)+ {\rm gcd}(m,n+1)-2& \  {\rm for \ case \ (a) }  \\
                         m+n+ {\rm gcd}(m+1,n+1)-1& \  {\rm for \ case \ (b) } 
                        \end{array}\right.  \label{rand}
\end{equation}

Let us give a short overview over local Calabi-Yau  geometries that arise as resolved 
orbifolds. We have seen that $\Delta$ has always an inner point which we called $\nu_{\cal O}$ 
and by (\ref{inner}, \ref{rand}) it is easy to see that in the case (a) the $\mathbb{Z}_3$ orbifold 
with $n=m=1$, the $\mathbb{Z}_4$ orbifold with $m=2,n=1$, and the $\mathbb{Z}_6$ orbifold with 
$m=3,n=2$ are the only orbifolds whose mirrors are related to elliptic curves, i.e.  
$I_2(\Delta)=1$. It is easy to see that $I_1(\Delta)$ is $0,1,3$ respectively. For $N\ge 6$ one 
has several choices of the exponents, e.g. for $\mathbb{Z}_6$ the choice  $m=1,n=4$ leads to a genus 
two mirror curve. In the case (b) orbifolds with genus one mirror curves are the 
$\mathbb{Z}_3 \times \mathbb{Z}_3$ orbifold with $m=n=2$ and $I_1(\Delta)=6$, the 
$\mathbb{Z}_2 \times \mathbb{Z}_4$  orbifold with  $m=1,n=3$ and $I_1(\Delta)=5$, and 
the $\mathbb{Z}_2 \times \mathbb{Z}_3$ with $m=1,n=2$ and $I_1(\Delta)=3$.

\subsection{The mirror construction of the spectral curves}  
\label{spectralcurves}

Above we described toric local Calabi--Yau threefolds $X$ that arise as resolved  
abelian orbifolds and can serve as A--model geometries for topological string. 
Let $\Sigma_X$ be, a bit more general, an arbitrary non-complete toric fan in 
$\mathbb{Z}_{\mathbb{R}}^3$, $\Delta$ not necessarily a simplicial 
trace, and $k=\bar I_2(\Delta)-3$. The Calabi--Yau condition is equivalent 
to the statement  that the 1--cone generators  $\nu^{(i)}, i=0,1,\ldots, k+2$, 
end on a hyperplane $H$ one unit distance away from the origin of 
$\mathbb{Z}_{\mathbb{R}}^3$, and $\Delta=H\cap\Sigma_X$. 
We choose the coordinate system of $\mathbb{Z}_{\mathbb{R}}^3$ such that 
the first coordinate of $\nu^{(i)}$ is always 1. 
The $k+3$ 1--cone generators $\nu^{(i)}$ satisfy $k$ linear relations. If the Mori cone is simplicial, we  
can choose them to be the Mori cone generators\footnote{See~\cite{Cox:2000vi,DelaOssa:2001xk} for an explanation 
on how to find the Mori cone generators. For all examples considered here the Mori cone generators have been determined in \cite{Klemm:2012sx}. Non-simplicial Mori-cones have more 
than $k$ generators. For the construction of the mirror geometry it is sufficient 
to chose $k$ of them. The calculation of large radius BPS invariants is more involved 
in this case.} 
$\ell^{(\alpha)} = (\ell^{(\alpha)}_0,\ell^{(\alpha)}_1,\ldots, \ell^{(\alpha)}_{k+2})$ 
with $\alpha=1,\ldots,k$, such that
\begin{equation}
	\sum_i \ell^{(\alpha)}_i \nu^{(i)} = 0 \ ,\quad \forall \alpha \ .
\end{equation}
Due to their interpretation in 2d $\cN=(2,2)$ supersymmetric gauged linear 
sigma models, $\ell^{(\alpha)}$ are also called the charge vectors. 
The triviality of the canonical bundle is ensured if 
\begin{equation}
	\sum_{i=0}^{k+2} \ell^{(\alpha)}_i = 0 \ , \quad \forall \alpha \ .
\end{equation}

To construct the Calabi--Yau threefold $\widehat{X}$ on which the mirror B--model 
topological string lives~\cite{Katz:1996fh}\cite{Hori:2000kt}, one introduces 
$k+3$  variables $Y_i$ in $\mathbb{C}$ satisfying the conditions
\begin{equation}\label{eq:Y1}
	\prod_{i=0}^{k+2} Y_i^{\ell^{(\alpha)}_i} = 1 \ , \quad \forall \alpha \ .
\end{equation}
Then the mirror manifold $\widehat{X}$ is given by
\begin{equation}\label{eq:hatXEqa}
	w^+w^- = W_X   \ ,\quad\quad w^+,w^- \in \mathbb{C}\ .
\end{equation}
where 
\begin{equation}\label{eq:Wa}
	W_X = \sum_{i=0}^{k+2} a_i Y_i \ .
\end{equation}
Due to the three independent $\mathbb{C}^*$ actions on the $Y_i$ 
subject to the constraints~\eqref{eq:Y1}, only the following combinations 
\begin{equation}\label{eq:Batyrev}
	\prod_{i=0}^{k+2} a_i^{\ell_i^{(\alpha)}} \equiv z_\alpha \ 
\end{equation}
are invariant deformations of the B--model geometry. If $l^{(\alpha)}$ are the Mori cone 
generators, the locus $z_\alpha=0$ is the large complex structure point, 
which corresponds to the large volume limit of the A--model geometry. 
The $z_\alpha$ parametrize the deformations of $\widehat{X}$. 
It is equivalent and often more convenient to replace (\ref{eq:Y1})  and (\ref{eq:Wa}) by 
\begin{equation} 
\prod_{i=0}^{k+2} Y_i^{\ell^{(\alpha)}_i} = z_\alpha
\label{relations}
\end{equation}
and 
\begin{equation}	
	W_X = \sum_{i=0}^{k+2} Y_i	
\end{equation}
respectively. Using (\ref{relations})  one eliminates 
$k$ of the $k+3$ $Y^i$ variables. One extra $Y^i$ variable can be set to $1$ using the 
overall $\mathbb{C}^*$ action. Renaming the remaining two $Y^i$ 
variables $e^x$ and $e^y$ the mirror geometry (\ref{eq:hatXEqa}) becomes 
\begin{equation}
	w^+w^- = W_X(e^x,e^y; {\underline z} ) \ ,
\end{equation}
which describes a hypersurface in $\mathbb{C}^2\times (\mathbb{C}^*)^2$. 
Note that all deformations of $\widehat{X}$ are encoded in $W_X(e^x,e^y; 
{\underline z})$. In fact the parameter dependence of all relevant amplitudes 
of the B--model on $\widehat{X}$ can be studied from the non-compact Riemann 
surface ${\cal C}_X$ given by the vanishing locus of the Newton--Laurent 
polynomial in $(\mathbb{C}^*)^2$ 
\begin{equation}\label{eq:WX}
	W_X(e^x, e^y; {\underline z}) = 0 \ 
\end{equation}
and the canonical meromorphic one form on ${\cal C}_X$, a differential of the third kind 
with non-vanishing residues, given as  
\begin{equation}
\lambda= x\, {\rm d } y \ . 
\label{eq:lambda} 
\end{equation} 
Because of its r\^ole in mirror symmetry and the matrix model 
reformulation of the B--model, ${\cal C}_X$ is called the mirror curve or 
the spectral curve  respectively, while $\lambda$ is the local limit 
of the holomorphic  $(3,0)$ Calabi--Yau form on the B--model geometry. 

\begin{figure}[h!] 
\begin{center} 
\includegraphics[angle=0,width=.6\textwidth]{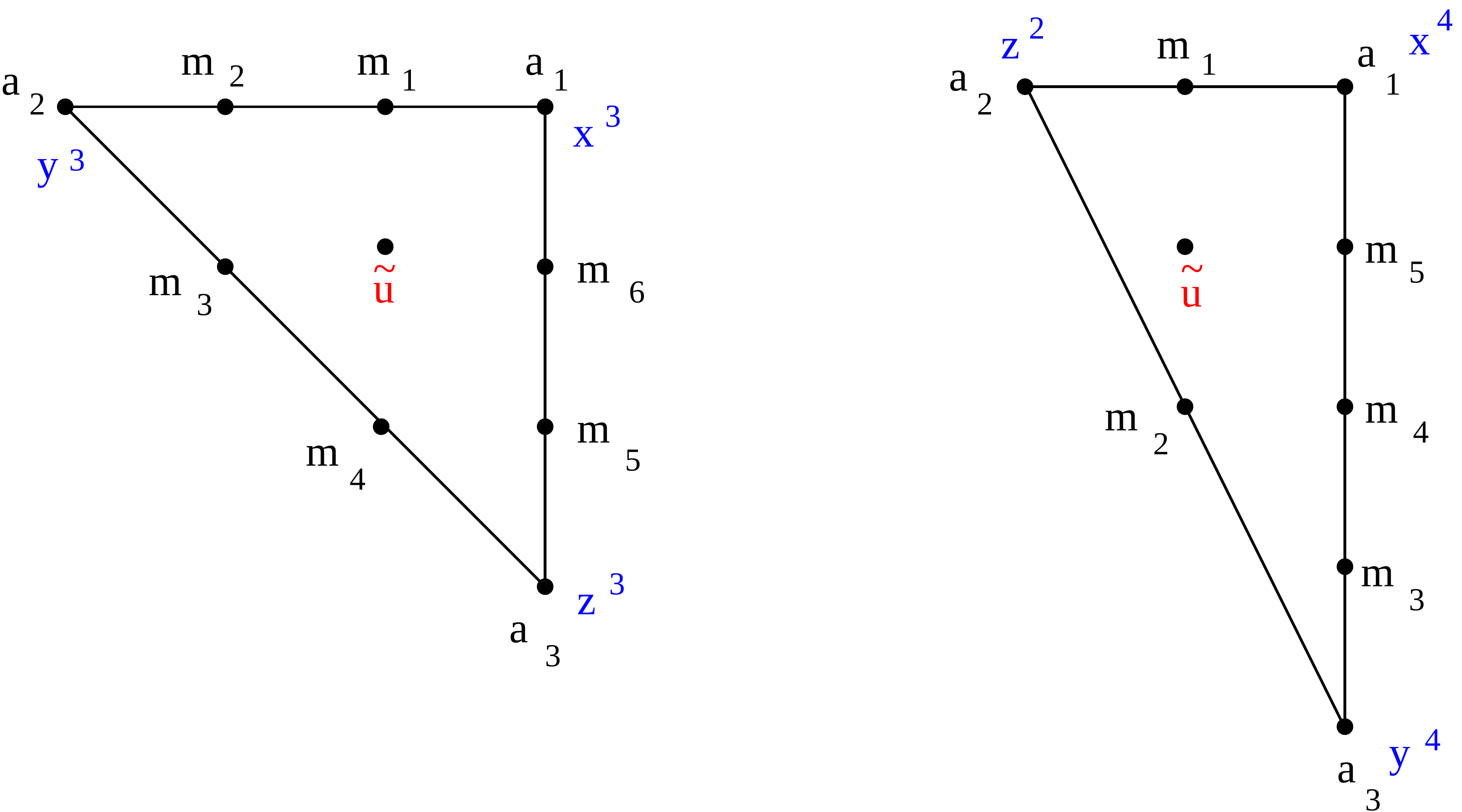}
\begin{quote} 
\caption{Toric traces for ${\widehat{\mathbb{C}^3/G}}$  with 
$G=\mathbb{Z}_3\times\mathbb{Z}_3$ and $G=\mathbb{Z}_2\times\mathbb{Z}_4$. They 
correspond  to case (b)  in Section~\ref{abelianorbifolds} with $(m=n=2)$  and $(m=1,n=3)$ 
respectively.}
\vspace{-1.2cm} \label{Z3Z3andZ2Z4}  \end{quote} 
\end{center} 
\end{figure}

The coefficients $\tu_i$, $i=1,\ldots, I_2(\Delta)$, of the monomials 
that correspond to inner points parametrize the complex structure of the 
family of mirror curves. To see this, note that all other coefficients can 
be set to one by automorphisms of a compactification of the mirror curve 
(\ref{eq:WX}), e.g. of ${\rm Aut}(\mathbb{P}_\Delta^*)$, which do not change the 
complex structure. However the other datum of the B--model, the meromorphic one 
form $\lambda$, is only invariant under the  three $\mathbb{C}^*$ actions on the 
coordinates of $\mathbb{P}_\Delta$. Therefore $\lambda$ depends on $\bar I_1(\Delta)-3$  
coefficients of the monomials on the boundary. We will set the 
coefficients of three points on the boundary to one, e.g. $a_i=1$, $i=1,\ldots,3,$ in Figure~\ref{Z3Z3andZ2Z4}. 
The coefficients of the other points on the boundary are then the mass parameters 
$m_i$, $i=1,\ldots, \bar I_1(\Delta)-3$. In this way the $z_\alpha({\underline \tu},{\underline m})$ 
can be seen as functions of the complex structure variables 
${\underline \tu}$  and the independent mass 
parameters ${\underline m}$.

Let us consider the $\mathbb{Z}_N$ orbifold geometry with the trace $\Delta$ given in (\ref{eq:DeltaZn}). 
To get the desired operator $\mO_{m,n}$ from the mirror curve, we associate 
$Y_1=e^{x}$ to the point $\nu_1=(1,0)$, $Y_2=e^{y}$ to the point $\nu_2=(0,1)$, and scale the 
$Y_{\nu_{\cal O}}$  coordinate that corresponds to the point $\nu_{\cal O}$  to $1$, while we denote 
the coefficient of  the $Y_{\nu_{\cal O}}$ coordinate by $u_1\equiv\tu$. This choice guarantees 
that the $Y_3$ coordinate associated to the point $\nu_3=(-m,-n)$  is expressed by solving 
(\ref{relations}) as $Y_3=e^{-m x - n y}$. Let us set all the other $\tu_i=0$ for $i=2,\ldots,I_2(\Delta)$, 
then the mirror curve has the shape         
\begin{equation}
	W_X(e^x,e^y) = e^x + e^y + e^{-mx-ny} + \sum_{i=1}^{I_1(\Delta)} f_i(\um) e^{\nu_1^{(i)}x+\nu_2^{(i)}y} +\tu \equiv \cO_X(x,y) + \tu \ ,
\end{equation}
where $f_i(\um)$ are monomials of mass parameters. Note that the function 
$\cO_X(x,y)$ can be regarded as a ``perturbation" of the function 
\be
\label{of-mn}
\cO_{m,n}(x,y)= e^x + e^y + e^{-mx-ny} 
\ee
and $\log(\tilde u)$ will be identified with the energy of the quantum system discussed 
below. (\ref{of-mn}) is the function which, upon quantization, leads to the operator (\ref{omn}). 
 If $I_2(\Delta)>1$, then the limit $u_i=0$, $i=2,\ldots, I_2(\Delta)$,
corresponds to a partial blow up of the  orbifold $\mathbb{C}^3/\mathbb{Z}_N$. 
Recall that all points on the trace $\Delta$ and the corresponding bounding 
fans as coordinate patches  have to be included to define 
$\widehat {\mathbb{C}^3/\mathbb{Z}_N}$ as a smooth variety.

In the rest of the paper we will only be concerned with the cases where $I_2(\Delta)=1$. This corresponds to smooth 
toric local Calabi--Yau threefolds whose spectral curves are elliptic curves. 
In particular, we consider  the anti-canonical bundles of almost del Pezzo surfaces $S$
\begin{equation}
	X = \cO(-K_S) \rightarrow S \ ,
\end{equation}
which have toric descriptions in terms of traces $\Delta$, which are one of the 16 
2-d reflexive polyhedra\footnote{They are toric del Pezzo if $I_1(\Delta)=0$ and 
almost toric del Pezzo otherwise.}. All of these except one, which involves a blow 
up, can be obtained by blow downs from the orbifold geometries discussed in the 
last section. In order to treat the toric cases in one go, we consider the 
largest polyhedra $\Delta$ for abelian group quotients with $I_2(\Delta)=1$ 
depicted in Figure~\ref{Z3Z3andZ2Z4}.
We compactify the corresponding mirror curves (\ref{eq:WX}) in 
$\mathbb{P}_{\Delta^*}$, but do not use the automorphism 
${\rm Aut}(\mathbb{P}_{\Delta^*})$ to eliminate the $m_i$. Rather we 
bring the corresponding mirror curves to the Weierstrass form
\begin{equation} 
y^2=4 x^3-g_2(u,{\underline m})  x- g_3 (u,{\underline m})\ ,
\label{eq:weierstrass}
\end{equation} 
using Nagell's algorithm, see Appendix~\ref{weierstrass}. In 
particular in that appendix we give in (\ref{WE6}) and (\ref{WE7}) 
the $g_2(u,{\underline m})$ and  $g_3 (u,{\underline m})$ for 
the mirror geometries of ${\widehat {\mathbb{C}^3/
\mathbb{Z}_3\times\mathbb{Z}_3}}$ and  ${\widehat {\mathbb{C}^3/
\mathbb{Z}_2\times\mathbb{Z}_4}}$. They can be specialized to the 
corresponding data of all examples discussed in detail in 
the paper, by setting parameters in these formulae to zero or one according to the embedding of the smaller traces $\Delta$ 
into the traces depicted in Figure~\ref{Z3Z3andZ2Z4}.

Let us introduce some conventions, which are usefull latter on. 
After gauging three coefficients of the boundary monomials to one by the $(\mathbb{C}^*)^3$ action, \eqref{eq:Batyrev} 
becomes $z_\alpha = \tu^{\ell^{(\alpha)}_0}\prod_{j=1}^{k-1} m_j^{\ell^{(\alpha)}_j}.$ 
The charge $-\ell_0^{(\alpha)}$ is the intersection number of the anti canonical class
$-K_S$ and the curve in the curve class $[C_\alpha]$ that bound the corresponding Mori cone generator 
on $X$.  Any such curve has a finite volume and lies entirely in $S$. Since $S$ is almost 
del Pezzo
\begin{equation}
	\dca \equiv -K_S \cap C_\alpha = -\ell^{(\alpha)}_0\ge 0\ .
\end{equation}
We define 
\begin{equation}
	r \equiv \textrm{gcd}(\dc_1, \ldots, \dc_k) \ ,
\end{equation}
and the reduced curve degrees
\begin{equation}
	\tca \equiv \dca/r \ , \label{eq:tcAlpha}
\end{equation}
as well as
\begin{equation}
	u \equiv \tu^{-r} \ .
\end{equation}
Then \eqref{eq:Batyrev} implies  
\begin{equation}\label{eq:zum}
	z_\alpha = \tu^{-\dca}\prod_{j=1}^{k-1} m_j^{\ell^{(\alpha)}_j} = u^{\tca}\prod_{j=1}^{k-1} m_j^{\ell^{(\alpha)}_j} \ .
\end{equation}
In \cite{Huang:2013yta,Huang:2014nwa} $u$ is used as the default elliptic modulus instead 
of $\tu$, because $u=0$ is the large complex structure point (\texttt{LCP}) in the moduli space of 
$W_X(e^x,e^y)$, and therefore convenient for computations around the \texttt{LCP}. In the following we will 
use the two variables interchangeably, preferring $\tu$ for the formal discussions related to the 
spectral problems, and $u$ for computations around the \texttt{LCP}.

Both data (\ref{eq:WX},\ref{eq:lambda}) are only fixed up to 
symplectic transformations
\begin{equation}
	\begin{aligned}
		x &\mapsto a\, x + b\, y + e \\
		y &\mapsto c\, x + d\, y + f
	\end{aligned} \ ,
	\quad \begin{pmatrix}
	a & b \\
	c & d
	\end{pmatrix} \in SL(2,\mathbb{Z})
\end{equation}
which preserve $\rd x \wedge \rd y$. In the rest of the paper, we will often call \eqref{eq:WX} the spectral curve of $X$ as well.

\subsection{Weierstrass data, Klein and Fricke theory and the B--model solution}
\label{ssc:BSolution}

According to the theory of  Klein and Fricke we get all the information about the periods and the Picard--Fuchs equations 
for the holomorphic differential, which reads
\[
	\omega=\frac{\rd x}{y} =\frac{\rm d }{{ \rm d} u}\lambda + \textrm{exact} \ ,
\]
in the Weierstrass coordinates $x,y$ of an elliptic curve, 
from properly normalized $g_2$ and $g_3$ and the $J$-invariant of the elliptic curve
\begin{equation}
\frac{j}{1728}=J=\frac{g_2^3}{g_2^3 - 27 g_3^2}=\frac{g_2^3}{\Delta_c} 
=\frac{E_4^3}{E_4^3-E_6^2}=\frac{1}{1728}\left(\frac{1}{q}+744+ 1926884 q +
\ldots \right) \ .
\label{J} 
\end{equation} 
A key observation in the treatment of Klein and Fricke is that \emph{any} modular
form $\phi_k(J)$ of weight $k$, w.r.t.\hspace{-0.4em} $\Gamma_0=SL(2,\mathbb{Z})$ (or a finite index subgroup $\Gamma_u$), 
fulfills as a function of the corresponding total modular invariant $J$ (or $u$) a linear differential 
equation of order $k+1$, see for an elementary proof~\cite{zagier}. In particular $\phi_k(J)$ can be meromorphic 
and the basic example~\cite{kleinfricke} is that $\sqrt[4]{E_4}$ can be written as the solution 
to the standard hypergeometric differential equation as 
\begin{equation} 
\sqrt[4]{E_4}={}_2F_1\left(\frac{1}{12},\frac{5}{12};1;1/J \right)\ . 
\end{equation} 
While solutions to the hypergeometric equation transform like weight one forms, 
other such objects such as \emph{in particular} the periods can be obtained by 
multiplying them with (meromorphic) functions of the total invariant $J$ (or $u$, 
which is a finite Galois cover of $J$). For example the unnormalized period $\Omega$ is a weight one form that fulfills the 
second order differential equation  
\begin{equation} 
\frac{{\rm d}^2 \Omega}{ {\rm d}^2 J}+\frac{1}{J} \frac{{\rm d} \Omega}{ {\rm d} J}+\frac{31 J-4}{144 J^2(1-J)^2}\Omega  =0\ , \qquad {\rm where} \  \Omega=\sqrt{\frac{E_6}{E_4}},
\label{PFequationuniversal} 
\end{equation}    
which is simply to be interpreted as the Picard--Fuchs equation for $\Omega$. It is easy to see that another way to write a 
solution to (\ref{PFequationuniversal}) is $\Omega_1=\sqrt[4]{\frac{1-J}{J}} {_2}F{_1}(\frac{1}{12},\frac{5}{12};1;1/J)$. 
These $u$ or $J$ dependent meromorphic factors can be fixed by global and boundary properties of 
the periods. In particular one can get the normalized solutions of the vanishing periods of 
$\omega$ at a given cusp as 
\begin{equation}
\label{aperiod} 
	\frac{\rd}{\rd u} t \equiv \frac{\rd}{\rd u}\int_a \lambda  = \int_a \omega = \sqrt{\frac{g_2}{g_3}} \Omega 
\end{equation}
for properly normalized $g_2(u,{\underline m}), g_3( u,{\underline m})$. Note that the mass 
parameters  ${\underline m}$ appear in this theory as deformation parameters, which 
are generically isomonodronic\footnote{I.e.~the nature of the Galois covering changes 
only at a few critical values of ${\underline m}$. Generically $t(\tu, {\underline m})$ is a transcendental 
function of $\tu$, while the corresponding flat coordinates $t_{m_j}({\underline m})$  are rational functions 
of  ${\underline m}$. More on the distinction between moduli and mass parameters of a B--model can be found in \cite{Huang:2013yta,Klemm:2015iya}.}.  
Similarly the normalized dual period to (\ref{aperiod}) is for $|J|>1$ and $|{\rm arg}(1-J) |<\pi$ 
\begin{equation}
\frac{{\rm d} }{{\rm d} u} F^{(0)}_t \equiv \frac{{\rm d} }{{\rm d} u}\int_b \lambda = \int_b \omega=\sqrt{\frac{g_2}{g_3}}\left(\sqrt{\frac{E_6}{E_4}}\log(1/j)-w_1\right) \ ,  
\label{bperiod} 
\end{equation}
where 
\begin{equation}
 w_1(J)=\sqrt[4]{\frac{1-J}{J}}\sum_{n=1}^\infty \frac{\left(\frac{1}{12}\right)_n \left(\frac{5}{12}\right)_n}{(n!)^2} h_n J^{-n},      
\end{equation}
with
\[
	h_n=2 \psi(n+1)-\psi\left(\tfrac{1}{12}+n\right)- \psi\left(\tfrac{5}{12}+n\right)+ \psi\left(\tfrac{1}{12}\right)+ \psi\left(\tfrac{5}{12}\right)-2\psi(1)
\]
as readily obtained from the Frobenius 
method for hypergeometric functions.

The monodromy group for loops on the $u$-plane acts on \eqref{aperiod}, \eqref{bperiod} as  a subgroup 
$\Gamma_u$ of index $K$ inside $\Gamma_0=SL(2,\mathbb{Z})$, where $K$ is the branching 
index of the Galois cover  of $u$ to $J$ defined by \eqref{J} and $\Gamma_u=\Gamma_0/G_{Galois}$, 
where $G_{Galois}$ is the Galois group of the covering \eqref{J}.

In \eqref{aperiod}, \eqref{bperiod} $t$ is the flat coordinate and $F^{(0)}_t$ the derivative of 
the prepotential $F^{(0,0)}\equiv F^{(0)}$ w.r.t. the former near the corresponding cusp\footnote{Which can be either 
the large complex structure point or the conifold. The formulae are related by a transformation in $\Gamma_0$ identifying 
the cusps and apply to both cusps.}. These  structures exist due to rigid special geometry and the fact that near the large 
complex structure point $F^{(0)}(t,\underline{m} )$ is a generating function for geometric invariants of 
holomorphic curves of genus zero in the Calabi-Yau $X$.

The refined amplitudes  $F^{(0,1)}(t,\underline{m})\equiv F_1(t,\underline{m})$ 
and $F^{(1,0)}(t,\underline{m})\equiv  F^{NS}_1(t,\underline{m})$ are given 
in  (\ref{defF1}) and (\ref{defF1NS}) respectively. The refined higher 
amplitudes $F^{(n,g)}(t,\underline{m})$  can be defined 
recursively by the refined holomorphic anomaly equation~\cite{hk,Krefl:2010fm} 
\begin{equation}
\frac{\partial F^{(n,g)}}{\partial \hat E_2}=\frac{c_0}{24} 
\left( \frac{\partial^2 F^{(n,g-1)}}{\partial^2 t} +\sum'_{m,h}\frac{\partial F^{(m,h)}}{\partial t} \frac{\partial F^{(n-m,g-h)}}{\partial t}\right) \ ,
\label{refha1}
\end{equation}
where $\hat E_2(\tau)=E_2-\frac{3}{\pi {\rm Im}(\tau)}$ is the almost 
holomorphic second Eisenstein series, which is a weight two form under $\Gamma_0$, and the prime 
on the sum means that $(m,h)=(0,0)$ and $(m,h)=(n,g)$ are omitted. $c_0$ is a model dependent constant.
It is convenient to define the an-holomorphic generator 
$\hat S= \left(\frac{{\rm d}u }{{\rm d}t}\right)^2  \hat E_2$, 
as well as $A=2 g_2 \partial_u g_3-3 g_2 \partial_u g_3$ and $B=g_2^2 \partial_u g_2- 18 g_3 \partial_u g_3$, so 
that by virtue of the Ramanujan relations 
\begin{equation}
\begin{gathered}
	\frac{{\rm d}^2u }{{\rm d}^2t}= \left(\frac{{\rm d}u }{{\rm d}t}\right)^2\frac{1}{4 \Delta_c}(A + 9 B \hat S) \ ,\\
	\frac{{\rm d} \hat S}{{\rm d}u}=\frac{1}{12 \Delta_c} ( g_2 A + 6 B \hat S + 27 A  \hat S^2) \ ,
\end{gathered}
\end{equation}
 the r.h.s. of (\ref{refha1}) becomes 
a polynomial in $\hat{S}$, while the derivatives w.r.t. $t$ can be converted to derivatives w.r.t. $u$
\begin{equation}
\frac{\partial F^{(n,g)}}{\partial \hat S}=\frac{c_0}{24} 
\left( \frac{\partial^2 F^{(n,g-1)}}{\partial^2 u} + \frac{A + 9 B \hat S}{4 \Delta_c}\frac{\partial F^{(n,g-1)}}{\partial u} + 
\sum'_{m,h}\frac{\partial F^{(m,h)}}{\partial u} \frac{\partial F^{(n-m,g-h)}}{\partial u}\right) \ .
\label{refha2}
\end{equation}
It follows that 
\begin{equation}
F^{(n,g)}=\frac{1}{\Delta_c^{2 (g+n)-2}(u,{\underline m})}\sum_{k=0}^{3 g+ 2n -3} \hat S^{k} p_k^{(n,g)}( u,{\underline m})  
 ,
\end{equation}
in other words, $F^{(n,g)}$ is a polynomial of degree $3g+2n-3$ in $\hat{S}$, where $p_{k>0}^{(n,g)}( u,{\underline m})$ 
is determined by  (\ref{refha2}), while $p_{k=0}^{(n,g)}( u,{\underline m})$ is determined from the regularity conditions on 
$F^{(n,g)}$ and the gap behaviour at the conifold divisor~\cite{hk}. The refined BPS states can be ontained 
from the large radius expansion of the $F^{(n,g)}({\underline t})$.     

\subsection{The mass deformed $E_8$ geometry}    
\label{mass-e8}

Let us exemplify  this construction with the function ${\cal O}_{2,3}$, leading to the operator $\mO_{2,3}$.      
The polyhedron $\Delta$ is depicted below. 
\begin{figure}[h!] 
\begin{center} 
\includegraphics[angle=0,scale=0.17]{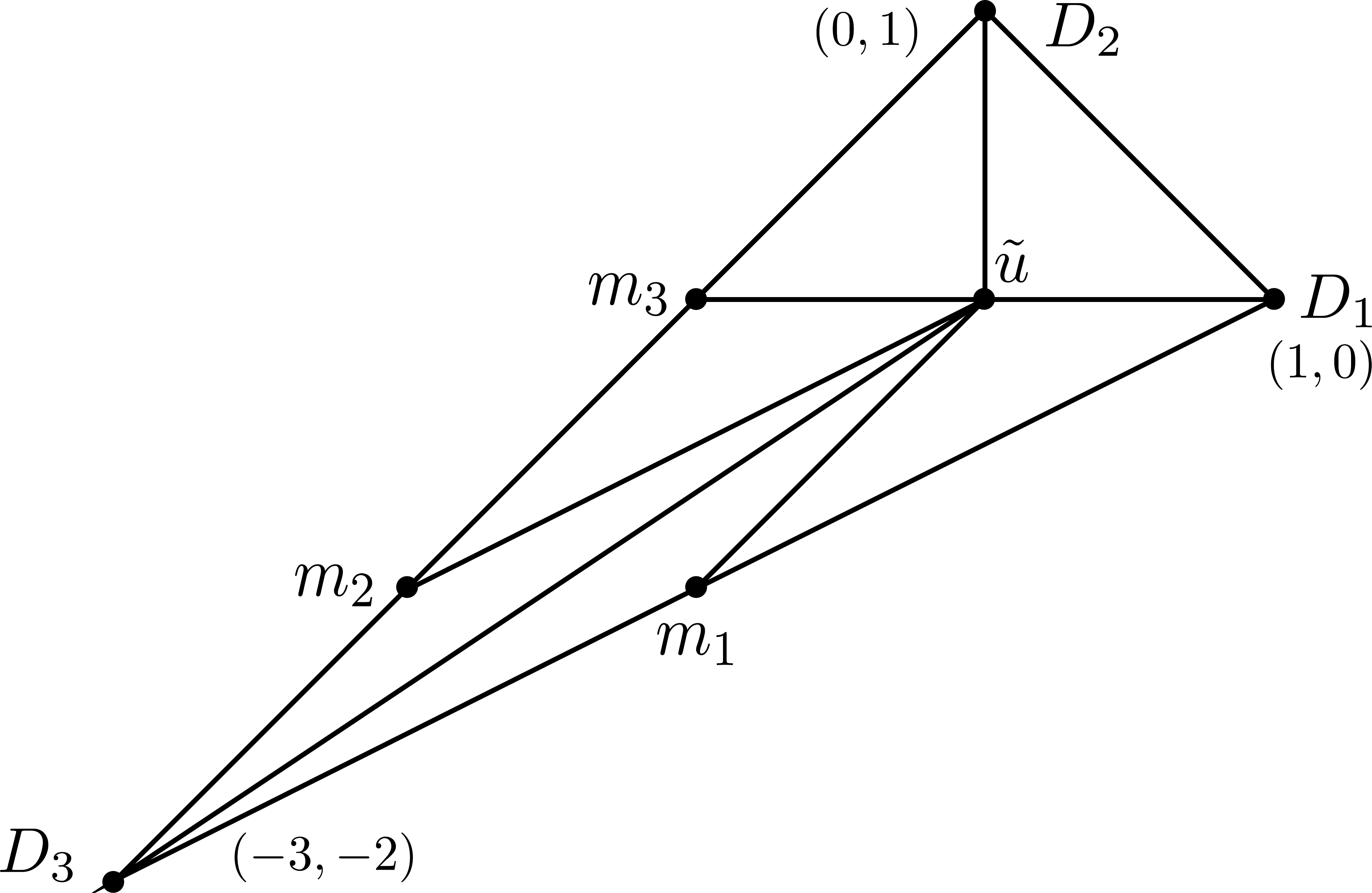}
\begin{quote} 
\caption{The polyhedron 10 with the choice of the mass parameters $m_1,m_2,m_3$  
and the modulus $\tilde{u}$.}
\vspace{-1.2cm} \label{poly10} \end{quote} 
\end{center} 
\end{figure}

The Mori cone vectors, which correspond to the depicted triangulation, are given below 
\begin{align}\label{dataf1}
 \begin{array}{ l rrr rrrr } \toprule
  \multicolumn{4}{c}{\nu_i} &l^{(1)}&l^{(2)}&l^{(3)}&l^{(4)}\\ \cmidrule{1-8}
   D_u    &    (\; 1&     0&   0 \;)&           0&-1& 0& 0 \\ 
   D_1    &    (\; 1&     1&   0 \;)&           1& 0& 0& 0 \\ 
   D_2    &    (\; 1&     0&   1 \;)&           0& 0& 0& 1 \\ 
   D_{m_3}&    (\; 1&    -1&   0 \;)&           0& 0& 1&-2 \\ 
   D_{m_2}&    (\; 1&    -2&  -1 \;)&           0& 1&-2& 1 \\ 
   D_3    &    (\; 1&    -3&  -2 \;)&           1&-1& 1& 0 \\ 
   D_{m_1}&    (\; 1&    -1&  -1 \;)&          -2& 1& 0& 0 \\ \bottomrule
  \end{array}
\end{align}
Following the procedure described in (\ref{relations}), one obtains the standard form 
of the Newton--Laurent polynomial as 
\begin{equation} 
W_{E_8}=\tu + e^x + \frac{m_1}{m_2 m_3^2} e^{-x-y}+ \frac{1}{m_2^2 m_3^4}e^{-3x-2y}+ \frac{1}{m_3^2} e^{-2 x -y}+ e^{-x} + e^{y} \, .
\end{equation} 
The monomials are ordered as the points in the figure and we rescaled $e^x\rightarrow e^x/\tu$ and $e^y\rightarrow e^y/\tu$ 
and  multiplied $W_{E_8}$ by $\tu$.  

With the indicated  three mass parameters  and the parameter $\tilde{u}$,
the Mori vectors determine the following large volume B--model coordinates
\begin{equation} 
 z_1=\frac{1}{m_1^2},\ \quad z_2=\frac{m_1 m_2}{\tilde{u}},\ z_3=\frac{m_3}{m_2^2},\ z_4=\frac{m_2}{m_3^2}\ .
\label{zcoordinatesp10}
\end{equation}

The anti-canonical class of the $E_8$ del Pezzo corresponds 
to an elliptic curve, which in turn has the following Mori vector  
\begin{equation} 
l_e= 3 l^{(1)} + 6 l^{(2)} + 4 l^{(3)} + 2 l^{(4)}=\sum_i a_i l^{(i)} \ .   
\label{ellipticMori}
\end{equation} 
This equation implies that $z_e=1/u^6 = z_1^3 z_2^6 z_3^4 z_4^2$ is the correct large volume 
modulus for this curve independent of the masses. By specializing the expression in Appendix \ref{weierstrass} 
as $m_1=0, m_2=0, m_3 =1,m_4=m_1, m_5=m_2, m_6= m_3, a_1= 1, a_2 = 0, a_3= 1, 
\tilde u =\frac{1}{u}$ and scaling $g_i\rightarrow \lambda^i g_i$ with $\lambda=  18 u^4 $  
we get the following coefficients of the Weierstrass form:
\begin{equation} 
\begin{array}{rl}
g_2=& 27 u^4 (24 m_1 u^3-48 m_2 u^4+16 m_3^2 u^4-8 m_3 u^2+1)\, ,\\ [3 mm]
g_3=&27 u^6 (216 m_1^2 u^6+12 m_3 u^2 (-12 m_1 u^3+24 m_2 u^4-1)+\\ 
& 36 m_1 u^3-72 m_2 u^4-64 m_3^3 u^6+48 m_3^2 u^4-864 u^6+1)\, .
\label{g2g3}
\end{array}
\end{equation}
Note there is a freedom of rescaling $g_2, g_3$ by an arbitrary function $\lambda(u,m)$
 \[
	g_i \mapsto \lambda^i(u,m) g_i
\]
without changing the Weierstrass form, if the coordinates $x, y$ of the Weierstrass form are also rescaled accordingly. Our particular choice of scaling makes sure that $\frac{dt}{du} = \frac{1}{u} +{\cal O}(1)$ and $t(u,m)$ 
becomes the logarithmic solution $t(u,m)=\log(u)+ {\cal O}(u)$ at the large complex structure point at $z_e=0$, 
which corresponds to $\frac{1}{j}\sim q \sim u^6$. We get as the transcendental mirror 
map $u= Q_t- m_3 Q_t^3+ {\cal O}(Q_t^4)$, with $Q_t \equiv e^{-t} =(Q_e)^{1\over 6}=\sqrt{Q_1} Q_2 Q_3^{2\over 3} Q_4^{1\over 3}$.  
The non-transcendental rational mirror maps are 
\begin{equation} 
z_1=\frac{ Q_1}{(1 + Q_2)^2}, \ \ z_3=Q_3 \frac{1 + Q_4 + Q_3 Q_4}{(1 + Q_3 + Q_3 Q_4)^2},\ \ z_4=Q_4\frac{1 + Q_3 + Q_3 Q_4}{(1 + Q_4 + Q_3 Q_4)^2}\ . 
\label{mirrormapp10} 
\end{equation}
The existence of these rational solutions for the mirror maps can be 
proven from the system of differential equations that corresponds to the Mori 
vectors listed above. With the knowledge of these rational solutions the 
system of differential equations can be reduced to a single third order differential equation in $u$ 
parametrized by the $m_i$, which is solved by the periods $t=\int_{a} \lambda$ and $F_t=\int_{b} \lambda$. 
Alternatively we can convert (\ref{g2g3}) into a second order differential equation in $u$ for $\int_{a,b} \omega$  
and integrate there later to find the desired third order Picard--Fuchs equation. For the mass deformed $E_8$ del 
Pezzo we obtain the following form 
\begin{equation} 
f_{9,8}(u,m)\frac{{\rm d} t(u,m) }{ {\rm d} u} + u g_{9,8}(u,m) \frac{{\rm d}^2 t(u,m)}{ {\rm d}^2 u}+u^2 
\Delta_o \Delta_c  \frac{{\rm d}^3 t(u,m)}{ {\rm d}^3 u}=0 \ ,  
\label{PFequationE8} 
\end{equation} 
where
\begin{equation}
\Delta_o=6 + 8m_2^2u^2 - 24m_3u^2 + 8m_1^2m_3u^2 - 9m_1^3u^3 + m_1(36u^3 - m_2u(7 + 4m_3u^2))
\end{equation} 
and 
\begin{equation}
\begin{array}{rl}
\Delta_c =& 1 - 12m_3u^2 + 48m_3^2u^4 - 432u^6 - 27m_1^4u^6 - 64m_2^3u^6 - 64m_3^3u^6 + m_2^2u^2(1 - 4m_3u^2)^2
       \\ & - 72m_2u^4(1 - 4m_3u^2) - m_1^3u^3(1 - 36m_3u^2) + m_1^2u^2(m_3 - 30m_2u^2 - 8m_3^2u^2 + 216u^4
        \\ & - 72m_2m_3u^4 + 16m_3^3u^4) + m_1(96m_2^2u^5 + 36u^3(1 - 4m_3u^2) - m_2u(1 - 4m_3u^2)^2)
\end{array} 
\end{equation} 
Furthermore $f_{9,8}(u,m)$ and $g_{9,8}(u,m)$ are polynomials of the indicated degrees in $u$ 
and the $m_i$. They can be simply derived from (\ref{g2g3},\ref{J},\ref{aperiod},\ref{bperiod}) and (\ref{PFequationuniversal}), 
or found in Appendix~\ref{app:FrickeE8}. The combinations that correspond to the actual periods can be 
obtained by analysing the behaviour of the solutions near the cuspidal points where the $a$ or $b$ 
cycle vanishes respectively. 

The more remarkable thing is the reduction to two special cases. The first is the massless $E8$ del Pezzo, which is obtained when
\begin{equation} 
m_1=m_2=m_3=0, \qquad Q_1=-1,\ \ Q_3=Q_4  = e^{\frac{2 \pi i}{3}}  \ .
\end{equation} 
In this case (\ref{PFequationE8}) simplifies to 
\begin{equation} 
\frac{{\rm d} t(u) }{ {\rm d} u} + u (3-5184 u^5-3888 u^6) \frac{{\rm d}^2 t(u)}{ {\rm d}^2 u}+u^2 
(1-432 u^6)\frac{{\rm d}^3 t(u)}{ {\rm d}^3 u}=0 \ .
\end{equation} 
The second case are the blow downs of the $A_1$ and $A_2$ types Hirzebruch sphere trees 
\begin{equation} 
m_1=2,\ m_2=m_3=3, \qquad Q_1=Q_3=Q_4=1   \ ,
\end{equation} 
in which case (\ref{PFequationE8}) simplifies to 
\begin{equation} 
\ba
& (1 + 2u - 96u^2 + 216u^3)\frac{{\rm d} t(u) }{ {\rm d} u}+   
u(3 + 4u - 120u^2 + 216u^3)\frac{{\rm d}^2 t(u)}{ {\rm d}^2 u}\\
& +
u^2(1 - 2u)(1 - 3u)(1 + 6u) \frac{{\rm d}^3 t(u)}{ {\rm d}^3 u}=0 \ . 
\ea
\end{equation}

Finally, we comment on the rational solutions to the Picard--Fuchs equation, see for instance \eqref{mirrormapp10}. 
They exist for the differential operators associated 
to Mori vectors that describe  the linear relations of points on an 
(outer) edge of a toric diagram. One can understand their existence 
from the fact that this subsystem describes effectively a non-compact 
two-dimensional CY geometry, whose compact part is a Hirzebruch sphere 
tree, which has no non-trivial mirror maps.               

This defines the K\"ahler parameters of the A--model geometry and 
relates them to the $u,m_j$. They allow to extract the BPS invariants 
for this mass deformation of the $E_8$ del Pezzo.

\section{Complete solutions to quantum spectral curves}

\subsection{Spectral curves and spectral problems}

In this section, we review the spectral problems corresponding to spectral curves in local mirror symmetry 
presented in \cite{ghm}. 

\begin{table}
\centering
\begin{tabular}{l >{$}l<{$} >{$}c<{$}}\toprule
$X$ & \cO_X(x,y) & r \\\midrule
local $\IP^2$ & e^x+e^y+e^{-x-y} & 3 \\
local $\IF_0$ & e^x+e^y+ e^{-y}+me^{-x} & 2 \\
local $\IF_1$ & e^x+e^y+e^{-x-y}+me^{-x} & 1 \\
local $\IF_2$ & e^x+e^y+e^{-2x-y}+ m e^{-x} & 2 \\
local $\CB_2$ & e^x+e^y+e^{-x-y}+m_1 e^{-y} +m_2 e^{-x} & 1\\
local $E_8$ del Pezzo & e^x+e^y+e^{-3x-2y}+m_1 e^{-x-y} +m_2 e^{-2x-y} +m_3e^{-x} & 1\\\bottomrule
\end{tabular}
\caption{The principal parts $\cO_X(x,y)$ of the spectral curves of some local del Pezzo surfaces, together with their $r$ values. }\label{tb:delPezzos}
\end{table}

The quantum operator $\mathsf{O}_X$ associated to $\cO_X(x,y)$ can be obtained by promoting the variables $x,y$ to quantum operators $\sx,\sy$ subject to the commutation relation 
\eqref{heis-cr}, where the (reduced) Planck constant is real. The ordering ambiguity is removed through Weyl's prescription
\begin{equation}
	e^{rx+sy} \mapsto e^{r\sx+s\sy} \ .
\end{equation}
We are interested in the spectral problem of $\mathsf{O}_X$. It was shown in \cite{ghm} that for local del Pezzo surfaces, $\mathsf{O}_X$ has a positive discrete spectrum
\begin{equation}\label{eq:OxSpec}
	\mathsf{O}_X |\psi_n\rangle = e^{E_n} |\psi_n\rangle \ ,\quad n = 0,1,\ldots \ .
\end{equation}
Note that after changing $\tu\mapsto -\tu$, the above spectral problem is equivalent to the quantum spectral curve problem considered in \cite{acdkv} in the Nekrasov--Shatashvili limit
\begin{equation}
	W_X(e^{\sx},e^{\sy}) |\psi_n\rangle = 0 \ ,
\end{equation}
where $|\psi_n\rangle$ is interpreted as a wavefunction on the moduli space of the branes of ``Harvey--Lawson'' type \cite{Harvey:1982xk,Aganagic:2000gs} in $X$ \cite{Aganagic:2003qj}, given that 
\begin{equation}
	\tu = e^{E} \ .
\end{equation}

In fact, it is more appropriate to study the operator
\begin{equation}
	\rho_X = \mathsf{O}^{-1}_X(\sx,\sy) \ ,
\end{equation}
as it was postulated \cite{ghm} and then proved rigorously \cite{km} that $\rho_X$ is a trace--class operator for a large category of geometries, including all those listed in Table \ref{tb:delPezzos}. As a consequence, both the \emph{spectral trace}
\begin{equation}\label{eq:sTrace}
	Z_\ell = \Tr_{\cH} \rho_X^\ell = \sum_{n=0}^\infty e^{-\ell E_n} \ , \quad \ell = 1,2,\ldots
\end{equation}
and the \emph{fermionic spectral trace}
\begin{equation}\label{eq:fsTrace}
	Z(N,\hbar) = \Tr_{\wedge^N \cH} \wedge^N \rho_X 
\end{equation}
are well--defined. Here $\cH$ is the Hilbert space discussed in detail in \cite{ghm}. The two spectral traces are related by
\begin{equation}
	Z(N,\hbar) = {\sum_{ \{ m_{\ell} \} } }' \prod_\ell \frac{(-1)^{(\ell-1)m_\ell}Z_\ell^{m_\ell}}{m_\ell! \ell^{m_\ell}} \ ,
\end{equation}
where $\sum'$ sums over all the integer vectors $\{m_\ell\}$ satisfying
\begin{equation}
	\sum_\ell \ell m_\ell = N \ .
\end{equation}
Furthermore, the \emph{spectral determinant} (also known as the Fredholm determinant)
\begin{equation}\label{eq:sDet}
	\Xi_X(\kappa,\hbar) = \det(1+\kappa\,\rho_X) = \prod_{n=0}^\infty \left(1+\kappa e^{-E_n}\right) = 1 + \sum_{N=1}^\infty Z(N,\hbar)\kappa^N
\end{equation}
is an entire function of the fugacity $\kappa$ in $\mathbb{C}$ \cite{simon}.

In the same spirit as \cite{Marino:2011eh}, the fermionic spectral trace $Z(N,\hbar)$ can be interpreted as the canonical partition function of an ideal fermi gas of $N$ particles, whose density matrix is given by the kernel of the $\rho_X$ operator
\begin{equation}
	\rho_X(x_1,x_2) = \langle x_1 | \rho_X | x_2 \rangle \ .
\end{equation}
Then $\Xi(\kappa,\hbar)$ is interpreted as the grand canonical partition function, and the fugacity $\kappa$ is the exponentiated chemical potential $\mu$,
\begin{equation}
	\kappa = e^\mu \ .
\end{equation}
It is then natural to consider the grand potential 
\begin{equation}
	\cJ_X(\mu,\hbar) = \log \Xi_X(\kappa,\hbar) \ ,
\end{equation}
from which the canonical partition functions can be recovered through taking appropriate residues at the origin
\begin{equation}\label{eq:ZJXold}
	Z(N,\hbar) = \int_{-\pi\ri}^{\pi\ri} \frac{\rd\mu}{2\pi\ri} e^{\cJ_X(\mu,\hbar) - N\mu} \ .
\end{equation}
Note that because $\Xi_X(\kappa,\hbar)$ is defined in terms of $\kappa$, $\cJ_X(\mu,\hbar)$ is a periodic function of $\mu$, being invariant under the shift
\begin{equation}
	\mu \mapsto \mu+ 2\pi \ri \ .
\end{equation}

\subsection{The conjecture}
\label{ssc:conjecture}

Directly solving the spectral problem of $\mathsf{O}_X$, including the calculation of $Z(N,\hbar)$ and $\Xi_X(\kappa,\hbar)$, is very difficult, although there has been great progress for some geometries \cite{mz,Kashaev:2015wia} by the use of quantum dilogarithm \cite{Faddeev:1993rs,Faddeev:1995nb} as well as identifying $Z(N,\hbar)$ as a (generalized) 
$O(2)$ matrix model integral, see (\ref{zmn-bis}) for an example. On the other hand, since the spectral curve $W_X(e^x,e^y)$ contains all the perturbative information of the B--model on $\widehat{X}$, and equivalently through mirror symmetry also the perturbative information of the A--model on $X$, there should be a deep connection between the spectral problem and the topological string theory on $X$. This is reflected in the conjecture  presented systematically in \cite{ghm}, drawing on previous results in 
\cite{Kallen:2013qla,hmmo,Huang:2014eha}. It provides a complete solution to the spectral problem using primarily the data of standard topological string and the refined topological string in the Nekrasov--Shatashvili limit on the target space $X$. We review the salient points of the conjecture here.

We first introduce the effective chemical potential $\tmu$. Let the quantum flat coordinate associated to the modulus $u$ be $t$. 
It is related to $u$ via a quantum mirror map \cite{acdkv},
\begin{equation}
	-t = \log u + \widetilde{\Pi}_A(u,\um,\hbar) \ .
\end{equation}
Then the effective chemical potential is defined to be
\begin{equation}\label{eq:mueff}
	\tmu = \mu- \frac{1}{r}\widetilde{\Pi}_A \left ((-1)^r e^{-r\mu},\um,\hbar \right ) \ .
\end{equation}


Next, we define the \emph{modified grand potential} $J_X(\mu,\um,\hbar)$ \cite{ghm}
\begin{equation}
	J_X(\mu,\um,\hbar) = J^{\rm (p)}(\tmu,\um,\hbar) + J_{\rm M2}(\tmu,\um,\hbar) + J_{\rm WS}(\tmu,\um,\hbar) \ ,
\end{equation}
including a perturbative piece $J^{\rm (p)}$, a M2 brane instanton piece $J_{\rm M2}$, and a worldsheet instanton piece $J_{\rm WS}$. These names come from the interpretation of their counterparts in the ABJM theory analog (see for instance \cite{hmmo}).

The perturbative piece $J^{\rm (p)}$ is given by
\begin{equation}
	J^{(p)}(\mu,\um,\hbar) = \frac{C(\hbar)}{3}\mu^3 + \frac{D(\um, \hbar)}{2}\mu^2 + B(\um, \hbar) \mu + A(\um,\hbar) \ .
\end{equation}
Of the four coefficient functions, the first three have finite WKB expansions
\begin{gather}
	C(\hbar) = \frac{C}{2\pi\hbar} \ , \\
	D(\um, \hbar) = \frac{D_0(\um)}{2\pi\hbar} \ , \\
	B(\um,\hbar) = \frac{B_0(\um)}{\hbar}  + B_1 \hbar \ , \label{eq:Bh}
\end{gather}
where the coefficients $C, D_0(\um), B_0(\um), B_1$ can be obtained as follows.

\begin{figure}
\centering
\subfloat[phase space]{\includegraphics[width=0.25\linewidth]{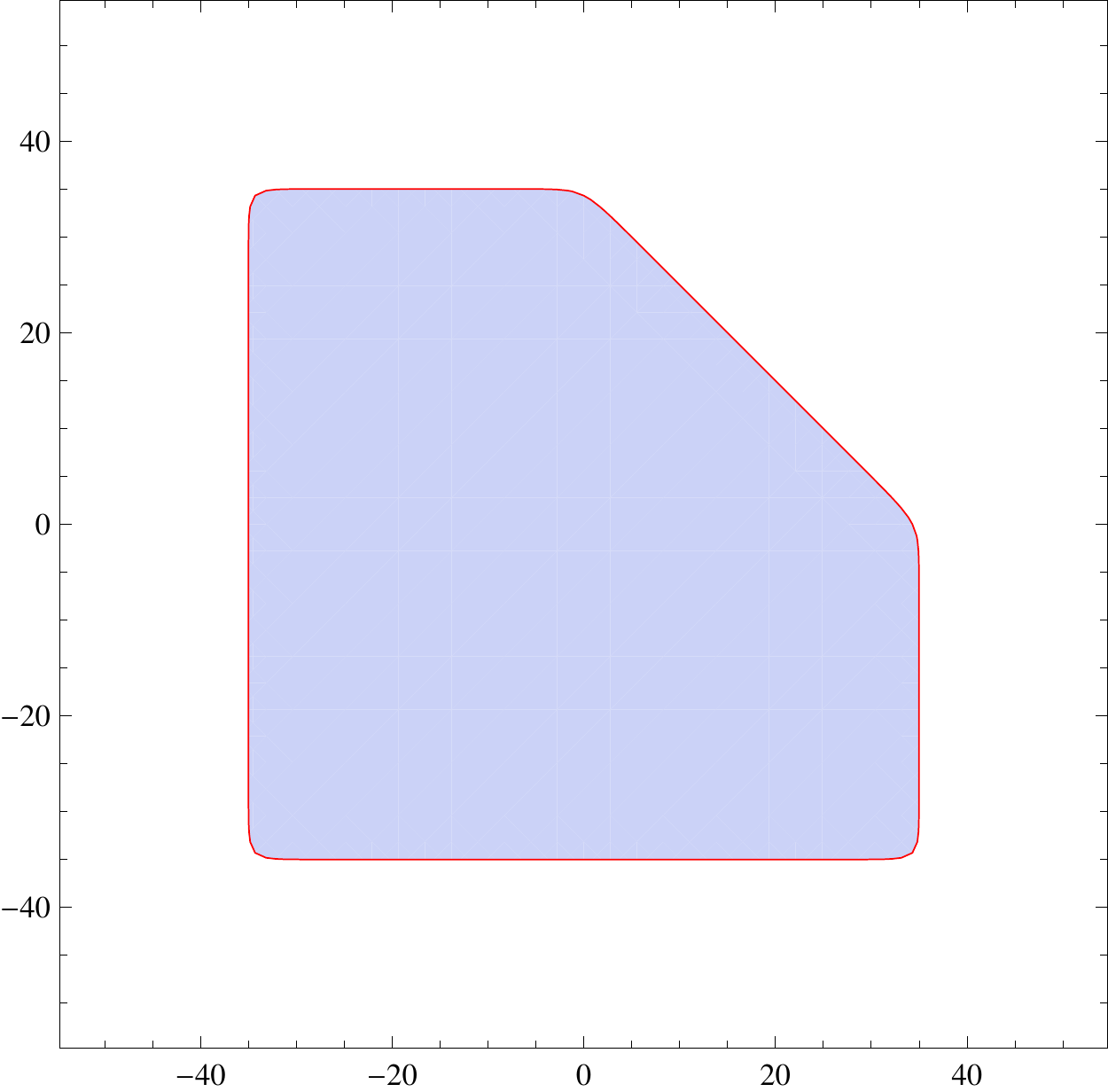}} \hspace{5em}
\subfloat[toric diagram]{\includegraphics[width=0.25\linewidth]{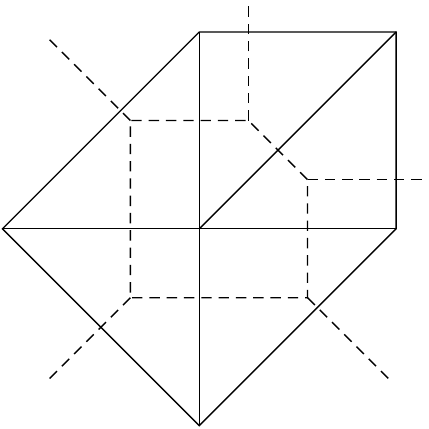}}
\caption{The bounded region $\cR$ in phase space for the quantum operator $\mathsf{O}_{\CB_2}$ associated to local $\CB_2$ with $E=35$ and $m_1 = m_2 =1$ (a) (figure taken from \cite{ghm}), and the toric diagram $\Upsilon_{\CB_2}$ (dashed) superimposed on the toric fan $\Sigma_{\CB_2}$ (solid) of local $\CB_2$ projected onto the supporting hyperplane $H$ (b).}\label{fg:B2_fs}
\end{figure}

In the semiclassical limit, the phase space of the system with energy no greater than $E$ is given by the bounded region
\begin{equation}
	\cR(E) = \{ (x,y) \in \mathbb{R}^2 : \cO_X(x,y) \leqslant e^E \} \ .
\end{equation}
In the high energy limit $E\geqslant 1$, the phase space has approximately the shape of the compact part of the dual toric diagram $\Upsilon_X$ 
projected onto the hyperplane $H$ in $\mathbb{R}^3$ where the endpoints of 1--cone generators of $\Sigma_X$ lie (see Figure~\ref{fg:B2_fs} for an example). Note that the boundary $\partial\cR(E)$ of $\cR(E)$ is the skeleton of the spectral curve $\cC_X$ with the punctures removed. Furthermore, in this limit, the volume of the phase space has the following asymptotic form \cite{ghm}
\begin{equation}\label{eq:vol0E}
	\vol_0(E) \approx CE^2 + D_0(\um) E + 2\pi\left( B_0(\um) - \frac{\pi}{6} C\right) + \cO(e^{-E}) \ , \quad E \geqslant 1 \ .
\end{equation}
Therefore we can use the approximation techniques used for instance in \cite{Huang:2014eha} to derive the leading contributions to $\vol_0(E)$ in the limit $E\geqslant 1$, and then extract the three coefficients $C, D_0(\um), B_0(\um)$. On the other hand, let $x_L$ and $x_R$ be the left and right limiting values of $x$ in $\cR(E)$. Between $x_L$ and $x_R$ the line of constant $x$ cuts through the boundary $\partial \cR(E)$ of $\cR(E)$ at two points with $y=y_+$ (up) and $y=y_-$ (down). Then the semiclassical phase space volume is
\begin{equation}
	\vol_0(E) = \int_{x_L}^{x_R} (y_+(x)-y_-(x)) dx = \oint_{\partial\cR(E)} ydx
\end{equation}
which coincides with the B--period of the elliptic spectral curve $\cC_X$. It is then natural to identify the total phase volume $\vol(E)$ including quantum corrections with the quantum B--period \cite{acdkv}. The first quantum correction $\vol_1(E)$ in the WKB expansion of $\vol(E)$
\begin{equation}
	\vol(E) = \sum_{k\geqslant 0} \hbar^{2k} \vol_k(E) \ ,
\end{equation}
can then be obtained from $\vol_0(E)$ through the differential operator $\cD_2$ which relates the first order quantum corrections in quantum periods to classical periods, with the following identification
\begin{equation}
	 u = e^{-r E} \ .
\end{equation}
In other words, we have
\begin{equation}\label{eq:vol1}
	\vol_1(E) = \cD_2 \vol_0(E)\ .
\end{equation}
For many local del Pezzo surfaces, this differential operator $\cD_2$ has been computed in \cite{Huang:2014nwa}, although when they are applied here, an extra minus sign is needed, because the $\hbar$ there differs from our convention by a factor of $\ri$. We find that $\cD_2$ in \cite{Huang:2014nwa} for local del Pezzo surfaces all have the following asymptotic form
\begin{equation}
	\cD_2 = \beta \partial^2_E + \cO(e^{-rE}) \ ,
\end{equation}
where $\beta$ is a constant. Therefore we generally find (with the aforementioned ``$-$'' sign)
\begin{equation}\label{eq:vol1E}
	\vol_1(E) = -2\beta C + \cO(e^{-E}) \ .
\end{equation}
and one can easily read off the constant $B_1$
\begin{equation}\label{eq:B1}
	B_1 = -\frac{\beta C}{\pi} \ .
\end{equation}
Finally, the coefficient function $A(\um,\hbar)$ is in general difficult to compute, although recently conjectures have 
been made for $A(\um,\hbar)$ in some special cases \cite{Hatsuda:2015oaa, hun}. On the other hand, later we will see in Section~\ref{ssc:genMass} that $A(\um,\hbar)$ does not enter into the quantization conditions, and furthermore it can be fixed by the normalization condition $Z(0,\hbar) = 1$.

Now we turn to the M2 brane instanton piece. It can be obtained from the instanton part of the refined topological string free energies in the Nekrasov--Shatashvili limit. We write the latter as
\begin{equation}
	\Fns(\bt, \hbar) = \sum_{j_L,j_R}\sum_{w,\bd} \BPSN \frac{\sin\tfrac{\hbar w}{2}(2j_L+1)\sin\tfrac{\hbar w}{2}(2j_R+1)}{2w^2\sin^3\tfrac{\hbar w}{2}} e^{-w\bd\cdot\bt} \ .
\end{equation}
Here $\bt$ is the vector of K\"{a}hler moduli, and $\bd$ the vector of degrees. We follow the convention of \cite{acdkv} and in contrast to the usual convention in the topological string literature, absorb a phase of $(-1)^{2j_L+2j_R}$ in $\BPSN$. We now introduce a variable $\lambda_s$ with
\begin{equation}\label{eq:lams}
	\lambda_s = \frac{2\pi}{\hbar} \ ,
\end{equation}
and a vector $\bT = \{T_\alpha\}$ with
\begin{equation}\label{eq:Ta}
	T_\alpha = \frac{2\pi}{\hbar} t_\alpha \ .
\end{equation}
The Nekrasov--Shatashvili free energy can be written as
\begin{equation}
	\Fns(\bt,\hbar) = \sum_{j_L,j_R}\sum_{w,\bd} \BPSN \frac{\sin\tfrac{\pi w}{\lambda_s}(2j_L+1)\sin\tfrac{\pi w}{\lambda_s}(2j_R+1)}{2w^2\sin^3\tfrac{\pi w}{\lambda_s}} e^{-w\bd\cdot\bT/\lambda_s} = \Fns\left(\frac{\bT}{\lambda_s},\frac{2\pi}{\lambda_s} \right) \ .
\end{equation}
Then $J_{\rm M2}(\tmu,\um,\hbar)$ is given by
\begin{equation}\label{eq:JM2}
	J_{\rm M2}(\tmu,\um,\hbar) = -\frac{1}{2\pi}\frac{\partial}{\partial \lambda_s} \left(\lambda_s \Fns\left(\frac{\bT}{\lambda_s},\frac{2\pi}{\lambda_s}\right)\right) \ .
\end{equation}

We still need to make the connection between $\tmu$ and $T_\alpha$ or $t_\alpha$. The flat coordinates $t_\alpha$ associated to the Batyrev coordinates $z_\alpha$ are related to the flat coordinate $t$ and the mass parameters by
\begin{equation}\label{eq:tat}
	t_\alpha = \tca t - \sum_j \alpha_{\alpha j} \log Q_{m_j} \ .
\end{equation}
Here $Q_{m_j}$ can be identified with the mass parameters $m_j$ in some geometries like local $\IF_0$, local $\IF_1$, and local $\CB_2$, but are rational functions of $m_j$ in some other geometries like local $\IF_2$ and the mass deformed local $E_8$ del Pezzo surface (see \cite{Huang:2014nwa} for more discussion on this distinction). For this reason, \eqref{eq:tat} is not a straightforward lift of \eqref{eq:zum}, although the exponent of $u$ in \eqref{eq:zum} can always be identified with the coefficient of $t$ in \eqref{eq:tat}. Now we relate $t_\alpha$ to $\tmu$ and the mass parameters by
\begin{equation}\label{eq:tamu}
	t_\alpha = \dca \tmu - \sum_j \alpha_{\alpha j} \log Q_{m_j} \ .
\end{equation}

With \eqref{eq:tamu} plugged in \eqref{eq:JM2}, and using \eqref{eq:lams} and \eqref{eq:Ta}, the M2 piece of the modified grand potential $J_{\rm M2}(\tmu,\um,\hbar)$ can be separated to two pieces
\begin{equation}\label{eq:JM2Split}
	J_{\rm M2}(\tmu,\um,\hbar) = \tmu \tJ_b(\tmu,\um,\hbar) + \tJ_c(\tmu,\um,\hbar) \ ,
\end{equation}
where
\begin{align}
	\tJ_b(\tmu,\um,\hbar) &= -\frac{1}{2\pi}\sum_{j_L,j_R}\sum_{w,\bd}(\bc\cdot \bd)\BPSN \frac{\sin\tfrac{\hbar w}{2}(2j_L+1)\sin\tfrac{\hbar w}{2}(2j_R+1)}{2w\sin^3\tfrac{\hbar w}{2}} e^{-w\bd\cdot\bt} \ , \label{eq:tJb}\\
	\tJ_c(\tmu,\um,\hbar) &=\frac{1}{2\pi}\sum_{\alpha,j}\sum_{j_L,j_R}\sum_{w,\bd}d_\alpha \alpha_{\alpha j}\log Q_{m_j} \BPSN \frac{\sin\tfrac{\hbar w}{2}(2j_L+1)\sin\tfrac{\hbar w}{2}(2j_R+1)}{2w\sin^3\tfrac{\hbar w}{2}} e^{-w\bd\cdot\bt} \nonumber\\
	&+\frac{1}{2\pi}\sum_{j_L,j_R}\sum_{w,\bd}\hbar^2\frac{\partial}{\partial \hbar}\left[ \frac{\sin\tfrac{\hbar w}{2}(2j_L+1)\sin\tfrac{\hbar w}{2}(2j_R+1)}{2\hbar w^2\sin^3\tfrac{\hbar w}{2}} \right] \BPSN e^{-w\bd\cdot \bt} \ .\label{eq:tJc}
\end{align}
Here $\mathbf{\dc}=\{\dca\}$ is the vector of the degrees of the Mori cone generators.

The last piece $J_{\rm WS}(\tmu,\um,\hbar)$ is related to the standard topological string free energies. We write the instanton part of the topological string free energy as
\begin{equation}
	\Ftop(\bt,g_s) = \sum_{j_L,j_R}\sum_{v,\bd}\BPSN \frac{(2j_R+1)\sin v g_s(2j_L+1)}{v(2\sin\tfrac{1}{2}v g_s)^2\sin v g_s} e^{-v\bd\cdot \bt} \ .
\end{equation}
Then the worldsheet instanton piece is given by
\begin{equation}
	J_{\rm WS}(\tmu,\um,\hbar) = \Ftop(\bT+\pi\ri\bB,2\pi\lambda_s) \ ,
\end{equation}
in other words
\begin{equation}\label{eq:JWS}
	J_{\rm WS}(\tmu,\um,\hbar) = \sum_{j_L,j_R}\sum_{v,\bd} \BPSN \frac{(2j_R+1)\sin \tfrac{4\pi^2 v}{\hbar}(2j_L+1)}{v(2\sin\tfrac{2\pi^2 v}{\hbar})^2\sin \tfrac{4\pi^2 v}{\hbar}} e^{-w\bd\cdot(\bT+\pi\ri\bB) } \ .
\end{equation}
It is crucial here to turn on the $B$--fields $\bB=\mathbf{\dc}$. It is easy to see from \eqref{eq:JM2}, \eqref{eq:tJb}, \eqref{eq:tJc} and \eqref{eq:JWS} that when $\hbar$ is $2\pi$ times a rational number, both $J_{\rm M2}(\tmu,\um,\hbar)$ and $J_{\rm WS}(\tmu,\um,\hbar)$ have poles. It was proved in \cite{ghm} as a direct generalization of \cite{hmmo} that these poles cancel against each other when $\bB=\mathbf{\dc}$, as in the HMO mechanism of pole cancellation in the ABJM model \cite{Hatsuda:2012dt}. For this pole cancellation mechanism to work, all nonzero BPS numbers $\BPSN$ have to satisfy
\begin{equation}\label{eq:HMO}
	2j_L+2j_R+1 \equiv \bd\cdot \bc \mod 2 \, ,
\end{equation}
which was proved in \cite{hmmo}.

Once $J_X(\mu,\um,\hbar)$ is given, the spectral determinant can be computed by
\begin{equation}\label{eq:XiJX}
	\Xi_X(\kappa,\um,\hbar) = \sum_{n\in\mathbb{Z}} e^{J_X(\mu+2\pi\ri n,\um, \hbar)} \ .
\end{equation}
Note that $J_X(\mu,\um,\hbar)$ differs from the genuine grand potential $\cJ_X(\mu,\um,\hbar)$ in that the former is not periodic in $\mu$. Nevertheless, the summation over the integral shift $n$ on the right hand side of \eqref{eq:XiJX} makes sure that $\Xi_X$ is still invariant under $\mu \mapsto\mu+2\pi\ri$, so that it is a well--defined function of $\kappa$.

The energy spectrum $\{E_n\}$ can be inferred from the spectral determinant. From its definition in \eqref{eq:sDet}, one can see that the zeros of $\Xi(\kappa,\um,\hbar)$ are given by
\begin{equation}
	\kappa = -e^{E_n} \ ,
\end{equation}
in other words
\begin{equation}
	\mu = E_n+\pi\ri \ .
\end{equation}
To find the zeros of $\Xi_X(\kappa,\um,\hbar)$ and thus the discrete energies $E_n$, we split the spectral determinant in two factors
\begin{equation}
	\Xi_X(\kappa,\um,\hbar) = e^{J_X(\mu,\um,\hbar)}\Theta_X(\mu,\um,\hbar) \ .
\end{equation}
Since the first factor is always positive, we can only find zeros in the second factor $\Theta_X(\mu,\um,\hbar)$. It has the form
\begin{align}
	\Theta_X(\mu,\um,\hbar) = \sum_{n\in\mathbb{Z}} &\exp\left[ -4\pi^2n^2\left(C(\hbar)\tmu+ \frac{D(\um,\hbar)}{2}\right)-\frac{8\pi^3\ri n^3}{3}C(\hbar) \right.\nonumber\\
	&+2\pi\ri n (C(\hbar)\tmu^2+D(\um,\hbar)\tmu+B(\um,\hbar)+ \tJ_b(\tmu,\um,\hbar))\nonumber\\
	&\left.+J_{\rm WS}(\tmu+2\pi\ri n,\um,\hbar)-J_{\rm WS}(\tmu,\um,\hbar)) \phantom{\frac{1}{1}} \hspace{-0.5em}\right] \ ,
\end{align}
and is called the \emph{generalized theta function} associated to $X$ \cite{ghm}. The reason for this name is that, when $\hbar=2\pi$, it becomes a conventional theta function. By analyzing when $\Theta_X(\mu,\um,\hbar)$ vanishes, concrete quantization conditions for the energy can be obtained, as we will explain in detail in Section~\ref{ssc:genMass}.

\begin{figure}
\centering
\includegraphics[width=0.18\linewidth]{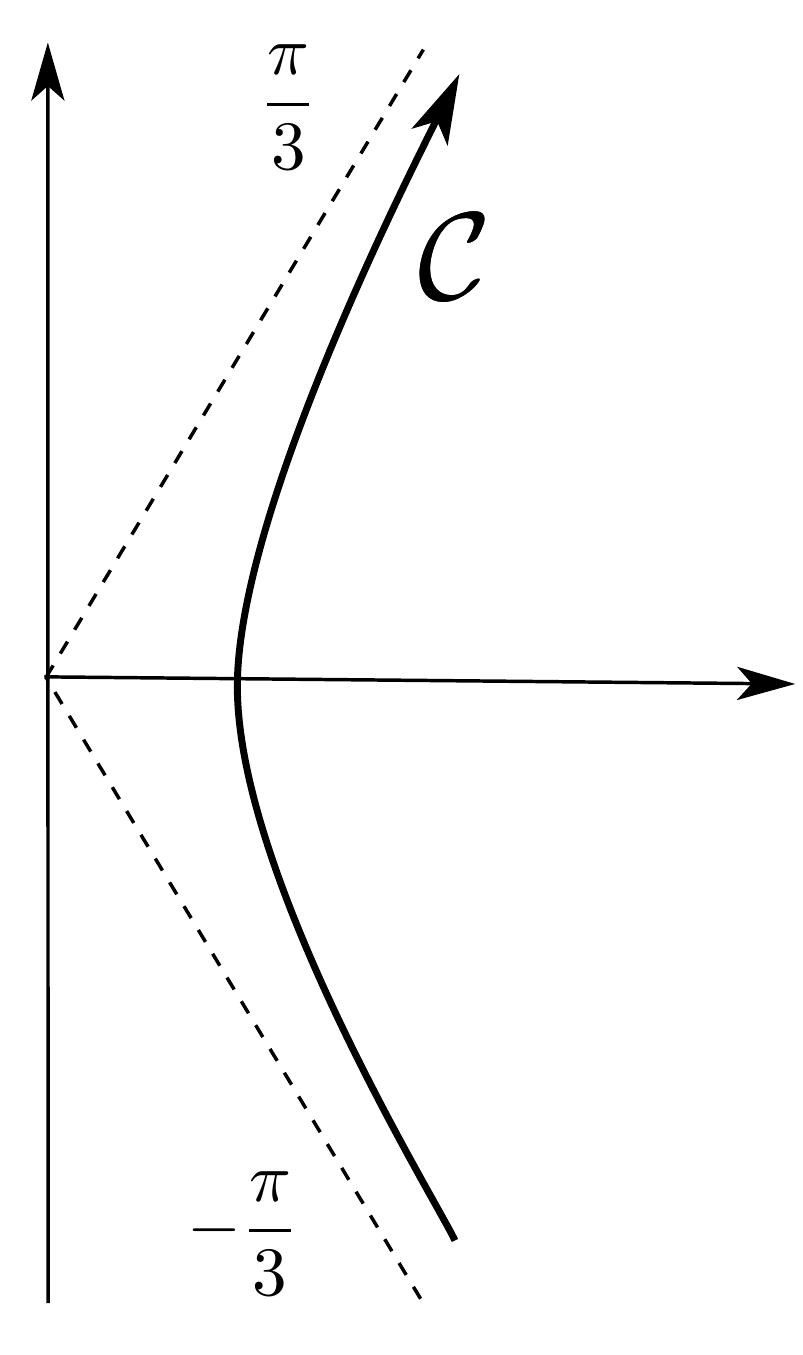}
\caption{The contour $\cC$ on the complex plane of $\mu$ in the integration of \eqref{eq:ZJXNew}. It is the same contour as the one in the integral definition of the Airy function.}
\label{airy-c}
\end{figure}

With the correct spectrum at hand, one can of course directly compute the fermionic spectral traces $Z(N,\um,\hbar)$ through the definition. However, one can compute them directly from $J_X(\mu,\um,\hbar)$ via a formula similar to \eqref{eq:ZJXold}. \eqref{eq:ZJXold} comes from taking residues of $\Xi_X(\kappa,\um,\hbar)$ at $\kappa=0$. Because of the sum over $n$ in \eqref{eq:XiJX}, when we replace $\cJ_X(\mu,\um,\hbar)$ by $J_X(\mu,\um,\hbar)$ in \eqref{eq:ZJXold}, the integral domain should be extended to infinity
\begin{equation}\label{eq:ZJXNew}
	Z_X(N,\um,\hbar) = \frac{1}{2\pi\ri} \int_\cC e^{J_X(\mu,\um,\hbar)-N\mu}\rd \mu \ .
\end{equation}
The integration path of the integral $\cC$ is chosen as in Fig.~\ref{airy-c} with the two ends asymptote to $e^{\pi\ri/3}\infty$ and $e^{-\pi\ri/3}\infty$ respectively so that the convergence of the integral is guaranteed. 

A third way to compute $Z(N,\um,\hbar)$ is to expand $\Xi_X(\kappa,\um,\hbar)$ around $\kappa=0$. Then the traces $Z(N,\um,\hbar)$ can just be read off as the series coefficients as in \eqref{eq:sDet}. Since the expansion is performed in the limit 
\begin{equation}
	\kappa \rightarrow 0, \quad \mu \rightarrow -\infty \ ,
\end{equation}
as seen in \eqref{eq:tamu}, we will need the expansion of the (refined) topological string free energies around the orbifold point.

\subsection{Generic mass parameters}
\label{ssc:genMass}

In \cite{ghm} the conjecture has been verified in some simple del Pezzo CYs for the cases where all mass parameters\footnote{To be precise the mass functions $Q_{m_j}$ are set to 1. But they coincide with $m_j$ in the examples studied in \cite{ghm}.}
are set to 1. In these cases, the formulae of the conjecture are greatly simplified. In particular, all the dependence on mass parameters drops out in the formulae. But by restricting mass parameters to one, it is difficult to probe the full scope of the conjecture. Furthermore, it is difficult to compare the results of \cite{ghm} with the results from operator analysis and matrix model computations in \cite{mz,Kashaev:2015wia}, where it is more natural to set all mass parameters to 0. It is the purpose of this paper to check the conjecture with arbitrary mass parameters, and 
for other examples of local del Pezzos beyond those considered in \cite{ghm}.

In the original conjecture, $J_{\rm M2}(\tmu,\um,\hbar)$ and $J_{\rm WS}(\tmu,\um,\hbar)$ are formulated in such a way that $\tmu$ and the mass parameters $m_j$ are treated on equal footings as in \eqref{eq:tamu}, and that the dependence on $\tmu$ and $m_j$ are realized in an indirect way through the variables $t_\alpha$ or $T_\alpha$. We would first like to reformulate $J_{\rm M2}(\tmu,\um,\hbar)$ and $J_{\rm WS}(\tmu,\um,\hbar)$ directly in terms of $\tmu, m_j$, and at the same time separate the different roles played by $\tmu$, the true modulus, and the $m_j$, the parameters of the system.


We introduce 
a function of mass parameters 
\begin{equation}
	\hQm(\bd) = \prod_j Q_{m_j}^{\sum_\alpha d_\alpha \alpha_{\alpha j}} \ .
\end{equation}
Then we find  that $\tJ_b(\tmu,\um,\hbar)$ and $\tJ_c(\tmu,\um,\hbar)$ can be written as 
\begin{align}
	\tJ_b(\tmu,\um,\hbar) &= \sum_{\ell} \tilde{b}_{\ell}(\um,\hbar) e^{-r\ell \tmu} \ , \label{eq:tJbNew}\\
	\tJ_c(\tmu,\um,\hbar) &= \sum_{\ell} \tilde{c}_\ell(\um,\hbar) e^{-r\ell \tmu} \ ,\label{eq:tJcNew}
\end{align}
where
\begin{equation}\label{eq:bl}
	\tilde{b}_{\ell}(\um,\hbar) = -\frac{r\ell}{4\pi} \sum_{j_L,j_R}\sum_{\ell=w \sum \tca d_\alpha} N^{\mathbf{d}}_{j_L,j_R} \frac{\sin \tfrac{\hbar w}{2}(2j_L+1) \sin \tfrac{\hbar w}{2}(2j_R+1)}{w^2 \sin^3 \tfrac{\hbar w}{2}} \hQm(\bd)^w \ , 
\end{equation}
\begin{equation}
\begin{aligned}
	\tilde{c}_\ell(\um,\hbar) = \frac{1}{2\pi} \sum_{j_L,j_R} \sum_{\ell=w\sum \tca d_\alpha} &N^{\mathbf{d}}_{j_L,j_R} \left\{ \log \hQm(\mathbf{d}) \frac{\sin \tfrac{\hbar w}{2}(2j_L+1) \sin \tfrac{\hbar w}{2}(2j_R+1) }{2w (\sin \tfrac{\hbar w}{2})^3}   \right. \\
	&\left. + \hbar^2 \frac{\partial}{\partial \hbar}\left[ \frac{\sin\tfrac{\hbar w}{2}(2j_L+1)\sin \tfrac{\hbar w}{2}(2j_R+1)  }{2\hbar w^2 (\sin\tfrac{\hbar w}{2})^3} \right] \right\} \hQm(\bd)^w \ .\label{eq:cl}
\end{aligned}
\end{equation}
In both $\tilde{b}_{\ell}(\um,\hbar)$ and $\tilde{c}_\ell(\um,\hbar)$, we have to sum over combinations of $w$ and $\{d_\alpha\}$ such that $\ell=w\sum \tca d_\alpha$ is satisfied. \ifthenelse{\equal{\CURVEDEGREE}{c}}{Do not confuse the $\tilde{c}_\ell(\um,\hbar)$ function defined here with the reduced curve degree $\tca$ defined in \eqref{eq:tcAlpha}.}{} Furthermore, $J_{\rm WS}(\tmu,\um,\hbar)$ can be written as 
\begin{equation}\label{eq:JWSNew}
	J_{\rm WS}(\tmu,\um,\hbar) = \sum_{m\geqslant 1} d_m(\um,\hbar) (-1)^{rm} e^{-2\pi rm \tmu/\hbar} \ ,
\end{equation}
where
\begin{equation}\label{eq:dm}
	d_m(\um,\hbar) = \sum_{j_L,j_R}\sum_{m=v\sum \tca d_\alpha} N^{\mathbf{d}}_{j_L,j_R} \frac{2j_R+1}{v\left(2\sin \tfrac{2\pi^2v}{\hbar}\right)^2} \frac{\sin \tfrac{4\pi^2v}{\hbar}(2j_L+1)}{\sin\tfrac{4\pi^2 v}{\hbar}}  \hQm(\bd)^{2\pi v/\hbar}\ .
\end{equation}

The reformulated $J_{\rm M2}(\tmu,\um,\hbar)$ and $J_{\rm WS}(\tmu,\um,\hbar)$ look very similar to their counterparts in \cite{ghm} where the dependence on the mass parameters is absent. The derivation of quantization conditions for energies then exactly parallels that in \cite{ghm}, and we just write down the final formulae here. 

Define the perturbative and non-perturbative quantum phase space volumes by
\begin{equation}\label{eq:qfVolumes}
\begin{aligned}
	\Omega_{\rm p}(E) &= C(\hbar)\tE^2+D(\um,\hbar) \tE + B(\um,\hbar) -\frac{\pi^2}{3}C(\hbar) + \tJ_b(\tE+\pi\ri,\hbar) \ ,\\
	\Omega_{\rm np}(E) &=-\frac{1}{\pi}\sum_{m\geqslant 1} d_m(\um,\hbar) \sin \frac{2\pi^2 rm}{\hbar} (-1)^{rm} e^{-2\pi rm\tE/\hbar} \ ,
\end{aligned}
\end{equation}
where $\tE$ is given by
\begin{equation}
	\tE = E - \frac{1}{r}\widetilde{\Pi}_A(e^{-rE},\um,\hbar) \ .
\end{equation}
Also define the auxiliary function $\lambda(E)$, which is the solution to
\begin{equation}\label{eq:gen_lambda}
\begin{aligned}
	\sum_{n=0}^{\infty} &e^{-4\pi^2n(n+1)(C(\hbar)\tE+D(\um,\hbar)/2)}(-1)^n e^{f_c(n)}\\
	&\phantom{==}\times\sin\left(\frac{4\pi^3n(n+1)(2n+1)}{3}C(\hbar)+f_s(n)+2\pi(n+1/2)\lambda(E))\right) = 0\ .
\end{aligned}
\end{equation}
In this equation we need $f_c(n)$ and $f_s(n)$, which are defined as
\begin{equation}
\begin{aligned}
	f_c(n) &= \sum_{m\geqslant 1} (-1)^{rm} d_m(\um,\hbar)\left(\cos\left( \frac{2\pi^2r m (2n+1)}{\hbar} \right)   -\cos\left( \frac{2\pi^2 r m}{\hbar} \right)\right)e^{-2\pi r m \tE/\hbar} \ , \\
	f_s(n) &= \sum_{m\geqslant 1} (-1)^{rm} d_m(\um,\hbar)\left(\sin\left( \frac{2\pi^2r m (2n+1)}{\hbar} \right)   -(2n+1)\sin\left( \frac{2\pi^2 r m}{\hbar} \right)\right)e^{-2\pi r m \tE/\hbar} \ .
\end{aligned}  
\end{equation}
Then the quantization condition is
\begin{equation}\label{eq:gen_qnz}
	\Omega_{\rm p}(E) + \Omega_{\rm np}(E) +\lambda(E) = s+\frac{1}{2} \ ,\quad s =0,1,2,\ldots  \ .
\end{equation}
Note that $A(\um,\hbar)$ does not enter the quantization condition. Although the above formulae look complicated, they are just obtained by requiring the vanishing of the 
spectral determinant, and in particular of the generalized theta function. It has been recently noted in \cite{wzh} that these conditions are equivalent to a simpler quantization 
condition involving only the NS refined free energy. The equivalence of the two conditions, the one above and the one in \cite{wzh}, leads to a non-trivial 
equivalence between the standard topological string free energy and the NS refined free energy. 

To calculate the fermionic spectral traces $Z(N,\um,\hbar)$ from \eqref{eq:ZJXNew} we note that $e^{J_X(\mu,\hbar)}$ appearing in the integrand of \eqref{eq:ZJXNew} always has the following expansion
\begin{equation}
	e^{J_X(\mu,\um,\hbar)} = e^{J^{\rm (p)}(\mu,\um,\hbar)} \sum_{\ell'>0} e^{-r\ell'\mu} \sum_{n=0}^{n_{\rm top}(\ell')}a_{\ell',n}(\um) \mu^n \ .
\end{equation}
Note here that the argument of $J^{\rm (p)}(\ldots)$ is $\mu$ instead of $\tmu$, i.e., we collect all the exponentially small corrections, including those originating from $\tmu$, in the double summation. The index $\ell'$ is not necessarily an integer, but any number which can be decomposed as
\begin{equation}
	\ell' = \ell + \frac{2\pi m}{\hbar} \ , \quad \ell,m\in \bZ_{\geqslant 0} \ .
\end{equation}
For a given $\ell'$, the integral index $n$ has an upper bound $n_{\rm top}(\ell')$, which depends on $\ell'$. If one can extract the coefficients $a_{\ell',n}(\um)$, the integral \eqref{eq:ZJXNew} can be rewritten as a sum of Airy functions ${\rm Ai}(z)$ and its derivatives
\begin{align}
	&Z(N,\um,\hbar) = \frac{1}{C(\hbar)^{1/3}} \exp\left( A(\um,\hbar)+\tfrac{D(\um,2\pi)}{2C(\hbar)}(N-B(\um,\hbar))+\tfrac{D(\um,\hbar)^3}{12C(\hbar)^2} \right)\times \nonumber\\
	  &\sum_{\ell'>0}\sum_{n=0}^{n_{\rm top}(\ell')} e^{\tfrac{D(\um,\hbar)}{2C(\hbar)}\,r\ell'} a_{\ell',n}(\um)\left(-\frac{\partial}{\partial N}-\frac{D(\um,\hbar)}{2C(\hbar)}\right)^n \textrm{Ai}\left(\frac{r\ell'+N-B(\um,\hbar)+\tfrac{D(\um,\hbar)^2}{4C(\hbar)}}{C(\hbar)^{1/3}}\right) \ .		\label{eq:ZNAi}
\end{align}
This formula is well--defined for $N=0$. Therefore, we can additionally use it to fix the value of $A(\um,\hbar)$ by 
the normalization condition $Z(0,\um,\hbar) = 1$, which is demanded by the definition of $Z(N,\um,\hbar)$.

\subsubsection{Rational Planck constants}

We will check our conjecture later in Section~\ref{sec:examples} for examples when the (reduced) Planck constant $\hbar$ is
\begin{equation}\label{eq:ratPlanck}
	\hbar = 2\pi\, \frac{p}{q} \ ,
\end{equation}
where $p, q$ are coprime positive integers. These are the cases when the pole cancellation mentioned in Section~\ref{ssc:conjecture} plays an important role. We call Planck constants of this type \emph{rational}. 

When $\hbar$ is rational, $\tJ_b(\tmu,\um,\hbar)$ and $\tJ_c(\tmu,\um,\hbar)$ have poles when the index $w$ in $\tilde{b}_\ell(\um,\hbar)$ and $\tilde{c}_\ell(\um,\hbar)$ is divisible by $q$, and $\tJ_{\rm WS}(\tmu,\um,\hbar)$ has poles when the index $v$ in $d_m(\um,\hbar)$ is divisible by $p$. Recall that 
\[
\begin{gathered}
	\tilde{b}_\ell(\um,\hbar) = \sum_{j_L,j_R}\sum_{w|\ell} \sum_{\sum \tca d_\alpha=\ell/w} \ldots \ ,\quad
	\tilde{c}_\ell(\um,\hbar) = \sum_{j_L,j_R}\sum_{w|\ell} \sum_{\sum \tca d_\alpha=\ell/w} \ldots \ ,\\
	d_m(\um,\hbar) = \sum_{j_L,j_R}\sum_{v|m} \sum_{\sum\tca d_\alpha = m/v} \ldots \ .
\end{gathered}
\]
We separate them by
\begin{align}
	\tilde{b}_\ell(\um,\hbar) &= \tilde{b}^{(0)}_\ell(\um,\hbar) + \tilde{b}^{(f)}_\ell(\um,\hbar) \ ,\label{eq:bSplit}\\
	\tilde{c}_\ell(\um,\hbar) &= \tilde{c}^{(0)}_\ell(\um,\hbar) + \tilde{c}^{(f)}_\ell(\um,\hbar) \ ,\label{eq:cSplit}\\
	d_m(\um,\hbar) &= d_m^{(0)}(\um,\hbar) + d_m^{(f)}(\um,\hbar) \ . \label{eq:dSplit}
\end{align}
according to
\begin{align}
	\tilde{b}_\ell^{(0)}(\um,\hbar) &= \sum_{j_L,j_R}\sum_{\substack{w|\ell \\ q| w}} \sum_{\sum \tca d_\alpha=\ell/w}\ldots ,\quad
	\tilde{b}_\ell^{(f)}(\um,\hbar) = \sum_{j_L,j_R}\sum_{\substack{w|\ell \\ q\nmid w}} \sum_{\sum \tca d_\alpha=\ell/w}\ldots ; \label{eq:b0bf} \\
	\tilde{c}_\ell^{(0)}(\um,\hbar) &= \sum_{j_L,j_R}\sum_{\substack{w|\ell \\ q| w}} \sum_{\sum \tca d_\alpha=\ell/w}\ldots ,\quad
	\tilde{c}_\ell^{(f)}(\um,\hbar) = \sum_{j_L,j_R}\sum_{\substack{w|\ell \\ q\nmid w}} \sum_{\sum \tca d_\alpha=\ell/w}\ldots ; \label{eq:c0bf} \\	
	d_m^{(0)}(\um,\hbar) &= \sum_{j_L,j_R}\sum_{\substack{v|m \\ p|v }} \sum_{\sum\tca d_\alpha = m/v}\ldots , \quad
	d_m^{(f)}(\um,\hbar) = \sum_{j_L,j_R}\sum_{\substack{v|m \\ p\nmid v}} \sum_{\sum\tca d_\alpha = m/v}\ldots .\label{eq:d0df}
\end{align}
We can split $\tJ_b(\tmu,\um,\hbar)$ to the singular summands and the regular summands
\begin{equation}
\begin{aligned}
	\tJ_b(\tmu,\um,\hbar) &= \sum_{\ell\geqslant 0} \tilde{b}^{(0)}_\ell (\um,\hbar) e^{-r \ell \tmu} + \sum_{\ell\geqslant 0} \tilde{b}^{(f)}_\ell (\um,\hbar) e^{-r \ell \tmu} \\
	&\equiv \tJ^{(0)}_b(\tmu,\um,\hbar) + \tJ^{(f)}_b(\tmu,\um,\hbar) \ ,\label{eq:tJbSplit}\\
\end{aligned}
\end{equation}
with the help of \eqref{eq:bSplit}. Similarly we can split $\tJ_c(\tmu,\um,\hbar)$, and $\tJ_{\rm WS}(\tmu,\um,\hbar)$ 
\begin{equation} \label{eq:tJcJWSSplit}
\begin{aligned}
	\tJ_c(\tmu,\um,\hbar) &= \tJ^{(0)}_c(\tmu,\um,\hbar) + \tJ^{(f)}_c(\tmu,\um,\hbar) \ ,\\
	J_{\rm WS}(\tmu,\um,\hbar) &= J^{(0)}_{\rm WS}(\tmu,\um,\hbar) + J^{(f)}_{\rm WS}(\tmu,\um,\hbar) 
\end{aligned}
\end{equation}
in the same spirit. 

Furthermore, for a function $f^{(0)}(\hbar)$ singular at $\hbar = 2\pi p/q$, we perturb $\hbar$ slightly away from its rational value
\begin{equation}
	\hbar = \frac{2\pi p}{q} + \epsilon \ ,
\end{equation}
and denote the principal part and the finite part of $f^{(0)}(\hbar)$ by
\begin{equation}
	\{ f^{(0)}(2\pi p/q) \} \ , \quad [f^{(0)}(2\pi p/q)]
\end{equation}
respectively. It can be checked that the poles in $J_X(\mu,\um,\hbar)$ cancel, i.e., 
\begin{equation}
	\tmu \{\tJ_b^{(0)}(\tmu,\um,\hbar)\} + \{\tJ_c^{(0)}(\tmu,\um,\hbar)\}+\{J_{\rm WS}^{(0)}(\tmu,\um,\hbar)\} = 0 \ ,
\end{equation}
if and only if the condition \eqref{eq:HMO} is satisfied. Furthermore, one finds that
\begin{align}
	[\tJ_b^{(0)}(\tmu,\um,\hbar)] =& 0 \ , \\
	[\tJ_c^{(0)}(\tmu,\um,\hbar)] =& - \sum_{j_L,j_R}\sum_{k,\bd} (-1)^{rdkp} \BPSN \frac{(1+2j_L)(1+2j_R)}{24kq^2} \nonumber\\
 &\phantom{\sum\sum}\times p (-1+4j_L+4j_L^2 +4j_R+4j_R^2) \hQm(\bd)^{kq} e^{-rdkq\tmu} \ ,\label{eq:tJc0}	\\
 	[J_{\rm WS}^{(0)}(\tmu,\um,\hbar)] =& - \sum_{j_L,j_R}\sum_{k,\bd} (-1)^{rdkp} \BPSN \frac{(1+2j_L)(1+2j_R)}{24k^3 p q^2\pi^2}  \Big\{ -3(rdkq)^2\tmu^2 \nonumber\\
&\phantom{\sum\sum} + 6(rdkq)\tmu (kq \log \hQm(\bd)-1) -3-3(kq\log \hQm(\bd)-1)^2 \nonumber\\
&\phantom{\sum\sum} + 2k^2q^2\pi^2(-1+8j_L+8j_L^2) \Big\} \hQm(\bd) e^{-rdkq\tmu} \ .\label{eq:Jws0}
\end{align}
Incidentally, let $F_1^{\rm NS,inst}(t,\um)$ be the instanton part of the genus one Nekrasov--Shatashvili limit topological string free energy, and $F_1^{\rm inst}(t,\um)$, $F_0^{\rm inst}(t,\um)$ be the instanton parts of genus one and genus zero unrefined topological string free energies, respectively. They have the following expansion
\begin{equation}\label{eq:F01instBPS}
\begin{aligned}
	F_0^{\rm inst}(t,\um) = \sum_{j_L,j_R}\sum_{w,\bd}\BPSN &\frac{(1+2j_L)(1+2j_R)}{w^3} \hQm(\bd)^w e^{-w d t}  \ ,\\
	F_1^{\rm inst}(t,\um) = \sum_{j_L,j_R}\sum_{w,\bd}\BPSN &\frac{(1+2j_L)(1+2j_R)}{12 w} (-1+8j_L+8j_R)\hQm(\bd)^w e^{-w d t}  \ ,\\
	F_1^{\rm NS, inst}(t,\um) = \sum_{j_L,j_R}\sum_{w,\bd}\BPSN &\frac{(1+2j_L)(1+2j_R)}{24 w}\nonumber\\
	&\times (-1+4j_L+4j_L^2+4j_R+4j_R^2)\hQm(\bd)^w e^{-w d t}  \ .
\end{aligned}
\end{equation}
Then it can be shown that
\begin{align}
	[\tJ_c^{(0)}(\tmu,\um,\hbar)] =& \frac{p}{q^2}F_1^{\rm NS,inst}(t-\ri r p\pi,\um) \Big|_{\substack{Q_{m_j}\rightarrow Q_{m_j}^q \\t\rightarrow rq\tmu }} \ , \label{eq:tJc0vonF}\\
	[J_{\rm WS}^{(0)}(\tmu,\um,\hbar)] =& \frac{1}{p} F_1^{\rm inst}(t-\ri r p \pi,\um) + \frac{1}{pq^2}\left(\frac{1}{4\pi^2} -\frac{1}{4\pi^2}\left(t\frac{\partial}{\partial t}+\log Q_{m_k}\frac{\partial}{\partial \log Q_{m_k}} \right) \right. \nonumber \\
	&  +\frac{1}{8\pi^2}\left( t^2\frac{\partial^2}{\partial t^2}+ 2t \log Q_{m_k} \frac{\partial^2}{\partial \log Q_{m_k} \partial t} \right. \nonumber \\
	&\left.\left.+\log Q_{m_k} \log Q_{m_l} \frac{\partial^2}{\partial \log Q_{m_k} \partial \log Q_{m_l}} \right)\right) F_0^{\rm inst}(t-\ri r p\pi ,\um) \Big|_{\substack{Q_{m_j}\rightarrow Q_{m_j}^q \\ t\rightarrow rq\tmu}} \ .\label{eq:Jws0vonF}
\end{align}

In summary, when the Planck constant is rational, we can compute the modified grand potential by
\begin{align}
	J_X(\mu,\hbar) = \frac{C(\hbar)}{3}\tmu^3 &+ \frac{D(\um,\hbar)}{2}\tmu^2 + B(\um,\hbar)\tmu+A(\um,\hbar) \nonumber\\
	&+ \tmu \tJ_b^{(f)}(\tmu,\um,\hbar) + \tJ_c^{(f)}(\tmu,\um,\hbar) + J_{\rm WS}^{(f)}(\tmu,\um,\hbar) \nonumber\\
	&+ [\tJ_c^{(0)}(\tmu,\um,\hbar)] + [J_{\rm WS}^{(0)}(\tmu,\um,\hbar)] \ ,
\end{align}
where $[\tJ_c^{(0)}(\tmu,\um,\hbar)]$ and $[J_{\rm WS}^{(0)}(\tmu,\um,\hbar)]$ are either given by \eqref{eq:tJc0} and \eqref{eq:Jws0} or by \eqref{eq:tJc0vonF} and \eqref{eq:Jws0vonF}, while $\tJ_b^{(f)}(\tmu,\um,\hbar)$, $\tJ_c^{(f)}(\tmu,\um,\hbar)$, and $J_{\rm WS}^{(f)}(\tmu,\um,\hbar)$ are defined through the decomposition in \eqref{eq:bSplit}--\eqref{eq:d0df}.

Let us also take a look at the quantization condition \eqref{eq:gen_qnz}, together with \eqref{eq:qfVolumes} and \eqref{eq:gen_lambda}, when $\hbar$ is rational. Other than $\tJ_b(\tE+\pi\ri,\hbar)$, $\Omega_{\rm np}(E)$, $f_s(n)$, and  $f_c(n)$ may also develop poles because of the coefficient function $d_m(\um,\hbar)$. Similar to \eqref{eq:tJbSplit} and \eqref{eq:tJcJWSSplit}, we split them according to \eqref{eq:dSplit}
\begin{equation}
\begin{aligned}
	\Omega_{\rm np}(E) &= \Omega^{(0)}_{\rm np}(E)+\Omega^{(f)}_{\rm np}(E)\ , \\
	f_c(n) &= f_c^{(0)}(n) + f_c^{(f)}(n) \ ,\\
	f_s(n) &= f_s^{(0)}(n) + f_s^{(f)}(n) \ ,
\end{aligned}
\end{equation}
insulating the poles in the pieces with superscript $(0)$, and then further decomposing the latter to singular components $\{\ldots\}$ and  finite components $[\ldots]$. It turns out reassuringly that $\{f_s^{(0)}(n)\}$ and $\{f_c^{(0)}(n)\}$ vanish, while $\{\tJ_b^{(0)}(\tE+\pi\ri,\hbar)\}$ and $\{\Omega^{(0)}_{\rm np}(E)\}$ cancel against each other when the condition \eqref{eq:HMO} is satisfied.

Furthermore we find
\begin{align}
&[\Omega^{(0)}_{\rm np}(E)] = \sum_{j_L,j_R}\sum_{k\geqslant 1,\mathbf{d}}(-1)^{rdk(p+q)} N^{\mathbf{d}}_{j_L,j_R} \nonumber\\
&\hspace{5em} \times\frac{(1+2j_L)(1+2j_R)(rd+r^2d^2kq \tE-rdkq \log \hQm(\mathbf{d}))}{4k^2 pq\pi^2} \hQm(\mathbf{d})^{kq} e^{-rdkq \tE} \ ,\label{eq:Omgnp0}\\
&[f_c^{(0)}(n)] = -\sum_{j_L,j_R}\sum_{k\geqslant 1,\mathbf{d}} (-1)^{rdk(p+q)} N^{\mathbf{d}}_{j_L,j_R} \nonumber\\
&\hspace{5em} \times\frac{(1+2j_L)(1+2j_R)r^2d^2n(1+n)}{2kp}  \hQm(\mathbf{d})^{kq} e^{-rdkq\tE}   \ ,\label{eq:fc0}\\
&[f^{(0)}_s(n)] = 0 \ .
\end{align}
Making use of \eqref{eq:F01instBPS}, we find that both $[\Omega^{(0)}_{\rm np}(E)]$ and $[f_c^{(0)}(n)]$ can be expressed in terms of the prepotential only
\begin{align}
	&[\Omega^{(0)}_{\rm np}(E)] = \frac{1}{4pq\pi^2} \left( -r\frac{\partial}{\partial t}+r^2 q \tE \frac{\partial^2}{\partial t^2} + r \sum_{k} \log Q_{m_k} \frac{\partial^2}{\partial \log Q_{m_k} \partial t} \right) F_0 \Big|_{\substack{Q_{m_j}\rightarrow Q_{m_j}^q \\ t\rightarrow rq \tE-r(p+q)\pi\ri}} \ ,\\
	&[f_c^{(0)}(n)] = -\frac{n(1+n)}{2p} r^2\frac{\partial^2}{\partial t^2} F_0 \Big|_{\substack{Q_{m_j}\rightarrow Q_{m_j}^q \\ t\rightarrow rq \tE-r(p+q)\pi\ri}} \ .
\end{align}

Therefore, when $\hbar$ is rational, we shall do the following replacement in the generic quantization conditions
\begin{equation}\label{eq:rat_h_qnz}
\begin{array}{c c l}
\tJ_b(\tE+\pi\ri,\hbar) & \longmapsto & \tJ^{(f)}_b(\tE+\pi\ri,\hbar) \\
\Omega_{\rm np}(E) & \longmapsto & [\Omega_{\rm np}^{(0)}(E)] + \Omega_{\rm np}^{(f)}(E) \\
f_c(n) & \longmapsto & [f_c^{(0)}(n)] + f_c^{(f)}(n)\\
f_s(n) & \longmapsto & f_s^{(f)}(n)
\end{array}
\end{equation}
where  $[\Omega_{\rm np}^{(0)}(E)]$ and $[f_c^{(0)}(n)]$ are given by \eqref{eq:Omgnp0} and \eqref{eq:fc0}.

\subsubsection{Maximal supersymmetry}

As emphasized in \cite{ghm}, the formulae of the conjecture become the simplest in the case of maximal supersymmetry when $\hbar = 2\pi$. In the ABJM theory analog, this is the scenario when the supersymmetry is enhanced from $\cN=6$ to $\cN=8$, hence the name ``maximal supersymmetry''. Note that this is a special case of rational $\hbar$ of \eqref{eq:ratPlanck}, where $p=q=1$.

In this special case, the quantum A--period in the definition of the effective chemical potential \eqref{eq:mueff} is reduced to the classical A--period in the unrefined topological string. Besides, the components of $\tJ_b$, $\tJ_c$, $J_{\rm WS}$ with superscript $(f)$ vanish, because the indices $w$ and $v$ are always divisible by $p=q=1$, while the remaining nonvanishing components $[\tJ_c(\tmu,\um,2\pi)]$, and $[J_{\rm WS}(\tmu,\um,2\pi)]$, as seen from \eqref{eq:tJc0vonF} and \eqref{eq:Jws0vonF}, only depend on genus 0 and genus 1 (refined) topological string free energies. Therefore it is possible to study $J_X(\mu,\um,\hbar)$ in different corners of the moduli space. In particular, we can expand $\Xi_X(\kappa,\um,\hbar)$ around $\kappa=0$ to compute $Z(N,\um,2\pi)$, as mentioned in the end of Section~\ref{ssc:conjecture}, by performing an analytic continuation of genus zero and genus one free energies to the orbifold point.

Let us first write down the modified grand potential. It has the form
\begin{align}
	J_X(\mu,\um, 2\pi) &= \frac{C(2\pi)}{3r^3}t^3+\frac{D(\um,2\pi)}{2r^2}t^2 + \frac{B(\um,2\pi)}{r} t + A(\um,2\pi) \nonumber\\
	  &+F_1^\textrm{inst}(t-r\pi\ri,\um) + F_1^{\rm NS,inst}(t-r\pi\ri,\um)+ \frac{1}{4\pi^2}F_0^{\rm inst}(t-r\pi\ri,\um)  \nonumber \\
	&+\frac{1}{8\pi^2} \left( t^2 \frac{\partial^2 }{\partial t^2} + 2t \frac{\log Q_{m_k}\partial^2}{\partial \log Q_{m_k} \partial t} +  \frac{\log Q_{m_k} \log Q_{m_l}\partial^2}{\partial \log Q_{m_k} \partial \log Q_{m_l}} \right) F_0^{\rm inst}(t-r\pi\ri,\um) \nonumber \\
	&-\frac{1}{4\pi^2} \left( t\frac{\partial}{\partial t} +  \frac{\log Q_{m_k}\partial}{\partial \log Q_{m_k}}\right)  F_0^{\rm inst}(t-r\pi\ri,\um) \Big|_{t\rightarrow r\tmu}\ , \label{eq:JXgeneric}
\end{align}
where the Einstein notation is used. To write it in a more compact form we split $B_0(\um)$ defined in \eqref{eq:Bh} to two pieces
\begin{equation}
	B_0(\um) = \frac{B_0^{(m)}(\um)}{2\pi} + B_0' \ ,
\end{equation}
where $B_0^{(m)}(\um)$ is a function of the mass parameters which vanishes when $Q_{m_j}=1$, and $B_0'$ is the remaining constant. Let us also define
\begin{equation}
	B'(\hbar) = \frac{B_0'}{\hbar} + B_1\hbar \ .
\end{equation}
Then we find that the full prepotential has the following form
\begin{equation}\label{eq:F0full}
	F_0(t,\um) = \frac{C}{3r^3} t^3 + \frac{D_0(\um)}{2r^2} t^2 + \frac{B_0^{(m)}(\um)}{r}t + F_0^{\rm inst}(t,\um) \ .
\end{equation}
Here the classical piece
\begin{equation}
	F_0^{\rm cls}(t,\um) = \frac{C}{3r^3} t^3 + \frac{D_0(\um)}{2r^2} t^2 + \frac{B_0^{(m)}(\um)}{r}t
\end{equation}
consist of Yukawa coupling terms, and therefore $D_0(\um)$ has to be a linear function of the flat coordinates $Q_{m_j}$ associated to the mass parameters, and $B_0^{(m)}(\um)$ a homogeneous function in $Q_{m_j}$ of degree two. Let us define the skewed prepotential $\tF_0(t,\um)$
\begin{equation}\label{eq:F0full_skew}
	\tF_0(t,\um) = \frac{C}{3r^3} t^3 + \frac{D_0(\um)}{2r^2} t^2 + \frac{B_0^{(m)}(\um)}{r}t + F_0^{\rm inst}(t-r\pi\ri,\um) \ .	
\end{equation}
Then using the aforementioned properties of $D_0(\um)$ and $B_0^{(m)}(\um)$, we find
\begin{align} 
	J_X(\mu,\um,2\pi) = A(2\pi) &+\frac{B'(2\pi)}{r}t+\frac{1}{8\pi^2}(D_t^2-3D_t+2)\tF_0(t,\um)\nonumber\\
	&+F_1^{\rm inst}(t-r\pi\ri,\um)+F_1^{\rm NS,inst}(t-r\pi\ri,\um)\big|_{t\rightarrow r\tmu} \ ,
\end{align}
where 
\begin{equation}
	D_t = t\frac{\partial}{\partial t} + \log Q_{m_k}\frac{\partial}{\partial \log Q_{m_k}} \ .
\end{equation}

The generalized theta function $\Theta_X(\mu, \um, 2\pi)$ has then the following compact expression
\begin{equation}
		\Theta_X(\mu, \um,2\pi) = \sum_{n\in\mathbb{Z}} \exp\left( \pi i n^2 \tau + 2\pi i n (\xi+B'(2\pi)) -\frac{2\pi i}{3}n^3 C \right) \ ,\label{equ:GeneralizedTheta}
\end{equation}
where
\begin{align}
	\tau  &= \frac{r^2}{4} \, \frac{2 i}{\pi} \frac{\partial^2}{\partial t^2} \tF_0(t,\um)\Big|_{t\rightarrow r\tmu}  \ , \label{eq:tauSim}\\
	\xi &= \frac{r}{4\pi^2} \left( t \frac{\partial^2}{\partial t^2} - \frac{\partial}{\partial t} + \log Q_{m_k} \frac{\partial^2}{\partial \log Q_{m_k}\partial t}  \right) \tF_0(t,\um)\Big|_{t\rightarrow r\tmu} \ . \label{eq:xiSim}
\end{align}
For those geometries whose $r$ is even so that $\tF_0$ coincides with $F_0$, $\tau$ is proportional to the elliptic modulus $\tau_0$ of the elliptic spectral curve $\cC_X$, since the latter is given by
\begin{equation}
	\tau_0 = -2\pi\ri\frac{\partial^2}{\partial t^2} F_0(t,\um) \ .
\end{equation}
As pointed out in \cite{ghm}, when $C$ is an integer or half--integer, $\Theta_X(\mu, \um,2\pi)$ is a conventional theta function, because
\[
	-\frac{2\pi\ri}{3}n^3 C = -\frac{2\pi\ri}{3} n C+ \frac{2\pi\ri}{3}n(n-1)(n+1) C \ ,
\]
where the last term is an integral multiple of $2\pi\ri$.

Finally, the quantization condition in the maximally supersymmetric case can be written as
\begin{align}
4\pi^2\left(s+\frac{1}{2}\right)=& C\tE^2 + D_0(\um) \tE + 4\pi^2 B(\um,2\pi) -\frac{\pi^2 C}{3} \label{eq:mSUSYqnz}\\
&+\left(-r \frac{\partial }{\partial t} + r^2 \tE \frac{\partial^2 }{\partial t^2} + r \log Q_{m_k} \frac{\partial^2 }{\partial \log Q_{m_k} \partial t} \right)F_0^{\rm inst}(t)\Big|_{t\rightarrow r\tE} \nonumber\ ,
\end{align}
with $s=0,1,\ldots$.

\subsection{Spectral traces and matrix models}
\label{st-mm}

In order to test the conjectural relation between spectral theory and topological strings, it is important to have as much information as possible on the operators 
$\rho_X$ obtained from the quantization of the spectral curves. In some simple cases, like the three-term operators (\ref{omn}), 
it was shown in \cite{km,Kashaev:2015wia} that one can compute the integral kernels of 
the $\rho_X$. This makes it also possible to write matrix integral representations for the fermionic spectral traces. We will review some of theses results here, as they will be 
used in the examples worked out in this paper. 

Let us consider the three-term operator (\ref{omn}). Note that $m, n$ can be {\it a priori} arbitrary positive, real numbers, although in 
the operators arising from the 
quantization of mirror curves they are integers. Let $ \fad(x)$ be Faddeev's quantum dilogarithm \cite{Faddeev:1995nb,Faddeev:1993rs} (for this function, 
we follow the conventions of \cite{km,Kashaev:2015wia}). We define as well 
\be
\label{mypsi-def}
\mypsi{a}{c}(x)= \frac{\re^{2\pi ax}}{\fad(x-\im(a+c))} \, , 
\ee
It was proved in \cite{km} that the operator 
\be
\rho_{m,n}=\mO_{m,n}^{-1}
\ee
 is positive-definite and of trace class. There is in addition a pair of operators $\mq$, $\map$, satisfying the normalized 
Heisenberg commutation relation
\be\label{nHeis}
[\map, \mq]=(2 \pi \ri)^{-1} \, . 
\ee
They are related to the Heisenberg operators $\mx$, $\my$ appearing in $\mO_{m,n}$ by the following linear canonical transformation: 
\begin{equation}
\mathsf{x}\equiv 2\pi\mathsf{b}\frac{(n+1)\mathsf{p}+n\mathsf{q}}{m+n+1} \, ,\quad \mathsf{y}\equiv -2\pi\mathsf{b}\frac{m\mathsf{p}+(m+1)\mathsf{q}}{m+n+1} \, , 
\end{equation}
so that $\hbar$ is related to $\mb$ by
\be
\label{b-hbar}
\hbar=\frac{2\pi\mathsf{b}^2}{m+n+1} \, . 
\ee
Then, in the momentum representation associated to $\map$, 
the operator $\rho_{m,n}$ has the integral kernel, 
\begin{equation}
\label{ex-k}
\rho_{m,n}(p,p')=\frac{\overline{\mypsi{a}{c}(p)}\mypsi{a}{c}(p')}{2\mathsf{b}\cosh\left({ \pi \over \mb} \left( p-p'+\ri C_{m,n} \right) \right)} \, ,
\end{equation}
where $a$, $c$ are given by 
\be
a =\frac{m \mb}{2(m+n+1)} \, , \qquad c=\frac{\mb}{2(m+n+1)} \, ,  
\ee
and 
\be
C_{m,n}=\frac{m-n+1}{2(m+n+1)} \, .
\ee
Once the trace class property has been established for the operators $\rho_{m,n}$, it can be easily established for operators $\rho_S$ whose inverse 
$\mO_S$ are perturbations of $\mO_{m,n}$ by a positive self-adjoint operator \cite{km}. This proves the trace class property for a large number of operators 
obtained through the quantization of mirror curves. This includes all the operators arising from the del Pezzo surfaces, except for the operator for local $\IF_0$. 
However, this operator can be also seen to be of trace class, and its kernel can be also 
computed explicitly \cite{km,Kashaev:2015wia}. The quantization of the curve for local $\IF_0$ leads to the operator
 \begin{equation}\label{f0}
\mathsf{O}_{\mathbb{F}_0}=\re^{\mathsf{x}} +m_{\IF_0} \re^{-\mathsf{x}} +\re^{\mathsf{y}} +\re^{-\mathsf{y}} \, .
\ee
Let us set
\begin{equation}
\label{hb-rel}
\hbar=\pi\mathsf{b}^2, \qquad  m_{\IF_0}=\re^{2\pi\mathsf{b}\mu} \, .
\end{equation}
Then, there are normalized Heisenberg operators $\map$, $\mq$ satisfying (\ref{nHeis}), related to $\mx$, $\my$ in (\ref{f0}) by a linear canonical 
transformation,  such that, 
\begin{equation}
\label{f0ker}
\rho_{\IF_0}(q_1, q_2)=\langle q_1| \mathsf{O}^{-1}_{\mathbb{F}_0}|q_2\rangle=
\re^{-\pi \mathsf{b} \mu/2}  {f(q_1) f^*(q_2) \over  2\mathsf{b} \cosh\left(\pi {q_1- q_2 \over \mathsf{b}}  \right) } \, , 
\end{equation} 
where
\be
f(q) =\re^{\pi \mathsf{b}q/2}{\fad(q-\mu/2+ \ri \mathsf{b}/4)\over \fad(q+\mu/2- \ri \mathsf{b}/4)} \, .
\ee

The above expression for the kernel of the trace class operator $\rho_{m,n}$ makes it also possible to obtain explicit results for the spectral traces $\tr \rho_{m,n}^\ell$, for low $\ell$. 
One finds, for example, 
\be
\label{two-traces}
\ba
\tr\, \rho_{m,n}&= {1\over 2\mathsf{b}\cos\left(\pi C_{m,n} \right)} \int_\IR \left| \Psi_{a,c}(p)\right|^2 \rd p \, , \\
\tr\, \rho^2_{m,n}&= \frac{\left|\fad\left(2\im(a+c)-c_{\mathsf{b}}\right)\right|^2}{2\mathsf{b}\sin\left(2\pi C_{m,n} \right)}\int_\IR \frac{\sinh(2\pi C_{m,n}\mb  s)}{\sinh(\pi\mathsf{b}s)}W_{\frac{\myh}2-a}(s)W_{\frac{\myh}2-c}(s)
\operatorname{d}\! s \, , 
\ea
\ee
where
\begin{equation}
W_a(x)\equiv\left|\mypsi{a}{a}(x)\right|^2
\end{equation}
and
\be
\myh=\frac{\mathsf{b}+\mathsf{b}^{-1}}2 \, . 
\ee
It turns out that these integrals can be evaluated analytically in many cases. Particularly important is the case in which $\mb^2$ is rational, since in that case, as recently shown in the 
context of state-integrals \cite{gk}, the quantum dilogarithm reduces to the classical dilogarithm and elementary functions, and the integrals (\ref{two-traces}) can be evaluated by residues. 
We will see various examples of this in the current paper. 

It turns out that the fermionic spectral traces $Z(N, \hbar)$ for the operator $\rho_{m,n}$ can be written in closed form, in terms of a matrix model \cite{mz}. By using Cauchy's inequality, as in 
the related context of the ABJM Fermi gas \cite{kwy2,Marino:2011eh}, one finds the representation 
\be
\label{zmn-bis}
Z_{m,n}(N,\hbar)=\frac{1}{N!}  \int_{\IR^N}  { \rd^N u \over (2 \pi)^N}  \prod_{i=1}^N  \left| \mypsi{a}{c}\left(  {\mb u_i \over 2 \pi} \right) \right|^2 
 \frac{\prod_{i<j} 4 \sinh \left( {u_i-u_j \over 2} \right)^2}{\prod_{i,j} 2 \cosh \left( {u_i -u_j \over 2} + \ri \pi C_{m,n} \right)} \, . 
\ee
The asymptotic expansion of the quantum dilogarithm makes it possible to calculate the asymptotic expansion of this integral in the 't Hooft limit
\be
N \rightarrow \infty \, , \qquad \hbar \rightarrow \infty \, , \qquad {N\over \hbar}= \lambda \quad \text{fixed} \, . 
\ee
It has the form, 
\be
\label{zmn-ex}
\log Z_{m,n}(N, \hbar)=\sum_{ g \ge 0} \CF_g^{(m,n)} (\lambda) \hbar^{2-2g} \, , 
\ee
and the functions $\CF_g^{(m,n)} (\lambda)$ can be easily computed in an expansion around $\lambda=0$ by using standard perturbation theory \cite{mz}. One finds, 
for the leading contribution, 
\be
\label{planar-f}
\CF^{(m,n)}_0(\lambda)=\frac{\lambda^2}{2} \left( \log \frac{ \lambda  \pi^3}{(m+n+1)a_{m,n}}-\frac{3}{2}\right)-c_{m,n} \lambda +\sum_{k=3}^\infty 
f_{0,k} \lambda^k \, . 
\ee
We have denoted
\be
a_{m,n}=2 \pi \sin \left( \frac{\pi m}{m+n+1}\right)\sin \left( \frac{\pi n}{m+n+1}\right) \sin \left( \frac{\pi}{m+n+1}\right) \, , 
\ee
while
\be
\label{cmnbw}
c_{m,n}=-\frac{m+n+1}{2\pi^2} D(-q^{m+1}\chi_m) \, .
\ee
In this equation, 
\begin{align}
q=\exp\left(\frac{\ri \pi}{m+n+1}\right) \, , \qquad \qquad \chi_k=\frac{q^k-q^{-k}}{q-q^{-1}} \, , 
\end{align}
and the Bloch-Wigner function is defined by, 
\begin{align}
	D(z)={\rm Im \, Li_2}(z)+{\rm arg}(1-z)\log|z| \, , 
\end{align}
where arg denotes the branch of the argument between $-\pi$ and $\pi$. The values of the coefficients $f_{0,k}$ can be calculated explicitly as functions of $m,n$, and 
results for the very first $k$ can be found in \cite{mz}.

\section{Examples}
\label{sec:examples}

\subsection{Local $\IF_2$}
\begin{figure}[h]
  \centering
  \includegraphics[width=0.25\linewidth]{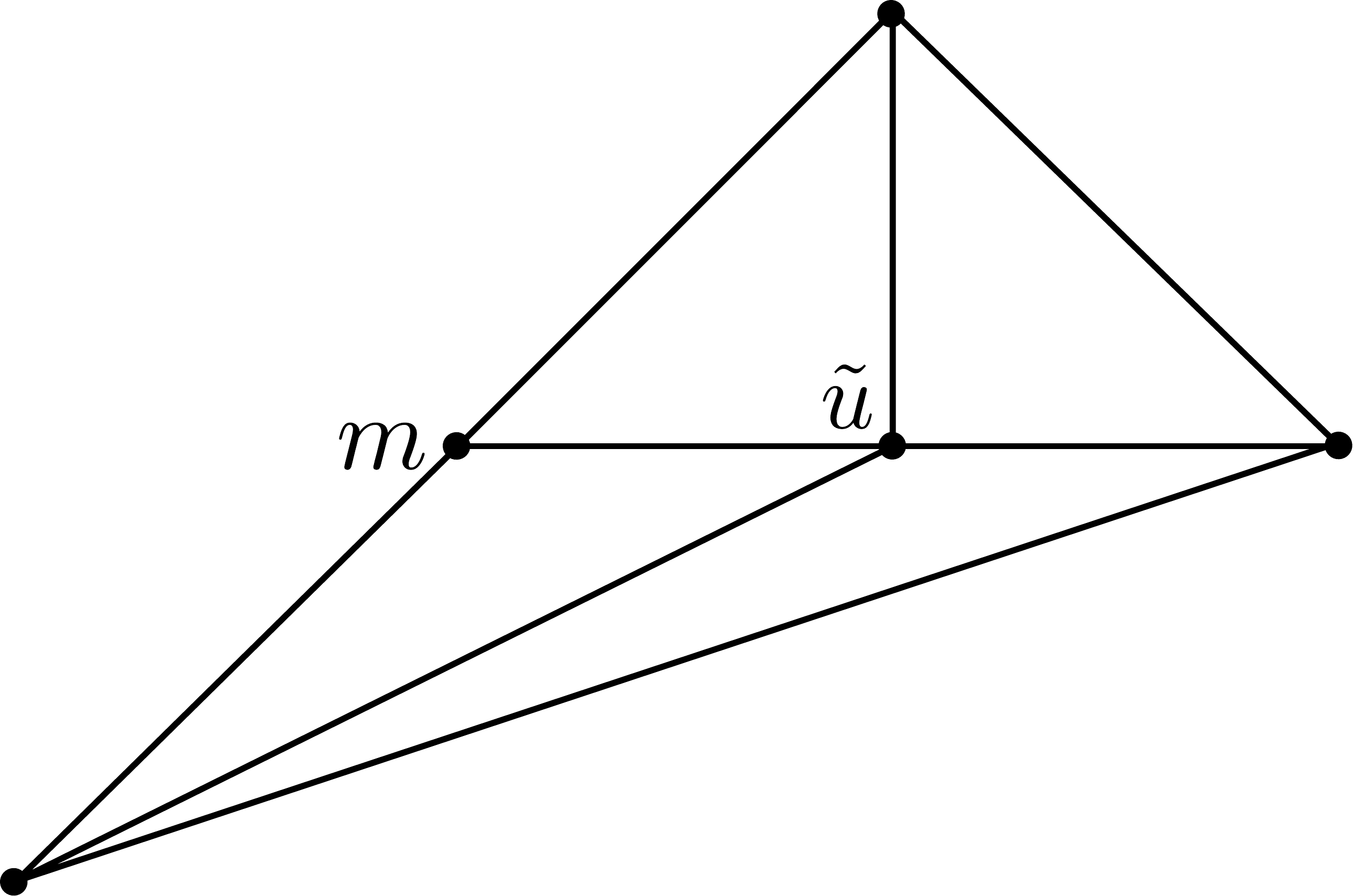}
  \caption{2d toric fan of $\mathcal{O}(-K_{\IF_2})\rightarrow \IF_2$.}
  \label{fig:toric_localf2}
\end{figure}
\begin{align}
 \label{eq:toric_localf2} 
 \begin{array}{l rrr rr } \toprule
    \multicolumn{4}{c}{\nu_i}    &l^{(1)}&l^{(2)}\\ \cmidrule{1-6}
    D_u    &    (\; 1&     0&   0 \;)&          -2& 0\\ 
    D_1    &    (\; 1&     1&   0 \;)&           1& 0\\ 
    D_2    &    (\; 1&     0&   1 \;)&           0& 1\\ 
    D_m    &    (\; 1&    -1&   0 \;)&           1&-2\\ 
    D_3    &    (\; 1&    -2&  -1 \;)&          0& 1\\ \bottomrule
  \end{array} 
\end{align}
The toric fan of local $\IF_2$ projected onto the supporting hyperplane $H$, which we will call the 2d toric fan of local $\IF_2$, is given in Figure \ref{fig:toric_localf2}. The toric data of local $\IF_2$ are given in \eqref{eq:toric_localf2}. From these toric data we can read off the Batyrev coordinates
\begin{align}
\label{eq:localf2_coordinates}
z_1=\frac{m}{\tilde{u}^2}=m u \,,\quad z_2=\frac{1}{m^2}\, ,
\end{align}
where we have used $r=2$ such that $\tilde{u}^{-2}=u$. Furthermore the spectral curve of this geometry is given by
\begin{align}
\label{curvef2}
W_{\IF_2}(e^x,e^y) = e^x + m e^{-x} + e^y + m^{-2} e^{-y+2x} + \tu \ .
\end{align}
This is the same spectral curve as the one in Table~\ref{tb:delPezzos} up to a symplectic transformation. For instance, let $X=e^{x}$, $Y=e^{y}$, then by using Nagell's algorithm \cite{Huang:2013yta, Huang:2014nwa} both curves can be converted to the Weierstrass form
\begin{equation}
	Y^2 = 4X^3 - g_2(u,m) X - g_3(u,m) \ ,
\end{equation}
where \cite{Huang:2013yta, Huang:2014nwa}
\begin{align}
\begin{split}
g_2(u, m) &= 27 u^4 \left(1-8m u+16 m^2 u^2-48u^2\right) \, ,\\
g_3(u, m) &=-27 u^6 \left(64 m^3 u^3-48 m^2 u^2-288 m u^3+12 m u+72 u^2-1\right) \, .
\end{split}
\end{align}
Analogous to the calculation in \cite{Huang:2014eha} we calculate the perturbative phase space volume in the large energy limit to read off the constants $C$, $D_0(m)$, and $B_0(m)$
\begin{align}\label{eq:f2_vol0}
\textrm{vol}_0(E) = 4E^2 - \frac{2}{3}\pi^2 - \left(\log \frac{m \pm \sqrt{m^2-4}  }{2}  \right)^2 + \mathcal{O}(e^{-E}) \, .
\end{align}
In this derivation we used the dictionary between the parameters $u_{\IF_0}$, $m_{\IF_0}$ of local $\IF_0$ and the parameters $u$, $m$ of local $\IF_2$ in Appendix \ref{rel-ff}
\begin{align}\label{eq:F0F2}
u=\sqrt{m_{\IF_0}} u_{\IF_0} \, , \quad m=\frac{1+m_{\IF_0}}{\sqrt{m_{\IF_0}}} \, .
\end{align}
It can be seen that this relation also holds at the level of quantum operators \cite{Kashaev:2015wia}. 
As mentioned in Section~\ref{ssc:conjecture}, the phase space volume can be identified with the B--period of the spectral curve, and we can use the quantum operators derived in \cite{Huang:2014nwa} to find the quantum corrections to the phase space volume. For local $\IF_2$ the first quantum operator with the substitution $u=e^{-rE}$ is given by
\begin{equation}
	\cD_2 = \frac{1}{48}\partial^2_E + \cO(e^{-E}) \ .
\end{equation}
Applying it to the perturbative phase space volume, and taking into account the extra ``$-$'' sign due to different conventions of $\hbar$, we find for the leading order of the first quantum correction to the phase space volume
\begin{align}\label{eq:f2_vol1}
\textrm{vol}_1 (E) = -\frac{1}{24} C + \mathcal{O}(e^{-E}) \, .
\end{align}
Comparing \eqref{eq:f2_vol0} and \eqref{eq:f2_vol1} to the general expressions \eqref{eq:vol0E} we find for the coefficients $C$, $D_0(m)$, $B_0(m)$ and $B_1$\footnote{In $B_0(m)$ the sign before the square root in the logarithm can be both positive and negative. This also happens in the mass function $Q_m$ which will be presented shortly. The final results are not affected by the sign as long as it is chosen consistently for $B_0(m)$ and $Q_m$. Here and later in $Q_m$ we choose a ``$+$'' sign.}
\begin{equation}
\begin{gathered}
C = 4 \ ,\quad  D_0(m) = 0\ ,\\
B_0(m) = \frac{\pi}{3} - \frac{1}{2\pi} \left( \log \frac{m+\sqrt{m^2-4}}{2} \right)^2 \ , \quad B_1 = -\frac{1}{12\pi} \ .
\end{gathered}
\end{equation}

\subsubsection{Maximal supersymmetry}\label{sec:f2_max_susy}

\paragraph{Energy spectrum}
We first work with the case of maximal supersymmetry with $\hbar=2\pi$, where the formulae are the simplest. We use \eqref{eq:mSUSYqnz} to calculate the energy spectrum. The coefficients $C, D_0(m), B_0(m), B_1$ have already been given in the previous section. As discussed in Section~\ref{ssc:BSolution}, the periods and the prepotential can be computed from \cite{Huang:2013yta,Huang:2014nwa}
\begin{equation}\label{eq:spGeom}
\begin{aligned}
\frac{\partial t}{\partial u} &= - \sqrt{\frac{E_6(\tau_0)g_2(u,m)}{E_4(\tau_0) g_3(u,m)}} \ ,\\
\frac{\partial^2F_0}{\partial t^2} &= -\frac{1}{2\pi\ri} \tau_0(t,m) \ ,
\end{aligned}
\end{equation}
where $\tau_0$ is the elliptic modulus of the elliptic spectral curve, and $E_4(\tau_0), E_6(\tau_0)$ are the Eisenstein series. Alternatively, we can use the formulae for A-- and B--periods for local $\IF_2$ given in \cite{Brini:2008rh}
\begin{align}
\begin{split}\label{eq:localf2_periods_lr}
\frac{\partial t}{\partial u}&= -\frac{2}{\pi u \sqrt{1-4(2+m) u}} K\left( \frac{16 u}{4(2+m) u -1} \right) \ , \\
\frac{\partial^2 F_0}{\partial u \partial t} &= - \frac{2}{u \sqrt{1- 4(m-2) u}} K \left( \frac{4(m+2) u -1}{4(m-2) u -1} \right) \, ,
\end{split}
\end{align}
from which the prepotential $F_0$ can be derived. Here $K(k^2)$ is the complete elliptic integral of the first kind. Near the \texttt{LCP}, the A--period has the expansion
\begin{equation}
	t = -\log u - 2m u - 3(2+m^2)u^2 -\frac{20}{3}m(6+m^2) u^3 - \frac{35}{2}(6+12m^2+m^4)u^4 + \cO(u^5) \ ,
\end{equation}
and the prepotential is
\begin{equation}
	F_0^{\rm inst} = -2m Q -\left(  \frac{7}{2} + \frac{m^2}{4} \right) Q^2 - \left( \frac{52m}{9}+\frac{2m^3}{27}\right)Q^3 + \cO(Q^4) \ ,
\end{equation}
where $Q = e^{-t}$. Furthermore, we notice that \cite{Huang:2014nwa}
\begin{equation}
	t_1=2t + \frac{1}{2} \log (Q_m) , \quad t_2=-\log (Q_m) \ ,\label{eq:F2ca}
\end{equation}
where
\begin{equation}
	Q_m = \left(\frac{m+ \sqrt{m^2-4}}{2} \right)^2 \ .
\end{equation}
So this is an example where the mass function $Q_m$ does not coincide with the mass parameter $m$. We can also read off the coefficients $c_\alpha, \alpha_{\alpha,j}$ from \eqref{eq:F2ca}.

Plugging all these data into the quantization condition \eqref{eq:mSUSYqnz}, we can compute the energy spectrum with an arbitrary mass parameter. We calculated the ground state energy $E_0$ for mass parameters $m=1, 2, 5/2$ respectively, with both the A--period in the definition of $\tmu$ and the prepotential expanded up to order 14. The results are listed in Table~\ref{tb:F2mSUSYEn} with all stabilized digits.

On the other hand, given the operator $\mathsf{O}_X$, we can also use the technique described in \cite{Huang:2014eha} to compute the energy spectrum numerically. We use wavefunctions of a harmonic oscillator as a basis of the Hilbert space, and calculate the Hamiltonian matrix $\langle n_1 |\mathsf{O}_X| n_2 \rangle$ truncated up to a finite size. After diagonalization the logarithms of the matrix entries give the energy eigenvalues, whose accuracy increases with increasing matrix size. We computed $E_0$ for $\mathsf{O}_X$ for $m=1, 2, 5/2$ with matrix size $500\times 500$. The results are given with all stabilized digits in the last column of Table~\ref{tb:F2mSUSYEn}. We find that the results computed with the conjecture match the numerical results in all stabilized digits.

\begin{table}
\centering
\begin{tabular}{>{$}c<{$} >{$}l<{$} >{$}l<{$}}\toprule
m & E_0\textrm{ from conjecture} & E_0\textrm{ from numerics}\\\midrule
1 & 2.828592218708195204728 & 2.828592218708\\
2 & 2.88181542992629678247 & 2.8818154299263\\
\tfrac{5}{2} & 2.9048366403731263260 & 2.904836640373\\\bottomrule
\end{tabular}
\caption{Ground state energy of $\mathsf{O}_{\IF_2}$ with $\hbar=2\pi$ computed from both the quantization condition \eqref{eq:mSUSYqnz} and with the numerical method with matrix size $500\times 500$ (see the main text). All stabilized digits are given in the results.}\label{tb:F2mSUSYEn}
\end{table}

\paragraph{Spectral determinant}
We can proceed to check the spectral determinant itself. Once we have the correct energy spectrum, we can compute the fermionic spectral trace $Z(N,\hbar)$ by its definition. We opt to use the quantization condition \eqref{eq:mSUSYqnz} to generate the spectrum as it is faster and the results have higher precision than the numerical method. We present the first two traces $Z(1,2\pi)$, $Z(2,2\pi)$ computed with $m=0,1,2$ respectively with this method in Table~\ref{tb:F2mSUSYZNEn}. 

On the other hand, the conjecture claims $Z(N,\hbar)$ can be calculated from the spectral determinant through \eqref{eq:ZNAi} in terms of Airy functions and its derivatives. Unlike the quantization condition \eqref{eq:mSUSYqnz}, this calculation requires the full expression of $J_X(\mu,m,\hbar)$ from the conjecture, and in addition to the prepotential $F_0$, genus one free energies $F_1$ and $F_1^{\rm NS}$ are also needed. 
The unrefined genus one free energy can be found in \cite{Haghighat:2008gw}. For elliptic toric geometry, it has the following generic form \cite{Huang:2013yta}\footnote{Here we are talking about the free energies in the holomorphic limit.}
\begin{equation}
	F_1 = -\frac{1}{12}\Delta + \log u^{a_0}\prod_j m_j^{a_j} +\frac{1}{2}\log\left( \frac{\partial t}{\partial u}\right) \ ,
\label{defF1}	
\end{equation}
where $\Delta$ is the discriminant, and the exponents $a_0, a_j$ can be fixed by constant genus one maps. In other words, near the \texttt{LCP}, the leading behavior of $F_1$ is\footnote{For some toric Calabi--Yau threefolds, the intersection numbers $c_2(X)\wedge J_\alpha$ are not well defined for some $J_\alpha$ because of the noncompact direction, and thus they can not be used to completely fix the exponents in $F_1$. Fortunately, these Calabi--Yau's can usually be converted to simpler ones $\tilde{X}$ by blowing down some divisors. Then one can fix the $F_1$ of $X$ by comparing BPS numbers of $X$ and $\tilde{X}$.}
\begin{equation}
	F_1 \rightarrow \frac{1}{24}\sum_\alpha t_\alpha \int_X c_2(X)\wedge J_\alpha + \cO(Q_\alpha )\ ,
\end{equation}
where $c_2(X)$ is the second Chern class, and $J_\alpha$ is the divisor dual to the Mori cone generator $C_\alpha$. For local $\IF_2$, the genus one free energy is
\begin{align}
	F_1 = -\frac{1}{12}\log \Delta - \frac{7}{12}\log u + \frac{1}{2}\log\frac{\partial t}{\partial u} \ ,
\end{align}
where
\begin{equation}
	\Delta = 1-8mu-64u^2+16m^2u^2 \ .
\end{equation}
The Nekrasov--Shatashvili genus one free energy can be computed following \cite{Huang:2013yta}. It generally has the form\footnote{This form of NS genus one free energy differs from that in \cite{Huang:2013yta} by a minus sign due to different conventions of $\hbar$.}
\begin{equation}
	F_1^{\rm NS} =  -\frac{1}{24}\log(\Delta u^{b_0}\prod_i m_i^{b_i}) \ .
	\label{defF1NS}
\end{equation}
where the exponents $b_0, b_i$ are fixed by requiring regularity in the limit $u\rightarrow \infty$. We find for local $\IF_2$
\begin{equation}
	F_1^{\rm NS} = -\frac{1}{24}\log(\Delta u^{-2}) \ .
\end{equation}

\begin{table}
\centering
\begin{tabular}{*{2}{>{$}c<{$}} *{1}{>{$}l<{$}} }\toprule
m & \multicolumn{2}{l}{$Z(N,2\pi)$}  \\\midrule
0 & Z(1,2\pi) & 0.08838834764831844055010 \\
  & Z(2,2\pi) & 0.0017715660567565415144447 \\
1 & Z(1,2\pi) & 0.08333333333333333333333  \\
  & Z(2,2\pi) & 0.00160191348951214746049  \\
2 & Z(1,2\pi) & 0.079577471545947667884 \\
  & Z(2,2\pi) & 0.0014799260223538892847 \\\bottomrule
\end{tabular}
\caption{$Z(1,2\pi)$, $Z(2,2\pi)$ of $\mathsf{O}_{\IF_2}$ computed from spectrum. In the results all the stabilized digits are listed.}\label{tb:F2mSUSYZNEn}
\end{table}

Now we have almost all the data to write down the complete expression of the modified grand potential $J_X(\mu,m,2\pi)$, except for $A(m,\hbar)$. This term can be fixed  by demanding $Z(0,\hbar)=1$. Once $A(m,\hbar)$ is known, we can compute again $Z(1,2\pi), Z(2,2\pi)$ with the Airy function method. In this process, we used free energies expanded up to order 14. We list the fermionic spectral traces as well as $A(m,\hbar)$ in Table~\ref{tb:F2mSUSYZNAi}. Comparing with Table~\ref{tb:F2mSUSYZNEn}, the results computed in this way have much higher precisions, i.e., they have far more stabilized digits. Due to space constraint, we only list the first 30 digits after the decimal point in the results. They agree with the results obtained from the spectrum in Table~\ref{tb:F2mSUSYZNEn}, giving strong support to the conjecture. 

Furthermore, the traces $Z(1,2\pi)$, $Z(2,2\pi)$ have been computed from operator analysis \cite{Kashaev:2015wia}. The results are
\begin{equation}
\begin{aligned}
	Z(1,2\pi) &= \frac{1}{4\pi}\frac{\cosh^{-1}(m/2)}{\sqrt{m-2}} \ , \\
	Z(2,2\pi) &= \frac{1}{32\pi^2}\left[\left(\frac{\cosh^{-1}(m/2)}{\sqrt{m-2}}\right)^2-\left(2\frac{\cosh^{-1}(m/2)}{\sqrt{m^2-4}}+1\right)^2+1+\frac{\pi^2}{m+2}\right] \ .
\end{aligned}
\end{equation}
In particular, when $m=0$
\begin{equation}\label{eq:F2mSUSYZN}
\begin{aligned}
	Z(1,2\pi) &=\frac{1}{8\sqrt{2}} \ , \\
	Z(2,2\pi) &=\frac{1}{256}\left( 3-\frac{8}{\pi}\right) \ .
\end{aligned}
\end{equation}
When $m=1$,
\begin{equation}
\begin{aligned}
	Z(1,2\pi) &=\frac{1}{12} \ , \\
	Z(2,2\pi) &=\frac{1}{216}\left( 2-\frac{3\sqrt{3}}{\pi}\right) \ .
\end{aligned}
\end{equation}
Finally, when $m=2$
\begin{equation}
\begin{aligned}
	Z(1,2\pi) &=\frac{1}{4\pi} \ , \\
	Z(2,2\pi) &=\frac{1}{128}\left( 1-\frac{8}{\pi^2}\right) \ .
\end{aligned}
\end{equation}
All these results agree with the predictions from the conjecture in Table~\ref{tb:F2mSUSYZNAi} in all the $125+$ stabilized digits. 

\begin{table}
\centering
\begin{tabular}{*{2}{>{$}c<{$}} *{1}{>{$}l<{$}} c }\toprule
m &   \multicolumn{2}{l}{$Z(N,2\pi)/A(m,2\pi)$}   & Precisions \\\midrule
0 & Z(1,2\pi) & 0.088388347648318440550105545263\ldots & 126\\
  & Z(2,2\pi) & 0.001771566056756541514444764789\ldots & 128\\
  & A(0,2\pi) & 0.182353979734290479565102066175\ldots & 125\\
1 & Z(1,2\pi) & 0.083333333333333333333333333333\ldots & 126\\
  & Z(2,2\pi) & 0.001601913489512147460490835671\ldots & 128\\
  & A(1,2\pi) & 0.238500357238526529618256294935\ldots & 125\\  
2 & Z(1,2\pi) & 0.079577471545947667884441881686\ldots & 132\\
  & Z(2,2\pi) & 0.001479926022353889284757533549\ldots & 133\\
  & A(2,2\pi) & 0.285676676163186113148112999786\ldots & 131\\\bottomrule
\end{tabular}
\caption{$Z(1,2\pi)$, $Z(2,2\pi)$, as well as $A(m,2\pi)$ for local $\IF_2$ computed with the Airy function method. The first 30 digits after the decimal point are given in the results. The column ``Precisions'' lists the number of stabilized digits after the decimal point for each result.}\label{tb:F2mSUSYZNAi}
\end{table}

In addition, the function $A(m,\hbar)$ for local $\IF_2$ has been conjectured in \cite{hun}, based on results for the ABJ matrix model:
\begin{equation}
 \label{amh}
	A(m,\hbar)=A_{\rm c} \left( {\hbar \over \pi} \right)
	-F_{\rm CS}\left( {\hbar \over \pi},M\right) \, , 
\end{equation}
where 
\begin{equation}
\label{ak}
A_{\rm c}(k)= \frac{2\zeta(3)}{\pi^2 k}\left(1-\frac{k^3}{16}\right)
+\frac{k^2}{\pi^2} \int_0^\infty \frac{x}{\re^{k x}-1}\log(1-\re^{-2x})\rd x
\end{equation}
is the $A(k)$ function of ABJM theory \cite{Marino:2011eh, hanada,ho-1} and $F_{\rm CS}\left(k,M\right)$ is the Chern--Simons (CS) 
free energy on the three-sphere for gauge group $U(M)$ and level $k$, 
\begin{equation}
F_{\rm CS} (k,M)= \log \, Z_{\rm CS}(k,M) \, , 
\end{equation}
where $M$ is related to the parameters of our problem as
\begin{equation}
\label{Mk}
M= {\hbar + \ri \log m_{\IF_0}\over 2 \pi} \, . 
\end{equation}
Recall that $m_{\IF_0}$ is related to $m$ via \eqref{eq:F0F2}. Since $M$ is a complex, arbitrary parameter and $k=\hbar/\pi$ is not necessarily an integer, we need 
an analytic continuation of the CS partition function. Such a continuation is not necessarily unique, but the spectral problem associated to $\IF_2$ requires a definite choice. Recently, a proposal for an analytic continuation of the CS free energy has been put forward in \cite{ho}\footnote{Similar, integral 
expressions for analytic continuations of the CS partition function have been obtained in \cite{universal,krefl-m}.}. The result can be written as 
\begin{equation}
\begin{aligned}
F_{\rm CS} (\hbar/\pi,M)&= {\hbar^2 \over 8 \pi^4} \left\{ {\rm Li}_3\left( \re^{2 \pi \ri \pi^2 M/\hbar}\right)+{\rm Li}_3\left( \re^{-2 \pi \ri \pi^2 M/\hbar}\right) -2 \zeta(3) \right \}\\
&+ \int_0^\infty {t\over \re^{2 \pi t}-1} \log \left[ {\sinh^2(\pi^2 t/\hbar) \over \sinh^2(\pi^2 t/\hbar)+ \sinh^2(\pi^2 M /\hbar)} \right] \rd t \, . 
\end{aligned}
\end{equation}
After plugging the value of $M$ in (\ref{Mk}), we find
\begin{equation}
\begin{aligned}
F_{\rm CS} (\hbar, m_{\IF_0})&= {\hbar^2 \over 8 \pi^4} \left\{ {\rm Li}_3\left(-m_{\IF_0}^{\pi\over \hbar} \right)+{\rm Li}_3\left( -m_{\IF_0}^{-{\pi\over \hbar}}\right) -2 \zeta(3) \right \}\\
&+ \int_0^\infty {t\over \re^{2 \pi t}-1} \log \left[ {4 \sinh^2(\pi^2 t/\hbar) \over 4 \sinh^2(\pi^2 t/\hbar)+ \left(m_{\IF_0}^{\pi\over 2\hbar} +m_{\IF_0}^{-{\pi\over 2 \hbar}} \right)^2}\right]\rd t \, . 
\end{aligned}
\end{equation}
In particular, in the maximally supersymmetric case $\hbar=2 \pi$, we have 
\begin{equation}
\begin{aligned}
F_{\rm CS} (2\pi, m_{\IF_0})&= {1\over 2 \pi^2} \left\{ {\rm Li}_3\left(-m_{\IF_0}^{1/2} \right)+{\rm Li}_3\left( -m_{\IF_0}^{-1/2}\right) -2 \zeta(3) \right \}\\
&+ \int_0^\infty {t\over \re^{2 \pi t}-1} \log \left[ {4 \sinh^2(\pi t/2) \over 4 \sinh^2(\pi t/2)+ \left(m_{\IF_0}^{1/4} +m_{\IF_0}^{-1/4} \right)^2}\right]\rd t \, . 
\end{aligned}
\end{equation}
By using that
\begin{equation}
A_{\rm c}(2)=-{\zeta(3) \over 2 \pi^2} \, ,
\end{equation}
we find the following expression, 
\begin{equation}
\begin{aligned}
A(m,2 \pi)&=- {1\over 2 \pi^2} \left\{ {\rm Li}_3\left(-\frac{m}{2}-\sqrt{\frac{m^2}{4}-1}\right)+{\rm Li}_3\left(\sqrt{\frac{m^2}{4}-1}-\frac{m}{2
   }\right)- \zeta(3) \right \}\\
&- \int_0^\infty {t\over \re^{2 \pi t}-1} \log \left[ {4 \sinh^2(\pi t/2) \over 4 \sinh^2(\pi t/2)+ m+2}\right]\rd t \, . 
\end{aligned}
\end{equation}
When $m=0, 1, 2$ are plugged in, this formula reproduces the values of $A(m,2\pi)$ in Table~\ref{tb:F2mSUSYZNAi} up to all the $125+$ stabilized digits. This confirms that 
the analytic continuation of the CS partition function put forward in \cite{ho} is the one needed to solve the spectral problem of local $\IF_2$.

\paragraph{Orbifold point expansion}

There is yet another way to compute the fermionic spectral traces $Z(N,2\pi)$ as indicated at the end of Section~\ref{ssc:conjecture}: Namely by expanding the spectral determinant around $\kappa = 0$, which corresponds to the orbifold point of the topological string theory. In other words, we need to analytically continue the topological string free energies used to construct $J_X(\mu,\um,\hbar)$ to the orbifold point. This is most convenient in the maximal supersymmetric case where only genus zero and genus one free energies are required. This method of calculating $Z(N,2\pi)$ is very interesting, as it reveals intriguing relations of Jacobi theta functions, as we will see at the end of the computations.

We are particularly interested in the locus
\begin{equation}
	1/u = 0, \quad m = 0
\end{equation}
in the moduli space, which is a $\bC^3/\bZ_4$ orbifold point. When $m$ is small, local $\IF_2$ has conifold points on the real axis of $u$ in both the positive and the negative directions. Therefore we wish to analytically continue the free energies along the imaginary axis to avoid the conifold points. To make this explicit, we perform a change of variables
\begin{equation}
	u = e^{-\pi\ri/2}\hu
\end{equation}
where now the new coordinate $\hu$ is real and positive. We rotate the mass parameter $m$ as well by
\begin{equation}
	m = e^{\pi\ri/2} \hm
\end{equation}
so that the power series part $\widetilde{\Pi}_A(u,m)$ of $t$ (as well as the instanton parts of free energies) remains real. We define the flat coordinate $\hht$ after the rotation
\begin{equation}\label{eq:tRotated}
	\hht = -\log \hu - \widetilde{\Pi}_A(e^{-\pi\ri/2}\hu, e^{\pi\ri/2} \hm)
\end{equation}
and thus
\begin{equation}
	t = \hht +\pi\ri/2 \ .
\end{equation}
We also introduce the free energies after the phase rotation
\begin{equation}\label{eq:F1Rotated}
\begin{aligned}
	\hF^{\rm inst}_0(\hht,\hm) \equiv & F_0^\textrm{inst}(t,m)\ ,\\ 
	\hF_1^\textrm{inst}(\hht,\hm)+\hF_1^\textrm{NS, inst}(\hht,\hm) \equiv & F_1^\textrm{inst}(t,m) + F_1^\textrm{NS, inst}(t,m) \ . 
\end{aligned}
\end{equation}
Similar to \eqref{eq:F0full}, the full prepotential after the phase rotation should be
\begin{equation}
	\hF_0 = \frac{1}{6} \hht^3 -\frac{1}{2}\left( \log\frac{\hm+\sqrt{\hm^2+4}}{2}\right)^2 \hht + \hF_0^{\rm inst}(\hht,\hm) \ ,
\end{equation}
where we have plugged in the coefficients $C, D_0(m)$, and $B_0^{(m)}(m)$ for local $\IF_2$. This implies that the B--period after the phase rotation is related to the B--period before the rotation by
\begin{equation}\label{eq:tDRotated}
	\frac{\partial F_0}{\partial t} = \frac{\partial \hF_0}{\partial \hht} + \frac{\pi\ri}{2}\hht -\frac{\pi\ri}{2}\log\frac{\hm+\sqrt{\hm^2+4}}{2} \ .
\end{equation}
Furthermore, similar to the example in \cite{ghm}, the phase rotation results in a shift of $1/8$ in $n$ in the spectral determinant. Explicitly, the spectral determinant after the phase rotation becomes
\begin{equation}
	\Xi(\mu,0,2\pi) = e^{\hJ_X(\hmu,0,2\pi)}\hTheta_X(\hmu,0,2\pi) \ ,
\end{equation}
where the rotated $\hmu$ is defined by
\begin{equation}
	\mu = \hmu + \pi\ri/4 \ ,
\end{equation}
and
\begin{align}
	\hJ_X(\hmu,0,2\pi) =& A(2\pi) +\hF_1+\hF_1^\textrm{NS} + \frac{\hF_0(\hht,\hm)}{4\pi^2}-\frac{1}{4\pi^2}\left( \hht\frac{\partial}{\partial \hht} + \pi\ri \frac{\partial}{\partial \hm} \right)\hF_0(\hht,\hm)  \nonumber\\
   +&\frac{1}{8\pi^2}\left( \hht^2\frac{\partial^2}{\partial \hht^2} +  2\pi\ri\,\hht \frac{\partial^2}{\partial \hm\partial \hht} - \pi^2 \frac{\partial^2}{\partial \hm^2}\right) \hF_0(\hht,\hm)\Big|_{\substack{\hht\rightarrow 2\hmu_{\rm eff}\\\hm\rightarrow 0}} \label{eq:JXRotated}\\
	\hTheta_X(\hmu,0,2\pi) =& \sum_{n\in\mathbb{Z}}\exp\left( \pi \ri (n+\tfrac{1}{8})^2\tilde{\tau} +2\pi i (n+\tfrac{1}{8})\tilde{\xi} -\frac{8\pi \ri }{3} (n+\tfrac{1}{8})^3 ) \right) 	\Big|_{\substack{\hht\rightarrow 2\hmu_{\rm eff}\\\hm\rightarrow 0}}	\label{eq:ThetaRotated}
\end{align}
where we have plugged in $\hm=m = 0$ whenever possible to simplify the expressions. In the formulae above, 
\begin{equation}
	\hmu_{\rm eff} = \hmu - \frac{1}{2}\widetilde{\Pi}_A(u)  = \hmu-\frac{1}{2}\widetilde{\Pi}_A(e^{-\pi\ri/2}e^{-2\hmu})\ .
\end{equation}
Besides, 
\begin{equation}\label{eq:tauxiRotated}
\begin{aligned}
	\hat{\tau} &= \frac{2i}{\pi} \frac{\partial^2}{\partial \hht^2} \hF_0 \ ,  \\
	\hat{\xi} &=\frac{1}{2\pi^2} \left(  \hht\frac{\partial^2}{\partial \hht^2} \hF_0 - \frac{\partial}{\partial\hht}\hF_0 + \pi\ri\, \frac{\partial^2 \hF_0}{\partial \hm \partial \hht}  \right) \ .
\end{aligned}
\end{equation}
Note that because 
\begin{align*}
	-\frac{8}{3}\pi i \left(n+\frac{1}{8}\right)^3 = -\pi i \left(n+\frac{1}{8}\right)^2-2\pi i \left(n+\frac{1}{8}\right) \frac{13}{48} +& \frac{5\pi i}{64}  \\
	&-\frac{\pi i\, 2n(2n+1)(2n-1)}{3}\ ,
\end{align*}
the generalized theta function becomes a conventional elliptic theta function
\begin{equation}
	\hTheta(\hmu,0,2\pi) = e^{\tfrac{5\pi i}{64}}\vartheta_{1/8}\left(\hat{\xi} - \frac{13}{48}; \hat{\tau}-1 \right) \ .
\end{equation}
where we have used the Jacobi theta function
\begin{equation}
	\vartheta_{1/8}(z;\tau) \equiv \vartheta\begin{bmatrix}
		\tfrac{1}{8} \\ 0
	\end{bmatrix}(z;\tau) = \sum_{n\in\mathbb{Z}}\exp \left(\pi i(n+\tfrac{1}{8})^2\tau + 2\pi i (n+\tfrac{1}{8})z  \right) \ .
\end{equation}

The expressions for the derivatives of the periods of local $\IF_2$ in \eqref{eq:localf2_periods_lr} can be translated through \eqref{eq:tRotated}, \eqref{eq:tDRotated} to the periods after the phase rotation\footnote{We cannot plug in the value of $\hm=0$ here because we will need derivatives of $\hm$ later in \eqref{eq:JXRotated}, \eqref{eq:tauxiRotated}.}
\begin{equation}
\begin{aligned}
	\frac{\partial \hht}{\partial \hu} &=  -\frac{2 K\left(\frac{16 \ri\hu}{1+4 \ri(\ri\hm+2)\hu}\right) }{\pi \hu\sqrt{1+4 \ri(\ri\hm+2)\hu}}		\	,	\\
	\frac{\partial^2 \hF_0}{\partial\hht\partial \hu} &= -\frac{2 K\left( \frac{1+4 \ri(\ri\hm+2)\hu}{1+4 \ri(\ri\hm-2)\hu} \right) }{\hu\sqrt{1+4 \ri (\ri\hm-2)\hu}} + \frac{ \ri  K\left( \frac{16 \ri\hu}{1+4 \ri(\ri\hm+2)\hu} \right) }{\hu\sqrt{1+4 \ri(\ri\hm+2)\hu}}	\	.
\end{aligned}
\end{equation}
From these formulae we can obtain the series expansion of the rotated periods near the \texttt{LCP}
\begin{equation}\label{eq:hPeriods}
\begin{aligned}
	\hht 	&= \int \frac{\partial \hht}{\partial \hu} \hu		, \quad\quad\quad\quad \textrm{(indefinite integral)}	 \\
			&= -\log \hu -2\hm\hu + 3(2-\hm^2) \hu^2 + \ldots		\\
	\hht_D 	&\equiv \frac{\partial \hF_0}{\partial \hu}= \int \frac{\partial^2 \hF_0}{\partial\hht\partial \hu} \hu  	\\
				&= \frac{1}{2}(\log \hu)^2 + \log \hu (2\hm\hu-3(2-\hm^2) \hu^2 + \ldots) +2\hm\hu-\frac{1}{2}(14-13\hm^2)\hu^2+\ldots \, .
\end{aligned}
\end{equation}

In order to analytically continue the rotated periods to the orbifold point $1/\hu = 0$, we use the reciprocal modulus formula for elliptic integrals, which implies
\begin{equation*}
	\frac{K\left(	\frac{1+4\ri(\ri\hm+2)\hu}{1+4 \ri(\ri\hm-2)\hu}\right)}{\sqrt{1+4 \ri(\ri\hm-2)\hu)}} = \frac{K\left(	\frac{1+4 \ri(\ri\hm-2)\hu}{1+4 \ri(\ri\hm+2)\hu}\right)}{\sqrt{1+4 \ri(\ri\hm+2)\hu)}} + \ri \frac{K\left( \frac{16 \ri\hu}{1+4 \ri(\ri\hm+2)\hu}\right)}{\sqrt{1+4 \ri(\ri\hm+2)\hu}}	\	.
\end{equation*}
Note the sign in front of the last term is positive because the imaginary part of
\[
	\frac{1+4\ri(\ri\hm+2)\hu}{1+4\ri(\ri\hm-2)\hu}	\	,
\]
in the argument of the elliptic integral $K$ on the left hand side is always positive, as long as $\hm\hu$  is kept small. Define the modulus $v$ around the orbifold point $1/\hu=0$
\begin{equation}
	\hv = 1/\sqrt{\hu} \ .
\end{equation}
The rotated periods after analytic continuation satisfy
\begin{align}
	\frac{\partial \hht}{\partial \hv} &= -\frac{4 \ri}{\pi}\left( \frac{K\left( \frac{4\ri(2+\ri\hm)+\hv^2}{4\ri(-2+\ri\hm)+\hv^2} \right)}{\sqrt{4\ri(-2+\ri\hm)+\hv^2}}  -	 \frac{K\left( \frac{4\ri(-2+\ri\hm)+\hv^2}{4\ri(2+\ri\hm)+\hv^2} \right)}{\sqrt{4\ri(2+\ri\hm)+\hv^2}}\right)	\	,\\
	\frac{\partial \hht_D}{\partial \hv} &=2 \left( \frac{K\left( \frac{4\ri(2+\ri\hm)+\hv^2}{4\ri(-2+\ri\hm)+\hv^2} \right)}{\sqrt{4\ri(-2+\ri\hm)+\hv^2}}  +	 \frac{K\left( \frac{4\ri(-2+\ri\hm)+\hv^2}{4\ri(2+\ri\hm)+\hv^2} \right)}{\sqrt{4\ri(2+\ri\hm)+\hv^2}}\right)	\	,
\end{align}
from which we can obtain the series expansions of the rotated periods around the orbifold point
\begin{equation}\label{eq:hPeriodsOrb}
\begin{aligned}
	\hht^\textrm{orb} &= \int_0^{\hv} \frac{\partial \hht}{\partial \hv'} \hv' 		, \quad\quad\quad\quad \textrm{(definite integral)}		\\
	&= \frac{\Gamma(\tfrac{1}{4})^2}{2\sqrt{2}\pi^{3/2}} \hv + \frac{\sqrt{\pi}(-1+\epsilon)}{12\sqrt{2}\Gamma(\tfrac{1}{4})^2}\hv^3 + \frac{(-3+40\epsilon-240\epsilon^2)\Gamma(\tfrac{1}{4})^2}{15360\sqrt{2}\pi^{3/2}} \hv^5+ \ldots \\
	\hht_D^\textrm{orb} &= \int_0^{\hv} \frac{\partial \hht_D}{\partial \hv'} dv' 	 \\
	&= \frac{\Gamma(\tfrac{1}{4})^2}{4\sqrt{2\pi}}\hv + \frac{\pi^{3/2}(1-12\epsilon)}{24\sqrt{2}\Gamma(\tfrac{1}{4})^2}\hv^3+\frac{(-3+40\epsilon-240\epsilon^2)\Gamma(\tfrac{1}{4})^2}{30720\sqrt{2\pi}} \hv^5 + \ldots \ ,
\end{aligned}
\end{equation}
where $\epsilon = \hm\hu$.

Note that the two sets of periods $(\hht, \hht_D)$ and $(\hht^{\textrm{orb}}, \hht_D^{\textrm{orb}})$ are not necessarily the same. As the analytical continuation was done at the level of their derivatives, a constant in $\hv$, which could be a function of $\hm$, can be missing. Let's call it a pure $\hm$ function. To disclose this term, we perform the definite integral in \eqref{eq:hPeriodsOrb} numerically for some large value of $\hv$, which corresponds to a diminutive $\hu$,  subtract from it the value of the (truncated) series expansion of $\hht$ in \eqref{eq:hPeriods}, and fit the difference as a function of $\hm$. The same exercise can be done for the pair of $\hht_D, \hht_D^{\textrm{orb}}$ as well. The pure $\hm$ functions are found to be,
\begin{equation}\label{eq:hPeriodsConst}
\begin{aligned}
	\hht^\textrm{orb} &= \hht+\log \left(\frac{\hm+\sqrt{\hm^2+4}}{2}\right)	\	, \\
	\hht_D^\textrm{orb} &=\hht_D+\frac{\pi^2}{6} -\frac{1}{2}\left( \log \frac{\hm+\sqrt{\hm^2+4}}{2} \right)^2 \ .
\end{aligned}
\end{equation}
These formulae together with \eqref{eq:hPeriodsOrb} give the expansion of the periods $\hht,\hht_D$ near the orbifold point, and we can proceed to compute the prepotential $\tilde{F}_0$ by integrating $\hht_D = \partial \tilde{F}_0/\partial \hht$, up to an integration constant. The latter, together with $A(m,2\pi)$, is fixed by normalizing $Z(0,2\pi)$ to 1.

We are finally in position to calculate $Z(N,2\pi)$ via the expansion of $\Xi_X(\mu,0,2\pi)$ around $\kappa = 0$. Noticing that
\begin{equation}
	\kappa = e^{\pi\ri/4}\hv \ ,
\end{equation}
the expansion takes the form
\begin{equation}
	\Xi(\kappa,0,\hbar) = 1+ \sum_{N\geq 1} Z(N,0,\hbar) e^{\tfrac{N\pi i}{4}} \hv^N \ . \label{eq:XiF2}
\end{equation}
In other words, the coefficients in the orbifold expansion of $\Xi(\kappa,0,\hbar)$ are
\begin{equation}
	e^{\tfrac{N\pi i}{4}}	Z(N,0,\hbar) \ .
\end{equation}

Rather than actually calculating $Z(N,0,2\pi)$, we assume the values of $Z(1,m=0,2\pi)$ and $Z(2,m=0,\pi)$ are given by \eqref{eq:F2mSUSYZN}, and extract the following relations of the elliptic Jacobi theta function $\vartheta_{1/8}(z;\tau)$
\begin{equation}
\begin{aligned}
	\frac{\partial_z\vartheta_{1/8}(\tfrac{-1+i}{8}; -1+i)}{\vartheta_{1/8}(\tfrac{-1+i}{8}; -1+i)} &= -\frac{\pi i}{4} - \frac{(1-i) \Gamma(\tfrac{1}{4})^2}{4\sqrt{2\pi}} \ , \\
	\frac{\pi i \partial_\tau\vartheta_{1/8}(\tfrac{-1+i}{8}; -1+i)}{\vartheta_{1/8}(\tfrac{-1+i}{8}; -1+i)}& - \frac{\partial^2_{z^2}\vartheta_{1/8}(\tfrac{-1+i}{8}; -1+i)}{16 \vartheta_{1/8}(\tfrac{-1+i}{8}; -1+i)}   \\
	&= -\frac{3(16\pi^2+\pi^3-(1+i)\sqrt{2}\pi^{3/2}\Gamma(\tfrac{1}{4})^2-i \Gamma(\tfrac{1}{4})^4)}{256\pi}\ .
\end{aligned}
\end{equation}
They can be verified numerically to arbitrarily high precision.

\subsubsection{Rational Planck constants}

\begin{table}
\centering
\resizebox{\linewidth}{!}{
\begin{tabular}{*{2}{>{$}c<{$}} c >{$}l<{$}  *{2}{>{$}c<{$}} }\toprule
 \hbar & m &  \multicolumn{2}{c}{Ground state energies} & \textrm{Errors} & \textrm{Deviations} \\\midrule
 3\pi & 2   & w/ $\lambda(E)$   & 3.5784100386973932885370276609 & 3.3\times 10^{-29} & - \\
      &     & w.o. $\lambda(E)$ & 3.5784100358696745628684580057 & 1.8\times 10^{-29} & 2.8\times 10^{-9} \\
      &     & numerical      & 3.5784100387 & & \\\cmidrule{2-6}
      & 5/2 & w/ $\lambda(E)$   & 3.596013630566028853057384426 & 2.2\times 10^{-28} & - \\
      &     & w.o. $\lambda(E)$ & 3.596013628010480256882055931 & 1.5\times 10^{-28} & 2\times 10^{-9} \\
      &     & numerical      & 3.596013630 & & \\\midrule      
 8\pi/3 & 2   & w/ $\lambda(E)$   & 3.3488711127605665243858784139 & 1.7\times 10^{-29} & - \\
        &     & w.o. $\lambda(E)$ & 3.3488711126985280038283987464 & 9.2\times 10^{-30} & 6\times 10^{-11} \\
        &     & numerical      & 3.34887111276 & & \\\cmidrule{2-6}
        & 5/2 & w/ $\lambda(E)$   & 3.367972636079200789494258599 & 1.2\times 10^{-28} & - \\
        &     & w.o. $\lambda(E)$ & 3.367972636018243377139143847 & 8.4\times 10^{-29} & 6\times 10^{-11} \\
        &     & numerical      & 3.36797263608 & & \\\bottomrule    
\end{tabular}
}
\caption{Ground state energies for $\mathsf{O}_{\IF_2}$ with rational $\hbar$, computed by the complete quantization condition \eqref{eq:gen_qnz} with $\lambda(E)$ (rows labeled by ``w/ $\lambda(E)$''), by incomplete quantization condition without $\lambda(E)$ (rows labeled by ``w.o. $\lambda(E)$''), and by numerical method with matrices of size $500\times 500$ (rows labeled by ``numerical''). All stabilized digits are listed in the results. ``Errors'' are estimated by dropping the highest order instanton corrections to the quantization condition (see the main text). The column ``Deviations'' gives the deviation from the numerical results. ``$-$'' means no deviation.}\label{tb:F2_spec_rat_h}
\end{table}

Here we wish to check the conjecture of the solution to the spectral operator $\mathsf{O}_X$ for local $\IF_2$ with generic rational Planck constants, i.e., $\hbar$ now takes the form of \eqref{eq:ratPlanck} with $(p,q)\neq (1,1)$. Unlike the case of maximal supersymmetry, the quantum A--period in the definition \eqref{eq:mueff} of $\tmu$ no longer reduces to the classical A--period. For local $\IF_2$, the quantum A--period can be found in \cite{Huang:2014nwa}. The leading contributions are
\begin{equation}
	\widetilde{\Pi}_A(u,m,q) = 2 m u+ \left(2+3m^2+ \frac{2}{q}+2q\right) u^2 + \cO(u^3) \ ,
\end{equation}
where $q = \exp(\ri\hbar)$. Furthermore, to construct the modified grand potential $J_X(\mu,\um,\hbar)$, we need (refined) topological string free energies with genera greater than one as well. It is not difficult to see from \eqref{eq:tJbNew}--\eqref{eq:dm} that the order of instanton corrections is controlled by
\begin{equation}\label{eq:d}
	d = \sum_\alpha \tca d_\alpha \ ,
\end{equation}
in the sense that if we want to compute $\tJ_b(\tmu,\um,\hbar)$, $\tJ_c(\tmu,\um,\hbar)$ up to $\ell = n$ or compute $J_{\rm WS}(\tmu,\um,\hbar)$ up to $m = n$, we need all the BPS numbers $N^{\rm d}_{j_L,j_R}$ with $d\leqslant n$. In the case of local $\IF_2$, we have BPS numbers up to $d = 2d_1 = 18$ (we have used $\tca$ from \eqref{eq:F2ca})\footnote{Here we partially use the data shared with us from Xin Wang. See the Acknowledgement.}.
The BPS numbers are too many even to fit into the appendix. 
Instead, we collect them as a \texttt{Mathematica} notebook in an ancillary file to this paper.

Using these data, we are able to compute the ground state energies $E_0$ from \eqref{eq:gen_qnz} together with \eqref{eq:rat_h_qnz} for $\hbar = 3\pi, 8\pi/3$ and mass parameters $m=2, 5/2$. The results are listed in Table~\ref{tb:F2_spec_rat_h} with all stabilized digits. To estimate errors of these results, we drop the highest order instanton corrections (corresponding to $d=18$) to the left hand side of \eqref{eq:gen_qnz}, and rerun the calculation. Furthermore, since the quantization condition described here and also first presented in \cite{ghm} improve the proposal in \cite{Kallen:2013qla} by the additional $\lambda(E)$ term in \eqref{eq:gen_qnz}, we calculate the ground state energies without the $\lambda(E)$ correction as well, which are also listed in Table~\ref{tb:F2_spec_rat_h}, to see how much the corresponding results differ from the results of the complete quantization condition. Finally, we calculate the ground state energies with same $\hbar$ and $m$ numerically by diagonalizing Hamiltonian matrices of size $500\times 500$, and list the results in the same table. We find that the ground state energies computed with the complete quantization condition always coincide with the numerical results in all stabilized digits, while the energies computed without $\lambda(E)$ correction always differ from the numerical results by margins much larger than the estimated errors.

\begin{table}[t]
\centering
\resizebox{\linewidth}{!}{
\begin{tabular}{*{3}{>{$}c<{$}} c >{$}l<{$} >{$}c<{$} } \toprule
\hbar & m & N & \multicolumn{2}{c}{Fermionic spectral traces $Z(N,m,\hbar)$} & \textrm{Precisions} \\\midrule
3\pi  & 2 & 1 & spectrum & 0.03556463950383471875248925 & 26 \\
      &   &   & Airy     & 0.03556463950383471875248925038476928\ldots & 133\\\cmidrule{3-6}
      &   & 2 & spectrum & 0.0002276038191693029375687476 & 28 \\
      &   &   & Airy     & 0.00022760381916930293756874760322638\ldots & 135\\\cmidrule{2-6}
      & 5/2 & 1 & spectrum & 0.034997174539993270348639863 & 27 \\
      &     &   & Airy     & 0.03499717453999327034863986301519796\ldots & 127\\\cmidrule{3-6}
      &     & 2 & spectrum & 0.0002214025998730489988354869 & 28 \\
      &     &   & Airy     & 0.00022140259987304899883548694708767\ldots & 130\\\midrule
 8\pi/3  & 2 & 1 & spectrum & 0.0460092453000601959288605673 & 28 \\
         &   &   & Airy     & 0.04600924530006019592886056730533817\ldots & 145\\\cmidrule{3-6}
         &   & 2 & spectrum & 0.000412702534221779106148301012 & 30 \\
         &   &   & Airy     & 0.00041270253422177910614830101201279\ldots & 147\\\cmidrule{2-6}
         & 5/2 & 1 & spectrum & 0.0452240112730311424402202520 & 28 \\
         &     &   & Airy     & 0.04522401127303114244022025207320863\ldots & 145\\\cmidrule{3-6}
         &     & 2 & spectrum & 0.00040068762952861763389783745 & 29 \\
         &     &   & Airy     & 0.00040068762952861763389783745215685\ldots & 147\\\bottomrule
\end{tabular}
}
\caption{First two fermionic spectral traces for $\mathsf{O}_{\IF_2}$ with rational $\hbar$, computed from the spectrum (rows labelled by ``spectrum'', with all stabilized digits), and by the Airy function method (rows labelled by ``Airy'', with the first 35 digits after the decimal point). The column ``Precisions'' gives the numbers of stabilized digits after the decimal point.}\label{tb:F2_Z_rat_h}
\end{table}

Next, we proceed to check the full spectral determinant by computing the fermionic spectral traces. As in the case of maximal supersymmetry, we first compute $Z(1,\hbar)$ and $Z(2,\hbar)$ by definition from the energy spectrum, which is generated using the complete quantization condition. The results for $\hbar=3\pi, 8\pi/3$ and $m=2, 5/2$ are given in Table~\ref{tb:F2_Z_rat_h} with all stabilized digits against both varying orders of instanton corrections and varying energy levels. Then we compute the same fermionic spectral traces through \eqref{eq:ZNAi} in terms of Airy functions and its derivatives. This formula makes use of the entire modified grand potential. Using BPS numbers up to $d=18$, the fermionic spectral traces can be computed with a precision of up to $127\sim147$ stabilized digits after the decimal point. Due to space constraint, we list the results with only the first 35 digits after the decimal point in Table~\ref{tb:F2_Z_rat_h}. They agree with the results computed from the energy spectrum. More importantly, for local $\IF_2$, the first few fermionic spectral traces $Z(1,\hbar), Z(2,\hbar)$ can be directly computed with rational $\hbar$ and arbitrary mass $m$ by integrating the kernel of the operator $\mathsf{O}_{\IF_2}$ given in (\ref{f0ker}). They agree with the results from the Airy function method in Table~\ref{tb:F2_Z_rat_h} up to all stabilized digits. 

In addition, the Airy function method fixes the value of $A(m,\hbar)$ as well by the normalization condition $Z(0,\hbar)=1$. We have thus computed $A(m,\hbar)$ for $\hbar=3\pi, 8\pi/3$ and $m=2, 5/2$. They agree with the predictions by \eqref{amh} consistently with up to $132\sim 147$ digits.

\subsection{Local $\IF_1$}
\begin{figure}[h]
\begin{center}
\includegraphics[scale=0.18]{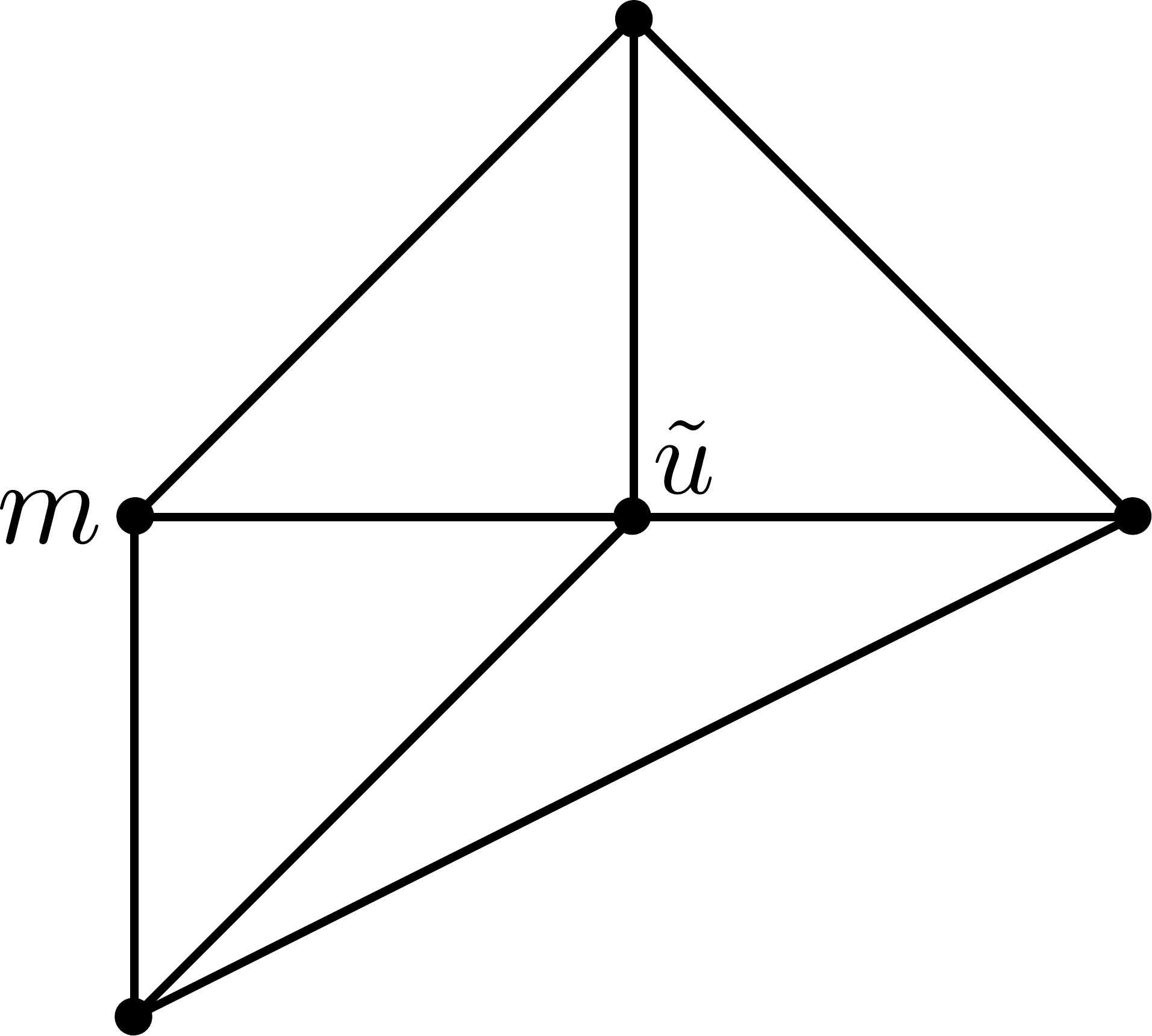}
\caption{\label{fig:toric_localf1} 2d toric fan of the geometry $\cO(-K_{\IF_1})\rightarrow \IF_1$.}
\end{center}
\end{figure}
\begin{align}\label{eq:toric_localf1} 
 \begin{array}{l rrr rr} \toprule
  \multicolumn{4}{c}{\nu_i}    & l^{(1)} & l^{(2)}\\ \cmidrule{1-6}
    D_u    &   (\; 1&     0&   0 \;)&           -2& -1\\ 
    D_1    &   (\; 1&     1&   0 \;)&            1&  0\\ 
    D_2    &   (\; 1&     0&   1 \;)&            0&  1\\ 
    D_m    &   (\; 1&    -1&   0 \;)&            1& -1\\ 
    D_3    &   (\; 1&    -1&  -1 \;)&            0&  1\\ \bottomrule
  \end{array} 
\end{align}
The local $\IF_1$ geometry is the anti--canonical bundle over the first Hirzebruch surface $\IF_1$. It is in fact the del Pezzo surface $\CB_1$ which is a blow up of $\IP^2$ at one generic point. The toric data of local $\IF_1$ is given in \eqref{eq:toric_localf1}, and its toric fan projected onto the supporting hyperplane is given in Figure \ref{fig:toric_localf1}. The Batyrev coordinates for this geometry are given by
\begin{align}
z_1=\frac{m}{\tilde{u}^2}=m u^2\, , \quad z_2=\frac{1}{m\tilde{u}}=\frac{u}{m}\, .
\end{align}
Here we have used $r=1$ leading to $\tilde{u}=\frac{1}{u}$. The B--model spectral curve for this geometry can be written as
\begin{align}
W_{\IF_1}(e^x, e^y) &= e^x+e^y+m e^{-x}+m^{-1} e^{x-y} +\tilde{u} \, .
\end{align}
It is identical with the curve in Table~\ref{tb:delPezzos} up to a symplectic transformation, since both curves have the same  Weierstrass form with
\begin{align}
\begin{split}
g_2(u, m) &= 27 u^4 (1 - 8 m u^2 + 24 u^3 + 16 m^2 u^4) \, ,\\
g_3(u, m) &= 27 u^6 (1 - 12 m u^2 + 36 u^3 + 48 m^2 u^4 - 144 m u^5 + 216 u^6 - 64 m^3 u^6) \, .
\end{split}
\end{align}

Next we compute the leading contributions to the semiclassical phase space volume in the large energy limit following \cite{Huang:2014eha}. We find
\begin{align}\label{eq:localf1_vol0}
\textrm{vol}_0(E) = 4 E^2 - E \log(m) - \frac{1}{2}(\log m)^2 - \frac{2}{3}\pi^2 + \mathcal{O}(e^{-E}) \, .
\end{align}
We obtain from \cite{Huang:2014nwa} the differential operator $\cD_2$ for the calculation of the first quantum correction to the phase space volume, with the substitution $u = e^{-rE}$. After applying it on the semiclassical phase space volume, we find
\begin{equation}\label{eq:localf1_vol1}
	\vol_1(E) = -\frac{1}{24} C +\cO(e^{-E}) \ .
\end{equation}
From \eqref{eq:localf1_vol0} and \eqref{eq:localf1_vol1} we can then read off the following constants 
\begin{equation}
\begin{gathered}
C=4\ , \quad \quad D_0(m) = -\log m \ ,\\
B_0(m)=\frac{\pi}{3}-\frac{1}{4\pi} \left (\log m\right)^2 \ , \quad B_1=-\frac{1}{12\pi} \ .
\end{gathered}
\end{equation}


\subsubsection{Maximal supersymmetry}

\paragraph{Energy spectrum}
We start with computing the energy spectrum in the maximal supersymmetric case, using the quantization condition \eqref{eq:mSUSYqnz}. We compute the periods and the prepotential through \eqref{eq:spGeom}. Near the \texttt{LCP}, the A--period has the expansion
\begin{equation}
	t = -\log u - m u^2 + 2u^3 - \frac{3m^2u^4}{2} + \cO(u^5) \ ,
\end{equation}
and the prepotential is
\begin{equation}
	F_0^{\rm inst} = \frac{Q}{m} +\left(  \frac{1}{8m^2}-2m \right)Q^2 + \left( 3+\frac{1}{27m^3}\right)Q^3 + \cO(Q^4) \ ,
\end{equation}
where $Q = e^{-t}$. The flat coordinates associated to Batyrev coordinates satisfy
\begin{equation}\label{eq:F1ca}
	t_1=2 t-\log m \ , \qquad t_2= t+\log m \ .
\end{equation}
Therefore we can identify the mass function $Q_m$ with $m$. We can also read off the coefficients $c_\alpha,\alpha_{\alpha,j}$ from these relations. Plugging these data into the quantization condition \eqref{eq:mSUSYqnz}, we can compute the energy spectrum with arbitrary $m$. The ground state energies $E_0$ have been calculated in this way for $m=1, 2, 16$ respectively with both the A--period and the prepotential expanded up to order 20. The results are listed in the second column of Table~\ref{tb:F1mSUSYEn}. To check these results, we compute the ground state energies numerically following \cite{Huang:2014eha} as in the case of local $\IF_2$, using Hamiltonian matrices of size $500\times 500$. The numerical results are given in the last column of the same table. We find that the ground state energies obtained with the two different methods agree in all stabilized digits.

\begin{table}
\centering
\begin{tabular}{>{$}c<{$} >{$}l<{$} >{$}l<{$}}\toprule
m & E_0\textrm{ from conjecture} & E_0\textrm{ from numerics}\\\midrule
1 & 2.864004259408190 & 2.8640042594081907\\
2 & 2.971234582260921 & 2.971234582260921\\
16 & 3.428058805696 & 3.42805880569628\\\bottomrule
\end{tabular}
\caption{Ground state energy of $\mathsf{O}_{\IF_1}$ with $\hbar=2\pi$ computed from both the quantization condition \eqref{eq:mSUSYqnz} and with the numerical method with matrix size $500\times 500$. All stabilized digits are listed in the results.}\label{tb:F1mSUSYEn}
\end{table}

\paragraph{Spectral determinant}
As in the case of local $\IF_2$, we proceed to check the spectral determinant. First we compute $Z(N,2\pi)$ according to its definition, using the energy spectrum, which is generated by the quantization condition. The first two traces $Z(1,2\pi)$, $Z(2,2\pi)$ with $m=1, 2$ computed this way are given in Table~\ref{tb:F1mSUSYZNEn}. 

Next, we compute the same traces with the help of \eqref{eq:ZNAi} in terms of Airy functions and its derivatives, utilizing the complete expression of $J_X(\mu,m,2\pi)$. For this purpose, 
we need the unrefined genus one free energy, which can be obtained from \cite{Haghighat:2008gw}
\begin{equation}
	F_1 = -\frac{1}{12}\log \Delta - \frac{2}{3}\log u + \frac{1}{2}\log \frac{\partial t}{\partial u} \ ,
\end{equation}
where
\begin{equation}
	\Delta = m-u-8m^2u^2+36mu^3-27u^4+16m^3u^4 \ ,
\end{equation}
as well as the Nekrasov--Shatashvili genus one free energy, which can be derived following \cite{Huang:2013yta}
\begin{equation}
	F_1^{\rm NS} =  -\frac{1}{24}\log(\Delta u^{-4}) \ .
\end{equation}
We also need $A(m,2\pi)$, which is obtained by demanding $Z(0,2\pi)=1$. Then using \eqref{eq:ZNAi}, with free energies expanded up to order 20, we have computed the same fermionic spectral traces $Z(1,2\pi)$, $Z(2,2\pi)$ with $m=1, 2$, albeit with much higher precisions. The results are given in Table~\ref{tb:F1mSUSYZNAi}. Due to space constraint, we only list the first 30 digits after the decimal point. They agree with the numerical results from Table~\ref{tb:F1mSUSYZNEn}.

\begin{table}
\centering
\begin{tabular}{*{2}{>{$}c<{$}} *{1}{>{$}l<{$}} }\toprule
m & \multicolumn{2}{l}{$Z(N,2\pi)$}  \\\midrule
1 & Z(1,2\pi) & 0.0806271202574356  \\
  & Z(2,2\pi) & 0.00150651698090292  \\
2 & Z(1,2\pi) & 0.0726979945606611 \\
  & Z(2,2\pi) & 0.00123301803769142 \\\bottomrule
\end{tabular}
\caption{$Z(1,2\pi)$, $Z(2,2\pi)$ of $\mathsf{O}_{\IF_1}$ computed from the spectrum. All the stabilized digits are listed in the results.}\label{tb:F1mSUSYZNEn}
\end{table}

\begin{table}
\centering
\begin{tabular}{*{2}{>{$}c<{$}} *{1}{>{$}l<{$}} c }\toprule
m &   \multicolumn{2}{l}{$Z(N,2\pi)/A(m,2\pi)$}   & Precisions \\\midrule
1 & Z(1,2\pi) & 0.080627120257435627781494805115\ldots & 75\\
  & Z(2,2\pi) & 0.001506516980902928703412802925\ldots & 77\\
  & A(1,2\pi) & 0.307577965374980255036479594884\ldots & 74\\  
2 & Z(1,2\pi) & 0.072697994560661149574438010102\ldots & 74\\
  & Z(2,2\pi) & 0.001233018037691426072124489653\ldots & 76\\
  & A(2,2\pi) & 0.310522603835097060481991044711\ldots & 73\\\bottomrule
\end{tabular}
\caption{$Z(1,2\pi)$, $Z(2,2\pi)$, as well as $A(m,2\pi)$ of local $\IF_1$ computed with the Airy function method. The results are listed with the first 30 digits after the decimal point. The column ``Precisions'' gives the number of stabilized digits after the decimal point in each result.}\label{tb:F1mSUSYZNAi}
\end{table}

\subsubsection{Rational Planck constants}

\begin{table}
\centering
\begin{tabular}{*{2}{>{$}c<{$}} c >{$}l<{$}  *{2}{>{$}c<{$}} }\toprule
 \hbar & m &  \multicolumn{2}{c}{Ground state energies} & \textrm{Errors} & \textrm{Deviations} \\\midrule
 3\pi   & 1 & w/ $\lambda(E)$   & 3.5607250021035 & 1.2\times 10^{-14} & - \\
        &   & w.o. $\lambda(E)$ & 3.5607249988919 & 9.2\times 10^{-15} & 3.2\times 10^{-9} \\
        &   & numerical         & 3.5607250021036 & & \\\cmidrule{2-6}
        & 2 & w/ $\lambda(E)$   & 3.6638398827159 & 5.0\times 10^{-14} & - \\
        &   & w.o. $\lambda(E)$ & 3.6638398798168 & 4.3\times 10^{-14} & 2.9\times 10^{-9} \\
        &   & numerical         & 3.663839882715 & & \\\midrule      
 8\pi/3 & 1 & w/ $\lambda(E)$   & 3.331429227013371 & 6.1\times 10^{-16} & - \\
        &   & w.o. $\lambda(E)$ & 3.331429227058059 & 5.5\times 10^{-16} & 4.5\times 10^{-11} \\
        &   & numerical         & 3.33142922701337 & & \\\cmidrule{2-6}
        & 2 & w/ $\lambda(E)$   & 3.43562063022815 & 2.2\times 10^{-15} & - \\
        &   & w.o. $\lambda(E)$ & 3.43562063022241 & 2.1\times 10^{-15} & 5.7\times 10^{-12} \\
        &   & numerical         & 3.4356206302281 & & \\\bottomrule    
\end{tabular}
\caption{Ground state energies for $\mathsf{O}_{\IF_1}$ with rational $\hbar$, computed by the complete quantization condition \eqref{eq:gen_qnz} with $\lambda(E)$ (rows labeled by ``w/ $\lambda(E)$''), by incomplete quantization condition without $\lambda(E)$ (rows labeled by ``w.o. $\lambda(E)$''), and by the numerical method with matrices of size $500\times 500$ (rows labeled by ``numerical''). The results are given with all stabilized digits. Other notations are the same as in Table~\ref{tb:F2_spec_rat_h}.}\label{tb:F1_spec_rat_h}
\end{table}

Here we check the conjecture for local $\IF_1$ with generic rational Planck constants. We will need the quantum A--period in the definition of $\tmu$, and higher genera free energies of unrefined topological string and refined topological string in the Nekrasov--Shatashvili limit. The quantum A--period for local $\IF_1$ can be found in \cite{Huang:2014nwa},
\begin{equation}
	\widetilde{\Pi}_A(u,m,q) = m u^2 -\left( \frac{1}{q^{1/2}} + q^{1/2}\right) u^3 + \frac{3m^2 u^4}{2} + \cO(u^5) \ ,
\end{equation}
where $q=\exp(\ri\hbar)$. As for (refined) topological string free energies, we have computed BPS numbers for local $\IF_1$ following \cite{Huang:2013yta} with
\begin{equation}
	d = 2d_1 + d_2 \leqslant 16 
\end{equation}
(See the definition of $d$ in \eqref{eq:d}. The coefficients are read off from \eqref{eq:F1ca}). As in the example of local $\IF_2$, we collect these BPS numbers in an ancillary \texttt{Mathematica} notebook attached to this paper.

Using the data above, we have computed the ground state energies using \eqref{eq:gen_qnz} and \eqref{eq:rat_h_qnz} for $\hbar=3\pi, 8\pi/3$ and mass parameters $m=1, 2$. We list the results with all stabilized digits in Table~\ref{tb:F1_spec_rat_h}. As in the example of local $\IF_2$, we have also computed the ground state energies with the same parameters but using the incomplete quantization condition without the $\lambda(E)$ correction, and give the results in the same table. Finally, Table~\ref{tb:F1_spec_rat_h} also contains the ground state energies computed by the numerical method with Hamiltonian matrices of size $500\times 500$. As in the case of local $\IF_2$, the results of the complete quantization condition agree with the numerical results\footnotemark, while the results of the incomplete quantization condition deviate by margins much greater than estimated errors.

\footnotetext{The ground state energy from the complete quantization condition with $\hbar=3\pi$ and $m=1$ seems to differ from the numerical result by a margin slighter larger than the estimated error. This probably can be explained by slow convergence of the numerical result.}

\begin{table}[t]
\centering
\begin{tabular}{*{3}{>{$}c<{$}} c >{$}l<{$} >{$}c<{$} } \toprule
\hbar & m & N & \multicolumn{2}{c}{Fermionic spectral traces $Z(N,m,\hbar)$} & \textrm{Precisions} \\\midrule
3\pi    & 1 & 1 & spectrum & 0.036084275689317 & 15 \\
        &   &   & Airy     & 0.03608427568931732310\ldots & 39\\\cmidrule{3-6}
        &   & 2 & spectrum & 0.00023208274513657 & 17 \\
        &   &   & Airy     & 0.00023208274513657062\ldots & 41\\\cmidrule{2-6}
        & 2 & 1 & spectrum & 0.03261451488440 & 14 \\
        &   &   & Airy     & 0.03261451488440392788\ldots & 39\\\cmidrule{3-6}
        &   & 2 & spectrum & 0.0001907533896520 & 16 \\
        &   &   & Airy     & 0.00019075338965202899\ldots & 42\\\midrule
8\pi/3  & 1 & 1 & spectrum & 0.046652991710186 & 15 \\
        &   &   & Airy     & 0.04665299171018683045\ldots & 44\\\cmidrule{3-6}
        &   & 2 & spectrum & 0.00042045881232812 & 17 \\
        &   &   & Airy     & 0.00042045881232812924\ldots & 45\\\cmidrule{2-6}
        & 2 & 1 & spectrum & 0.04213857928256 & 14 \\
        &   &   & Airy     & 0.04213857928256301516\ldots & 43\\\cmidrule{3-6}
        &   & 2 & spectrum & 0.0003451957488795 & 16 \\
        &   &   & Airy     & 0.00034519574887953481\ldots & 45\\\bottomrule
\end{tabular}
\caption{$Z(1,\hbar)$ and $Z(2,\hbar)$ for $\mathsf{O}_{\IF_1}$ with rational $\hbar$, computed from spectrum (rows labelled by ``spectrum'', with all stabilized digits), and by the Airy function method (rows labelled by ``Airy'', with the first 20 digits after the decimal point). The column ``Precisions'' gives the numbers of stabilized digits after the decimal point.}\label{tb:F1_Z_rat_h}
\end{table}

Next, we compute the first few fermionic spectral traces. This is first done by using the spectrum generated by the complete quantization condition. For $\hbar=3\pi, 8\pi/3$ and $m=1, 2$, the results are given in Table~\ref{tb:F1_Z_rat_h} including all stabilized digits. Then the fermionic spectral traces are computed by the Airy function method with \eqref{eq:ZNAi}, utilizing the entire modified grand potential. Using the available BPS numbers with $d\leqslant 16$, we can compute the fermionic spectral traces with up to $39\sim 45$ stabilized digits after the decimal point. We list the results with the first 20 digits in the same table, and they agree with the results obtained from spectrum.

\subsection{Local $\CB_2$}
\begin{figure}[h]
\begin{center}
  \includegraphics[scale=0.2]{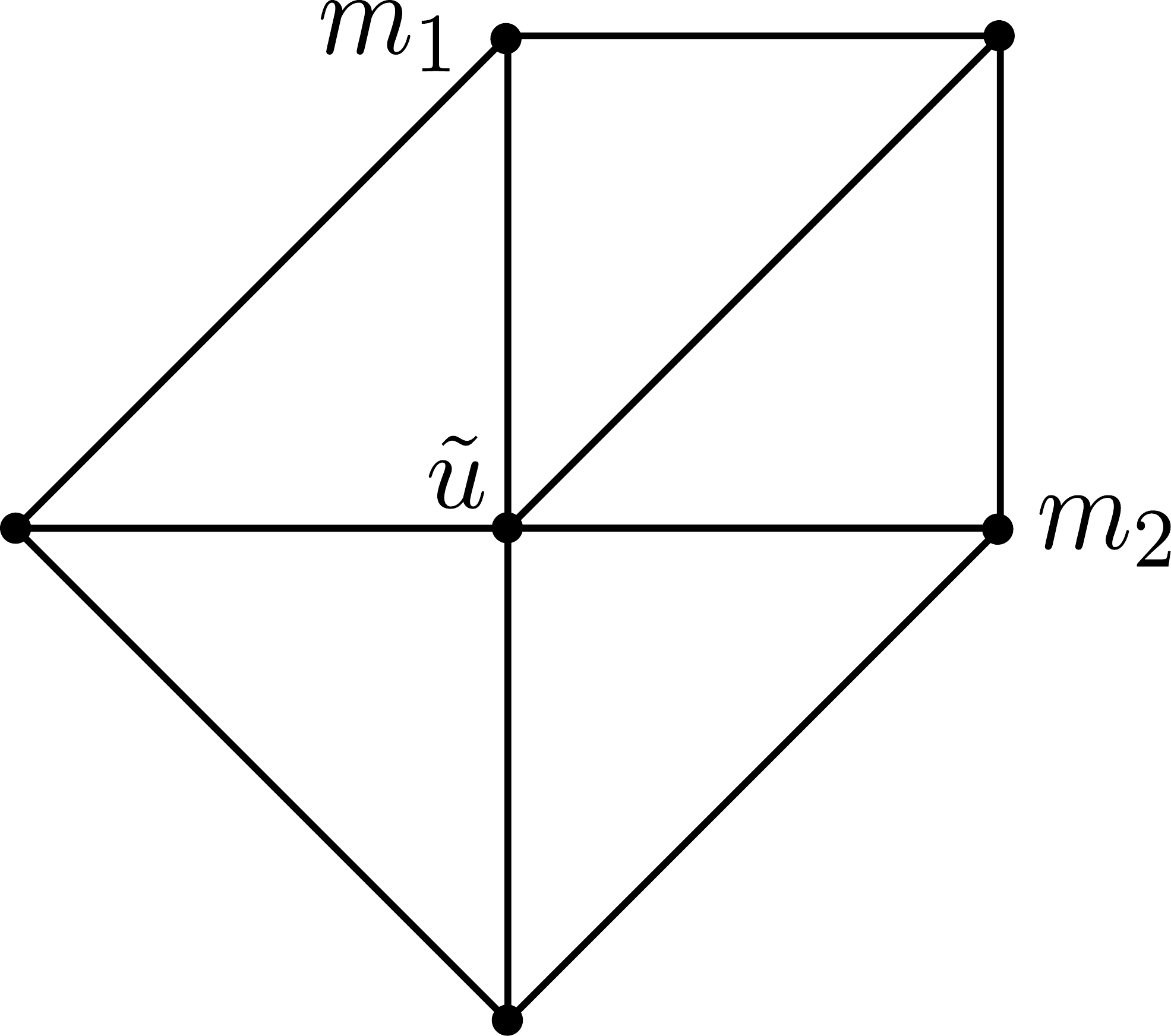}
  \caption{\label{fig:toric_localb2}2d toric fan of $\cO(-K_{\CB_2})\rightarrow \CB_2$.}
\end{center}
\end{figure}
This geometry is based on the del Pezzo surface $\CB_2$ which is a two--point blow--up of $\IP^2$. The toric data of local $\CB_2$ depicted in Figure \ref{fig:toric_localb2} are
\begin{align}\label{eq:toric_localb2} 
 \begin{array}{ l rrr rrr } \toprule
    \multicolumn{4}{c}{\nu_i}    &l^{(1)} & l^{(2)} &l^{(3)}\\  \cmidrule{1-7}
    D_u    & (\;1&  0&  0\;)&  -1& -1& -1\\ 
    D_{m_2}& (\;1&  1&  0\;)&  -1&  1&  0\\ 
    D_1    & (\;1&  1&  1\;)&   1& -1&  1\\ 
    D_{m_1}& (\;1&  0&  1\;)&   0&  1& -1\\ 
    D_2    & (\;1& -1&  0\;)&   0&  0&  1\\ 
    D_3    & (\;1&  0& -1\;)&   1&  0&  0\\ \bottomrule
  \end{array}
\end{align}
From the toric data we can read off the Batyrev coordinates
\begin{align}
z_1 = \frac{1}{\tilde{u} m_2}=\frac{u}{m_2} \, , \qquad z_2 = \frac{m_1 m_2}{\tilde{u}}=m_1 m_2 u \, , \qquad z_3 = \frac{1}{\tilde{u} m_1}=\frac{u}{m_1} \, .
\end{align}
Here we used $u=\frac{1}{\tilde{u}}$ which implies $r=1$. We can write the B--model spectral curve as
\begin{align}
	W_{\CB_2}(e^x,e^y)=  e^x + e^y + m_2 e^{-x} + m_1 m_2 e^{y-x} + m_2^{-1} e^{x-y} + \tilde{u} \ .
\end{align}
It can be identified with the curve in Table~\ref{tb:delPezzos} up to a symplectic transformation, since both of them can be converted to the same Weierstrass form with \cite{Huang:2013yta, Huang:2014nwa}
\begin{equation}\label{eq:B2g23}
\begin{aligned}
\begin{split}
g_2(u,m_1,m_2) & = 27 u^4 \left ( 1 - 8 \left(m_1+m_2\right) u^2 + 24 u^3 + 16 \left(m_1^2-m_1 m_2+m_2^2\right) u^4 \right ) \, ,\\
g_3(u,m_1,m_2) & = 27 u^6 \left(1 -12\left(m_1+m_2\right) u^2 +36 u^3 +24\left(2m_1^2+m_2 m_1+2m_2^2\right) u^4 \right .\\
 \phantom{ ={}} &\left . - 144 \left(m_1+m_2\right) u^5 + \left(-64 m_1^3+96 m_2 m_1^2+96 m_2^2 m_1-64 m_2^3+216\right) u^6 \right) \, .
\end{split}
\end{aligned}
\end{equation}

To find the coefficients $C$, $D_0(\um)$, $B_0(\um)$, and $B_1$, we calculate the semiclassical phase space volume as in \cite{Huang:2014eha}. In the large energy limit we find
\begin{align}\label{eq:localb2_phase_space0}
\vol_0(E) =& \frac{7}{2}E^2 - \left (\log m_1 + \log m_2 \right ) E - \frac{1}{2}\left[ \left (\log m_1 \right )^2 + \left (\log m_2 \right )^2\right] - \frac{5}{6} \pi^2 + \mathcal{O}(e^{-E}) \, ,
\end{align}
Using $u=e^{-E}$ we can translate the quantum operator $\cD_2$ from \cite{Huang:2014nwa} to a differential operator with respective to energy. Since the quantum phase space volume can be identified with the quantum B--period of the spectral curve, we can apply the operator $\cD_2$ to obtain the first quantum correction to the phase space volume
\begin{align}\label{eq:localb2_phase_space1}
\vol_1(E) =& -\frac{5}{24} + \mathcal{O}(e^{-E}) \, ,
\end{align}
up to exponentially suppressed corrections.
From the phase space volumes \eqref{eq:localb2_phase_space0} and \eqref{eq:localb2_phase_space1} we can read off the desired coefficients
\begin{equation}\label{eq:B2CDB}
\begin{gathered}
	C=\frac{7}{2}\ , \qquad  D_0(\um) = -(\log m_1 + \log m_2) \ , \\
	B_0(\um) = \frac{1}{6} \pi -\frac{1}{4\pi}\left[ \left (\log m_1 \right )^2 + \left (\log m_2 \right )^2\right] \, , \qquad B_1 = -\frac{5}{48\pi} \, .
\end{gathered}
\end{equation}

\subsubsection{Maximal supersymmetry}

\paragraph{Energy spectrum}
In the case of maximal supersymmetry we use the simplified quantization condition \eqref{eq:mSUSYqnz} to calculate the energy spectrum. In addition to the coefficients extracted in \eqref{eq:B2CDB}, we need the periods and the prepotential, which are computed by integrating out \eqref{eq:spGeom}, using $g_2(u,\um), g_3(u,\um)$ given in \eqref{eq:B2g23}. The classical A--period is
\begin{equation}\label{eq:B2t}
	t = -\log u -(m_1+m_2) u^2 + 2u^3 -\frac{3}{2}(m_1^2+4m_1m_2 m_2^2) u^4 + \cO(u^5) \ ,
\end{equation}
and the instanton part of the prepotential is
\begin{equation}\label{eq:B2F0}
	F_0^{\rm inst} = \left(\frac{1}{m_1}+\frac{1}{m_2} + m_1m_2\right) Q + \left( \frac{1}{8m_1^2} - 2m_1+ \frac{1}{8m_2^2}-2m_2 +\frac{m_1^2m_2^2}{8}\right)Q^2 + \cO(Q^3) \ ,
\end{equation}
where $Q = e^{-t}$. Furthermore, the flat coordinates $t_\alpha$ associated to the Batyrev coordinates satisfy
\begin{equation}
t_1 = t + \log m_2 \, , \qquad t_2 = t - \log m_1 - \log m_2 \, , \qquad t_3 = t + \log m_1 \, .
\end{equation}
Therefore we can choose $Q_{m_1} = m_1, Q_{m_2} = m_2$, and read off the coefficients $\tca,\alpha_{\alpha, j}$. We plug these data in the quantization condition \eqref{eq:mSUSYqnz}, and compute the ground state energy for combinations of mass parameters $(m_1, m_2) = (1,1)$, $(1,2)$, $(3,2)$, with the A--period and the prepotential expanded up to order 17. The results are listed in the second column of Table~\ref{tb:B2mSUSYEn}. Next, we compute the ground state energy numerically as in the previous examples along the line of \cite{Huang:2014eha}, using Hamiltonian matrices of size $500\times 500$, and list the results in the last column of Table~\ref{tb:B2mSUSYEn}. Again the results from the quantization condition and from the numerical calculation agree.

\begin{table}
\centering
\begin{tabular}{>{$}c<{$} >{$}l<{$} >{$}l<{$}}\toprule
(m_1,m_2) & E_0\textrm{ from conjecture} & E_0\textrm{ from numerics}\\\midrule
(1,1) & 3.1995075383598 & 3.19950753835985\\
(1,2) & 3.31222613819186 & 3.31222613819186\\
(3,2) & 3.4995746425315    & 3.49957464253155\\\bottomrule
\end{tabular}
\caption{Ground state energy of $\mathsf{O}_{\CB_1}$ with $\hbar=2\pi$ computed from both the quantization condition \eqref{eq:mSUSYqnz} and with the numerical method with matrix size $500\times 500$. All stabilized digits are listed in the results.}\label{tb:B2mSUSYEn}
\end{table}

\paragraph{Spectral determinant}

We follow the same computation as in the previous sections. First, we compute $Z(N,2\pi)$ from the energy spectrum, which is generated by using the quantization condition \eqref{eq:mSUSYqnz}. As examples we list $Z(1,2\pi)$, $Z(2,2\pi)$ for $(m_1,m_2)=(1,1)$, $(1,2)$, $(3,2)$ computed in this way in Table~\ref{tb:B2mSUSYZNEn}. Then we compute the same fermionic spectral traces by making use of the formula \eqref{eq:ZNAi} in terms of Airy functions and its derivatives. For this purpose, we need genus one free energies of local $\CB_2$. The unrefined genus one free energy can be obtained by the method described in Section~\ref{sec:f2_max_susy}, and we find
\begin{equation}
	F_1^{\rm inst} = -\frac{1}{12}\log \Delta - \frac{7}{12}\log u + \frac{1}{12}\log \frac{\partial t}{\partial u} \ ,
\end{equation}
where the discriminant $\Delta$ is
\begin{align}
	\Delta =& -m_1 m_2 +(m_1^2 m_2^2+m_1+m_2)u + (8 m_1^2 m_2+8 m_1 m_2^2-1)u^2-2 (4 m_1^3 m_2^2\nonumber\\
	 &+4 m_1^2 m_2^3 +4 m_1^2+23 m_1 m_2+4 m_2^2)u^3-4  (4 m_1^3 m_2-16 m_1^2 m_2^2+4 m_1 m_2^3 \nonumber\\
	 &-9 m_1-9 m_2)u^4 + (16 m_1^4 m_2^2-32 m_1^3 m_2^3+16 m_1^3+16 m_1^2 m_2^4 \nonumber\\
	 &-24 m_1^2 m_2-24 m_1 m_2^2+16 m_2^3-27)u^5 \ .
\end{align}
The Nekrasov--Shatashvili genus one free energy is derived following the prescription in \cite{Huang:2013yta}
\begin{equation}
	F^{\rm NS, inst}_1 = -\frac{1}{24}\log(\Delta u^{-5}) \ .
\end{equation}
With these free energies, together with the prepotential \eqref{eq:B2F0}, the A--period \eqref{eq:B2t}, and the coefficients in \eqref{eq:B2CDB}, we can first compute $A(\um,2\pi)$ by the normalization condition $Z(0,2\pi)=1$, and then proceed to compute $Z(1,2\pi)$, $(2,2\pi)$. In this process, we use free energies expanded up to order 17. The results with same mass combinations $(m_1,m_2)=(1,1), (1,2), (3,2)$ are listed in Table~\ref{tb:B2mSUSYZNAi}. They agree with Table~\ref{tb:B2mSUSYZNEn} in all stabilized digits.

\begin{table}
\centering
\begin{tabular}{*{2}{>{$}c<{$}} *{1}{>{$}l<{$}} }\toprule
(m_1,m_2) & \multicolumn{2}{l}{$Z(N,2\pi)$}  \\\midrule
(1,1) & Z(1,2\pi) & 0.056125936740909  \\
  & Z(2,2\pi) & 0.00069099641289523  \\
(1,2) & Z(1,2\pi) & 0.050226831674578  \\
  & Z(2,2\pi) & 0.00055527599449115  \\  
(3,2) & Z(1,2\pi) & 0.041756014419873 \\
  & Z(2,2\pi) & 0.00038586924264883 \\\bottomrule
\end{tabular}
\caption{$Z(1,2\pi)$, $Z(2,2\pi)$ of $\mathsf{O}_{\CB_2}$ computed from the spectrum. In the results all the stabilized digits are listed.}\label{tb:B2mSUSYZNEn}
\end{table}

\begin{table}
\centering
\begin{tabular}{*{2}{>{$}c<{$}} *{1}{>{$}l<{$}} c }\toprule
(m_1,m_2) &   \multicolumn{2}{l}{$Z(N,2\pi)/A(m,2\pi)$}   & Precisions \\\midrule
(1,1) & Z(1,2\pi) & 0.056125936740909740204777674742\ldots & 65\\
      & Z(2,2\pi) & 0.000690996412895238460921136534\ldots & 67\\
      & A(1,2\pi) & 0.525034841250017873187431756727\ldots & 64\\  
(1,2) & Z(1,2\pi) & 0.050226831674578744141567494273\ldots & 64\\
      & Z(2,2\pi) & 0.000555275994491156163380800220\ldots & 66\\
      & A(1,2\pi) & 0.552375136623000884957104480047\ldots & 63\\ 
(3,2) & Z(1,2\pi) & 0.041756014419873813182581770745\ldots & 59\\
      & Z(2,2\pi) & 0.000385869242648837520056887594\ldots & 61\\
      & A(2,2\pi) & 0.631961797812417591353288642144\ldots & 57\\\bottomrule
\end{tabular}
\caption{$Z(1,2\pi)$, $Z(2,2\pi)$, as well as $A(m,2\pi)$ of local $\CB_2$ computed with the Airy function method. The first 30 digits after the decimal point are listed for each result, while the total number of stabilized digits are given in the column ``Precisions''.}\label{tb:B2mSUSYZNAi}
\end{table}

\subsection{Mass deformation of local $E_8$ del Pezzo surface}
%
The toric data for the mass deformation of the local $E_8$ del Pezzo surface were given in Section \ref{mass-e8}.
The full $E_8$ del Pezzo surface is the blow--up of $\IP^2$ in eight generic points and can be constructed as a hypersurface in $\IP(1,1,2,3)$. The two geometries have identical prepotentials for vanishing masses. 
As explained in Section \ref{mass-e8}, the spectral curve can be written as a deformation of the function $\cO_{3,2}$
\begin{align}
W_{E_8}(x,y)=  e^x + e^p + e^{-3x-2y} + m_1 e^{-x-y} + m_2 e^{-2x-y} + m_3 e^{-x} +  \tilde{u}\ .
\end{align}

Analogous to the calculation in \cite{Huang:2014eha} we compute the semiclassical phase space volume which is the B--period of the spectral curve. In the large energy limit, we find
\begin{equation}\label{eq:locale8_phase_0}
\begin{aligned}
\textrm{vol}_0(E) = 3 E^2 + \frac{3}{2}(\log m_2 - \log m_3)^2 - &\left( \log\frac{m_1 + \sqrt{m_1^2-4}}{2} \right)^2 \\
&- \frac{1}{2}\sum_{i=1}^3 \left(\log(-e_i) \right)^2 - \pi^2 + \cO(e^{-E}) \, .
\end{aligned}
\end{equation}
Here $e_i, i=1,2,3$ are the three roots of the cubic equation
\begin{align}\label{eq:locale8_roots}
s^3 + \frac{m_2^2}{m_3} s^2 + \frac{m_2^2}{m_3} s + \frac{m_2^3}{m_3^3} = 0  \, .
\end{align}
Using $u=e^{-E}$ we can translate the quantum operator $\cD_2$ given in \cite{Huang:2014nwa} for the mass deformed $E_8$ geometry to a differential operator with respect to the energy. Applying this operator to the semiclassical phase space volume, we find the following first quantum correction to the phase space volume up to exponentially suppressed terms
\begin{align}\label{eq:locale8_phase_1}
\textrm{vol}_1(E) &= -\frac{1}{4} + \mathcal{O}(e^{-E}) \, .
\end{align}
Comparing \eqref{eq:locale8_phase_0} and \eqref{eq:locale8_phase_1} to the general expression for the phase space volume \eqref{eq:vol0E}, \eqref{eq:vol1E}, \eqref{eq:B1}, we can read off the coefficients $C$, $D_0(\um)$, $B_0(\um)$, and $B_1$
\begin{equation}\label{eq:E8CDB}
\begin{gathered}
C=3\, , \qquad D_0(\um) = 0 \ ,\qquad B_1 = -\frac{1}{8\pi} \, , \\
B_0 = \frac{1}{2\pi} \left\{ \frac{3}{2}(\log m_2 - \lg m_3)^2 - \left( \log\frac{m_1 + \sqrt{m_1^2-4}}{2} \right)^2 - \frac{1}{2}\sum_i \left(\log(-e_i) \right)^2 \right\} \, .
\end{gathered}
\end{equation}

\subsubsection{Maximal supersymmetry}

\paragraph{Energy spectrum}
We first use the quantization condition in the maximal supersymmetric case \eqref{eq:mSUSYqnz} to compute the energy spectrum. Other than the coefficients in \eqref{eq:E8CDB}, we need the periods and the prepotential, which are computed from \eqref{eq:spGeom} using $g_2(u,\um), g_3(u,\um)$ from \eqref{g2g3}. Near the \texttt{LCP}, the A--period has the expansion
\begin{equation}
	t = -\log u -m_2 u^2 + 2m_1 u^3 -\frac{3}{2}(2m_2 + m_2^2) u^4+ \cO(u^5) \ ,
\end{equation}
and the instanton part of the prepotential has the expansion
\begin{equation}
	F_0^{\rm inst} = m_1 m_2 Q - \left(  \frac{3 m_2}{2} + \frac{m_2^2}{4}+ \frac{m_1^2 m_2}{4} - \frac{m_1^2 m_2^2}{8}\right) Q^2 + \cO(Q^3) \ ,
\end{equation}
where $Q = e^{-t}$. Furthermore, the flat coordinates $t_\alpha$ satisfy
\begin{equation}\label{eq:E8ca}
\begin{gathered}
t_2 = t + \frac{1}{2} \log Q_{m_1} + \frac{2}{3} \log Q_{m_2} + \frac{1}{3}\log Q_{m_3} \, , \\
t_1 = -\log Q_{m_1} \, , \qquad t_3 = -\log Q_{m_2} \, , \qquad t_4 = -\log Q_{m_3} \, ,
\end{gathered}
\end{equation}
where the mass functions $Q_{m_j}$ are related to the mass parameters through the following rational relations
\begin{align}\label{eq:E8MassFunctions}
m_1 = \frac{1+Q_{m_1}}{\sqrt{Q_{m_1}}} \, , \quad m_2 = \frac{1+Q_{m_2} + Q_{m_2} Q_{m_3}}{Q_{m_2}^{2/3}Q_{m_3}^{1/3}} \, , \quad m_3 = \frac{1+Q_{m_3}+ Q_{m_2} Q_{m_3}}{Q_{m_2}^{1/3}Q_{m_3}^{2/3}} \, .
\end{align}
We can also read off the coefficients $\tca, \alpha_{\alpha, j}$ from \eqref{eq:E8ca}. With these data, we used \eqref{eq:mSUSYqnz} to compute the ground state energies for the mass combinations $(m_1,m_2,m_3) = (0,0,0), (2,3,3)$, with the A--period and the prepotential expanded up to order 20, and list the results in the second column of Table~\ref{tb:E8mSUSYEn}\footnotemark. These energies can be verified by numerical calculations similar to previous examples, and the corresponding results are listed in the last column of Table~\ref{tb:E8mSUSYEn} (we used Hamiltonian matrices of size $800\times 800$ here). Again the conjecture reproduces the numerical results in all stabilized digits.

\footnotetext{The mass combination $(m_1,m_2,m_3) = (0,0,0)$ is obtained by first setting $m_1=0, m_2=m_3 = m >0$, and then sending $m \rightarrow 0$. In this way, $\vol_0(E)$ in \eqref{eq:locale8_phase_0} and the cubic equation in \eqref{eq:locale8_roots} always remain finite.
}

\begin{table}
\centering
\begin{tabular}{>{$}c<{$} >{$}l<{$} >{$}l<{$}}\toprule
(m_1,m_2,m_3) & E_0\textrm{ from conjecture} & E_0\textrm{ from numerics}\\\midrule
(0,0,0) & 3.298393786995024728240 & 3.2983937\\
(2,3,3) & 3.59765161280909860 & 3.597651612809\\\bottomrule
\end{tabular}
\caption{Ground state energy $E_0$ of $\mathsf{O}_{E_8}$ with $\hbar=2\pi$ computed from both the quantization condition \eqref{eq:mSUSYqnz} and with the numerical method with matrix size $800\times 800$. In the results all stabilized digits are listed.}\label{tb:E8mSUSYEn}
\end{table}

\paragraph{Spectral determinant}

Similar to previous examples, we first compute the fermionic spectral traces $Z(N,2\pi)$ from the energy spectrum, which we generate through the quantizaton condition \eqref{eq:mSUSYqnz}. For the mass combinations $(m_1,m_2,m_3) = (0,0,0)$, $(2,3,3)$, the first two traces computed in this way are given in Table~\ref{tb:E8mSUSYZNEn}. Next, we compute the same traces from the spectral determinant, using formula \eqref{eq:ZNAi} in terms of Airy functions and its derivatives. For this, we need the unrefined genus one free energy  and the Nekrasov--Shatashvili limit genus one free energy. The former is computed by the method described in Section~\ref{sec:f2_max_susy}, and we find
\begin{equation}
	F_1 = -\frac{1}{12}\log\Delta -\frac{1}{2}\log u +\frac{1}{2}\log\frac{\partial t}{\partial u} \ ,
\end{equation}
where the discriminant $\Delta$ is
\begin{align}
	\Delta =& 1-m_1 m_2 u +(m_2^2-12 m_3+m_1^2 m_3)u^2 + (36m_1-m_1^3+8m_1m_2m_3)u^3 \nonumber\\
	&+(-72m_2-30m_1^2m_2-8m_2^2m_3+48m_3^2-8m_1^2m_3^2) u^4 +(96m_1m_2^2-144 m_1m_3 \nonumber\\
	&+36m_1^3m_3-16m_1m_2m_3^2)u^5 +(-432+216m_1^2-27m_1^4-64m_2^3+288m_2m_3 \nonumber\\
	&-72m_1^2m_2m_3+16m_2^2m_3^2-64m_3^3+16m_1^2 m_3^3) u^6 \ .\label{eq:E8Disc}
\end{align}
The latter is derived following \cite{Huang:2013yta} and the result is
\begin{equation}
	F^{\rm NS}_1 = -\frac{1}{24}\log(\Delta u^{-6}) \ .
\end{equation}
Then we can compute $A(2\pi)$ by the normalization condition $Z(0,2\pi)=1$, and furthermore proceed to compute $Z(1,2\pi), Z(2,2\pi)$. In this process, we always use free energies expanded up to order 20. The results are given in Table~\ref{tb:E8mSUSYZNAi}. We find yet again agreement with Table~\ref{tb:E8mSUSYZNEn} from the numerical method.

The example with mass combination $(m_1,m_2,m_3) = (0,0,0)$ is particularly interesting, as here the traces $Z(N,2\pi)$ can be directly computed from 
the kernel (\ref{ex-k}). One finds from (\ref{two-traces})
\begin{equation}
\begin{aligned}
	Z(1,2\pi) &= \frac{1}{12\sqrt{3}} \ ,\\
	Z(2,2\pi) &= \frac{7}{864} - \frac{1}{23\sqrt{3}\pi} \ .
\end{aligned}
\end{equation}
They agree with our results in all the 85 plus stabilized digits. Furthermore, when all the mass parameters are turned off, the form of $A(\um,2\pi)$ has been conjectured in \cite{Hatsuda:2015oaa}, and it translates to
\begin{equation}
	A(2\pi) = \frac{\log (2)}{4} + \frac{\log (3)}{6} - \frac{5\zeta(3)}{24\pi^2} \ .
\end{equation}
It also agrees with our result in all the stabilized digits.

\begin{table}
\centering
\begin{tabular}{*{2}{>{$}c<{$}} *{1}{>{$}l<{$}} }\toprule
(m_1,m_2,m_3) & \multicolumn{2}{l}{$Z(N,2\pi)$}  \\\midrule
(0,0,0) & Z(1,2\pi) & 0.04811252243246881370910  \\
        & Z(2,2\pi) & 0.0004445060821047400530834  \\
(2,3,3) & Z(1,2\pi) & 0.036508307084758465 \\
        & Z(2,2\pi) & 0.000271580920140099445 \\\bottomrule
\end{tabular}
\caption{$Z(1,2\pi)$, $Z(2,2\pi)$ of $\mathsf{O}_{E_8}$ computed from the spectrum. In the results all the stabilized digits are listed.}\label{tb:E8mSUSYZNEn}
\end{table}

\begin{table}
\centering
\begin{tabular}{*{2}{>{$}c<{$}} *{1}{>{$}l<{$}} c }\toprule
(m_1,m_2,m_3) &   \multicolumn{2}{l}{$Z(N,2\pi)/A(m,2\pi)$}   & Precisions \\\midrule
(0,0,0) & Z(1,2\pi) & 0.048112522432468813709095731708\ldots & 87\\
        & Z(2,2\pi) & 0.000444506082104740053083428264\ldots & 86\\
        & A(1,2\pi) & 0.331015129036010216936639294459\ldots & 86\\  
(2,3,3) & Z(1,2\pi) & 0.036508307084758465484352644702\ldots & 88\\
        & Z(2,2\pi) & 0.000271580920140099445491626397\ldots & 88\\
        & A(2,2\pi) & 0.794548079957547835370107278880\ldots & 88\\\bottomrule
\end{tabular}
\caption{$Z(1,2\pi)$, $Z(2,2\pi)$, as well as $A(m,2\pi)$ for mass deformed local $E_8$ del Pezzo surface computed with the Airy function method. Only the first 30 digits after the decimal point are listed for each result, while the total number of stabilized digits are given in the column ``Precisions''.}\label{tb:E8mSUSYZNAi}
\end{table}

\subsubsection{Conifold point prepotential}

Here we want to check the 't Hooft expansion of the logarithm of the fermionic spectral trace presented in Section \ref{st-mm}. Let the operator $\mathsf{O}_X$ be the perturbation of the operator $\mathsf{O}_{m,n}$. It was shown in \cite{mz} that in the 't Hooft limit
\[		N \rightarrow \infty\ , \quad \hbar \rightarrow \infty \ , \quad \frac{N}{\hbar} = \lambda \; \textrm{ finite,} \]
the mass parameters should also be scaled accordingly by
\begin{equation}
	 \frac{\log Q_{m_j}}{\hbar} \;\textrm{ finite} .
\end{equation}
In particular, we can choose $Q_{m_j} = 1$. In this case, \cite{ghm} implies the functions $\cF_g^{(m,n)}$ appearing in the 't Hooft expansion (\ref{zmn-ex}) 
coincide with the (unrefined) topological string free energies\footnote{These are actually the ``skewed'' free energies in the sense of \eqref{eq:F0full_skew}, i.e., $t$ in the instanton part of the free energy is shifted by $r\pi\ri$, while it remains unshifted in the classical part of the free energy. When $r$ is even, they coincide with the usual topological string free energies.\label{fn:skewed}}
 at a conifold point with the mass parameters $m_j$ set to proper values. In particular, the conifold prepotential is defined by
\begin{equation}\label{eq:conBPeriod}
	\frac{\partial \cF^{(m,n)}_0}{\partial \lambda} = -\frac{t}{2\pi} \ .
\end{equation}
Here $t$ is the flat coordinate near the large complex structure point (\texttt{LCP}), and $\lambda$ the flat coordinate which vanishes at the conifold point. Besides, \cite{mz} predicts $\lambda$ is given by
\begin{equation}\label{eq:lambCFP}
	\lambda = \frac{r}{8\pi^3}\left( \frac{\partial F_0}{\partial t} + \frac{8\pi^3}{r} B_1 \right) \ ,
\end{equation}
with $F_0$ the prepotential near the \texttt{LCP}.

In the case of the mass deformed local $E_8$ del Pezzo surface, $\mathsf{O}_X$ is a deformation of $\mathsf{O}_{3,2}$. According to \eqref{eq:E8MassFunctions}, $Q_{m_j}$ being one corresponds to the mass combination
\begin{equation}\label{eq:E8MassComb}
	(m_1,m_2,m_3) = (2, 3, 3) \ .
\end{equation}
In this case, we can find three conifold points from the discriminant
\begin{equation}
	u = -1/6 \ , \quad u = 1/2 \ , \quad u = 1/3 \ .
\end{equation}
Furthermore, the conifold point flat coordinate $\lambda$ takes the form
\begin{equation}\label{eq:E8lamb}
	\lambda = \frac{1}{8\pi^3}\left( \frac{\partial F_0}{\partial t} - \pi^2 \right) \ .
\end{equation}
Since the B--period $\partial F_0/\partial t$ takes the value of $\pi^2$ at $u=-1/6$, the functions $\cF_g^{(3,2)}$ should be the free energies around this conifold point. Following \eqref{planar-f}, the conifold point prepotential has the expansion
\begin{equation}\label{eq:F032}
	\cF_0^{(3,2)}(\lambda) = -c_{3,2} \lambda + \frac{\lambda^2}{2} \left(\log \frac{\pi^2\lambda}{3\sqrt{3}}-\frac{3}{2}\right) + \sum_{k=3}^\infty f_{0,k} \lambda^k \ ,
\end{equation}
where $c_{3,2}$ is given in (\ref{cmnbw}), 
\begin{equation}
	c_{3,2} = -\frac{3}{\pi^2}D(2 e^{\ri\pi/3}) = -\frac{3}{\pi^3}\textrm{Im}(\textrm{Li}_2(2e^{\ri\pi/3})) + \frac{3}{2\pi}\log(2) \ .
\end{equation}

The expansion \eqref{eq:F032} of the conifold prepotential together with \eqref{eq:conBPeriod} implies that
\begin{equation}
	-\frac{t}{2\pi} = -c_{3,2} + \lambda\left( \log\frac{\pi^2}{3\sqrt{3}} - 1\right) + \lambda\log\lambda + \sum_{k=3}^\infty
	k f_{0,k} \lambda^k \ .
\end{equation}
Since $t$ has to be a linear combination of the periods at the conifold point, we can write
\begin{equation}\label{eq:tCFP}
	-\frac{t}{2\pi} = -c_{3,2} + \lambda\left( \log\frac{\pi^2}{3\sqrt{3}} - 1\right) + S \ ,
\end{equation}
where $S$ is the conifold point period with the leading behavior $\lambda\log \lambda+ \ldots$.

We can verify this relation through numerical analytic continuation of the periods from the \texttt{LCP} to the conifold point (\texttt{CFP}) $u=-1/6$. Let the \texttt{LCP} periods be $(1, t, \partial F_0/\partial t)$ and 
\begin{equation}
	v = u + 1/6 
\end{equation}
be the modulus near the \texttt{CFP}. We solve the Picard--Fuchs equation of the mass deformed local $E_8$ del Pezzo with the mass combination \eqref{eq:E8MassComb} around the \texttt{CFP}, and choose the periods $(1,\Pi_A^{\rm C}, \Pi_B^{\rm C})$, where
\begin{equation}
\begin{aligned}
	\Pi_A^{\rm C} &= v+ \frac{9v^2}{2}+ \frac{43v^3}{2}+\ldots \ ,\\
	\Pi_B^{\rm C} &= \log(v)(v+\frac{9v^2}{2} + \frac{43v^3}{2}+ \ldots) + 4v^2+\frac{219 v^3}{8}+\frac{7697v^4}{48} +\ldots \ .
\end{aligned}
\end{equation}
The two sets of periods are related by the transition matrix $M$ as
\begin{equation}
	\begin{pmatrix}
	1 \\ t \\ \frac{\partial F_0}{\partial t}
	\end{pmatrix} 
	= M \cdot
	\begin{pmatrix}
	1 \\ \Pi_A^{\rm C} \\ \Pi_B^{\rm C}
	\end{pmatrix} , \textrm{ where }
	M = \begin{pmatrix}
	1 & 0 & 0\\
	m_{1,0} & m_{1,1} & m_{1,2} \\
	m_{2,0} & m_{2,1} & m_{2,2} 
	\end{pmatrix}  \ ,		\label{eq:transition}
\end{equation}
and the entries of $M$ can be computed numerically with very high precision\footnote{In fact, since $r=1$ for mass deformed local $E_8$ del Pezzo, $\cF^{(3,2)}_0$ corresponds to the skewed prepotential as explained in footnote \ref{fn:skewed}. The corresponding periods are also ``skewed''. As a consequence, when performing analytic continuation we need to flip the sign of $u$ in $\log u$ in both $t$ and $\partial F_0/\partial t$.}. For instance, with the periods expanded up to 1500 terms, the entries of $M$ can be computed with approximately 450 reliable digits.

We find that $m_{2,0} = \pi$ and $m_{2,2} = 0$. Combined with \eqref{eq:E8lamb}, we conclude
\begin{equation}
	\Pi_A^{\rm C} = \frac{8\pi^3}{m_{2,1}} \lambda \ .
\end{equation}
Furthermore, if we plug the above relation into the asymptotic expression of $S$
\begin{align*}
	S &= \lambda \log(\lambda) + \cdots \\
	&= \frac{m_{2,1}}{8\pi^3} \Pi_A^{\rm C} \log\left( \frac{m_{2,1}}{8\pi^3} \Pi_A^{\rm C}\right)+\cdots \\
	&= \frac{m_{2,1}}{8\pi^3} \Pi_A^{\rm C} \log\left(  \Pi_A^{\rm C}\right)+\frac{m_{2,1}}{8\pi^3}\log\left(\frac{3m_{2,1}}{8\pi^3}\right) \Pi_A^{\rm C}+\cdots
\end{align*}
By looking at the series expansions of $\Pi_A^{\rm C}$ and $\Pi_B^{\rm C}$, we find that
\begin{equation}
	S = \frac{m_{2,1}}{8\pi^3}\Pi_B^{\rm C}  +\frac{m_{2,1}}{8\pi^3}\log\left(\frac{m_{2,1}}{8\pi^3}\right) \Pi_A^{\rm C} \ .	\label{eq:SPi}
\end{equation}

Now let us express $t$ in terms of the \texttt{CFP} periods. Combining \eqref{eq:transition} and \eqref{eq:SPi}, we find
\begin{equation}
	-\frac{t}{2\pi} = - \frac{m_{1,0}}{2\pi} + \frac{4\pi^2}{m_{2,1}}\left(m_{1,2} \log\left(\frac{m_{2,1}}{8\pi^3}\right)- m_{1,1}\right) \lambda -\frac{4\pi^2}{m_{2,1}}m_{1,2}\cdot  S   \ .
\end{equation}
Comparing this with the conjecture \eqref{eq:tCFP}, three identities are implied
\begin{equation}
\begin{gathered}
	m_{1,0}  = 2\pi c_{3,2}\ ,\\
	\frac{4\pi^2}{m_{2,1}}m_{1,2}  =- 1 \ ,\\
	\frac{4\pi^2}{m_{2,1}}\left(m_{1,2}\log\left(\frac{m_{2,1}}{8\pi^3}\right)- m_{1,1}\right)  = \log\left(\frac{\pi^2}{3\sqrt{3}}\right)-1\ ,
\end{gathered}
\end{equation}
all of which are verified up to 449 or 450 digits.

\section{Conclusions and outlook}

The conjecture put forward in \cite{ghm} postulates an intimate relationship between the spectral theory of certain class operators, obtained 
by quantization of mirror curves, and the enumerative geometry of the underlying CY threefolds. 
In this paper we have performed an extensive test of this conjecture for many local del Pezzo geometries. In addition, we have obtained a better understanding 
on the geometric realization of the operators, which has led in particular to a conjecturally exact solution for the spectral problem of the $\mO_{2,3}$ operator. Many of our 
tests have been done away from the maximally supersymmetric case, where we use all the available data on higher genus invariants of the 
CY. This allows us to test the conjecture of \cite{ghm} with very high precision. 

There are clearly many avenues for further research. On a technical level, some of our results can be certainly improved. It would be interesting to have 
exact expressions for the integral kernels of the operators for arbitrary masses, as it happens for local $\IF_0$. This would allow to perform more analytic tests. 
It would be also important to better understand the structure of the spectral determinant. As noted in \cite{ghm}, in the maximally supersymmetric case it has the 
same structure of the blowup functions appearing in Donaldson--Witten theory, and for general $\hbar$ it is a quantum deformation thereof. This is an intriguing connection 
which should be further explored. Another direction to explore is the generalization to mirror curves of higher genus. Many of the results of \cite{ghm} can be 
extended to this setting, and one can introduce for example a generalized spectral determinant related to higher genus Riemann theta functions \cite{cgm}, but clearly 
much more work is needed along this direction. 

Of course, it would be important to make steps towards a proof of the conjecture. From the point of view of spectral theory, the conjecture of \cite{ghm} 
supplements the perturbative WKB analysis of \cite{acdkv} with an infinite series of quantum-mechanical instanton corrections. It would be of course very interesting to have some 
way to calculate these corrections directly in spectral theory. Since these corrections are encoded in the Gopakumar--Vafa invariants, this would shed light on the enumerative 
geometry of toric CY manifolds from an unexpected angle. 

It has been emphasized in \cite{ghm,mz} that the conjectural correspondence of topological strings and spectral problems provides in fact a non-perturbative 
realization of the topological string, in the spirit of the AdS/CFT correspondence. The implications of this non-perturbative definition have not been fully 
explored and they might lead to valuable insights on quantum geometry. It would also be of great importance to see whether the quantum mechanical problem is a manifestation of a more 
complex entity, like a theory of M2 branes or a gauge theory. Finally, an important challenge would be to generalize the correspondence between 
spectral theory and topological strings to the case of compact Calabi--Yau manifolds. 

\section*{Acknowledgements}
We would like to thank Andrea Brini, Santiago Codesido, 
Alba Grassi, Minxin Huang, Rinat Kashaev, Amir Kashani--Poor, Xin 
Wang and Szabolcs Zakany for useful discussions and correspondence. 
We are particular grateful to Xin Wang for sharing with us data for the 
refined BPS numbers, which he obtained from the refined vertex formalism. 
They matched our numbers obtained by holomorphic anomaly to high degree, 
but Xin Wang made them available to us in many cases to higher degree 
then we had calculated them. The work of M.M. is supported in part by the Fonds National 
Suisse, subsidies 200021-156995 and 200020-141329, and by the NCCR 51NF40-141869 
``The Mathematics of Physics" (SwissMAP). J.G. and J.R.~are partially supported 
by a scholarship of the Bonn-Cologne Graduate School BCGS. A.K. is 
supported by KL 2271/1-1 and DMS-11-59265.

\section{Appendix} 

\subsection{Weierstrass data for ${\widehat{\mathbb{C}^3/\mathbb{Z}_3\times\mathbb{Z}_3}}$ and 
${\widehat{\mathbb{C}^3/\mathbb{Z}_2\times\mathbb{Z}_4}}$.} 
\label{weierstrass}

The toric local Calabi--Yau with genus one mirror curves can be obtained, 
with one exception\footnote{The one exception is 
obtained by a blow up in an intermediate step instead.}, by blowing 
down orbifold geometries ${\widehat{\mathbb{C}^3/G}}$ with 
$G=\mathbb{Z}_3\times\mathbb{Z}_3$ and $G=\mathbb{Z}_2\times\mathbb{Z}_4$. 
The Weierstrass data are the invariants of the mirror curves. If 
the Weierstrass data are know for the above examples they follow 
for the other examples discussed in this paper simply by 
specialization of the mass parameters $m_j$ and the edge parameters $a_i$. 

As indicated on the edges of Figure~\ref{Z3Z3andZ2Z4} the Newton polynomial will be homogenized 
as cubic in $\mathbb{P}^2$ and as quartic in $\mathbb{P}_{\Delta^*}=\mathbb{P}^2(1,1,2)$. 
By  Nagell's algorithm\footnote{See \cite{Huang:2013yta} for a short description of this algorithm.} the latter can be brought into the Weierstrass form
\begin{equation} 
y^2=4 x^3-g_2(\tilde u,{\underline m})  x- g_3 (\tilde u,{\underline m})\ .
\end{equation}
For the cubic. i.e. the mirror of ${\widehat{\mathbb{C}^3/G}}$ with $G=\mathbb{Z}_3\times\mathbb{Z}_3$ one gets  
{\footnotesize 
\begin{equation}
\begin{array}{rl} 
g^{C}_2\!\!\!\! &=\frac{1}{12}(16 [m_1^2 m_4^2 + m_2^2 m_5^2 - m_2 m_3 m_5 m_6 + m_3^2 m_6^2 + 9 [a_1 a_3 m_2 m_3 + a_1 a_2 m_4 m_5 + a_2 a_3 m_1 m_6] -\\ &  
                       m_1 m_4 (m_2 m_5 + m_3 m_6) - 3 [a_1 (m_2 m_4^2 + m_3^2 m_5) + a_3 (m_1^2 m_3 + m_2^2 m_6) + a_2 (m_1 m_5^2 + m_4 m_6^2)]] + \\& 
                       24 [a_3 m_1 m_2 + a_1 m_3 m_4 + m_1 m_3 m_5 + m_2 m_4 m_6 + a_2 m_5 m_6-9 a_1 a_2 a_3] \tilde u - \\&  8 [m_1 m_4 + m_2 m_5 + m_3 m_6] \tilde u^2 + 
                       \tilde u^4)\ ,
\end{array}
\end{equation}
\begin{equation}
\label{WE6}
\begin{array}{rl} 
g^{C}_3\!\!\!\! &=\frac{1}{216}(8 [486 a_1 a_2 a_3 (a_3 m_1 m_2 + a_1 m_3 m_4 + a_2 m_5 m_6) - 8 (m_1^3 m_4^3 + m_2^3 m_5^3 + m_3^3 m_6^3) +\\ &  
         6 [2 m_1^2 m_4^2 (m_2 m_5 + m_3 m_6) + 2 m_2 m_3 m_5 m_6 (m_2 m_5 + m_3 m_6) + m_1 m_4 (2 m_2^2 m_5^2 + \\ & 
         m_2 m_3 m_5 m_6 + 2 m_3^2 m_6^2)] -108 [a_2^2 (a_1 m_5^3 + a_3 m_6^3) + a_2 (a_3^2 m_1^3 + a_1 m_4 (a_1 m_4^2 -\\ &  
         m_1 m_4 m_5 - m_2 m_5^2) - a_3 m_1 m_6 (m_1 m_4 + m_3 m_6)) + a_1 a_3 (a_3 m_2^3 + m_3 (a_1 m_3^2 - m_2^2 m_5 - \\ & 
         m_2 m_3 m_6))] + 27 [a_3^2 m_1^2 m_2^2 + a_1^2 (-27 a_2^2 a_3^2 + m_3^2 m_4^2) + m_1^2 m_3^2 m_5^2 - 6 a_2 a_3 m_1 m_2 m_5 m_6 + \\ & 
         m_2^2 m_4^2 m_6^2 + a_2^2 m_5^2 m_6^2 - 6 a_1 (a_2 m_3 m_4 m_5 m_6 + a_3 (m_1 m_2 m_3 m_4 + a_2 m_1 m_3 m_5 + a_2 m_2 m_4 m_6))] + \\ & 
         9 [a_1 (4 m_1 m_2 m_4^3 - 2 m_1 m_3^2 m_4 m_5 - 8 m_2^2 m_4^2 m_5 - 8 m_2 m_3^2 m_5^2 - 2 m_2 m_3 m_4^2 m_6 + 4 m_3^3 m_5 m_6) + \\ & 
         a_3 (4 m_1^3 m_3 m_4 - 2 m_1 m_2^2 m_4 m_6 + 4 m_2^2 m_6 (m_2 m_5 - 2 m_3 m_6) - 2 m_1^2 m_3 (m_2 m_5 + 4 m_3 m_6)) - \\ &  
         2 a_2 (4 m_1^2 m_4 m_5^2 + m_4 m_6^2 (m_2 m_5 - 2 m_3 m_6) - m_1 (2 m_2 m_5^3 - m_3 m_5^2 m_6 - 4 m_4^2 m_6^2))]] -\\ &  
         144 [m_1^2 m_3 m_4 m_5 + m_1 m_2 m_3 m_5^2 + m_1 m_2 m_4^2 m_6 + m_1 m_3^2 m_5 m_6 - 5 a_2 m_1 m_4 m_5 m_6 + m_2^2 m_4 m_5 m_6 +\\ &  
         a_2 m_2 m_5^2 m_6 + m_2 m_3 m_4 m_6^2 + a_2 m_3 m_5 m_6^2 + a_3 m_1 m_2 (m_1 m_4 + m_2 m_5 - 5 m_3 m_6) + a_1 m_3 m_4 (m_1 m_4 -\\ &  
         5 m_2 m_5 + m_3 m_6) + 9 a_1 a_2 a_3 (m_1 m_4 + m_2 m_5 + m_3 m_6) - 6 (a_1 (a_3 m_1 m_3^2 + a_3 m_2^2 m_4 + a_2 m_3 m_5^2 + \\ & 
         a_2 m_4^2 m_6) + a_2 a_3 (m_1^2 m_5 + m_2 m_6^2))] \tilde u +24 [2 m_1^2 m_4^2 + 2 m_2^2 m_5^2 + m_2 m_3 m_5 m_6 + 2 m_3^2 m_6^2 + \\ & 
         27 (a_1 (a_3 m_2 m_3 + a_2 m_4 m_5) + a_2 a_3 m_1 m_6) + m_1 m_4 (m_2 m_5 + m_3 m_6) - 3 (a_1 (m_2 m_4^2 + m_3^2 m_5) + \\ & 
         a_3 (m_1^2 m_3 + m_2^2 m_6) + a_2 (m_1 m_5^2 + m_4 m_6^2))] \tilde u^2 + 36 [15 a_1 a_2 a_3 + a_3 m_1 m_2 + a_1 m_3 m_4 +  \\ & 
         m_1 m_3 m_5 + m_2 m_4 m_6 + a_2 m_5 m_6] \tilde u^3 - 12 [m_1 m_4 + m_2 m_5 + m_3 m_6] \tilde u^4 + \tilde u^6)\ .
\end{array}
\end{equation}}
While for the quartic i.e. the mirror for $G=\mathbb{Z}_2\times\mathbb{Z}_4$ the coefficients are
{\footnotesize 
\begin{equation}
\label{WE7} 
\begin{array}{rl} 
g^{Q}_2\!\!\!\! &=\frac{1}{12}(1 - 8 (m_1 m_2 + a_2 m_4) \tilde u^2 + 24 a_2 (m_1 m_3 + m_2 m_5) \tilde u^3 + 16 (12 a_1 a_2^2 a_3 + m_1^2 m_2^2 - \\ &  
                  a_2 m_1 m_2 m_4 + a_2^2 m_4^2 - 3 a_2 (a_3 m_1^2 + a_1 m_2^2 + a_2 m_3 m_5)) \tilde u^4\ ,\\
g^{Q}_3\!\!\!\! &=\frac{1}{216}(1 - 12 (m_1 m_2 + a_2 m_4) \tilde u^2 + 36 a_2 (m_1 m_3 + m_2 m_5) \tilde u^3 +24 ( 2 m_1^2 m_2^2 + a_2 m_1 m_2 m_4 + \\ &  
                  2 a_2^2 m_4^2 -24 a_1 a_2^2 a_3- 3 a_2 (a_3 m_1^2 + a_1 m_2^2 + a_2 m_3 m_5)) \tilde u^4 + 144 (6 a_2^2 (a_1 m_2 m_3 + a_3 m_1 m_5) -\\ &  
                  a_2 (m_1 m_2 + a_2 m_4) (m_1 m_3 + m_2 m_5)) \tilde u^5 + 8 (12 a_2 m_1 m_2 m_4 (m_1 m_2 + a_2 m_4) - 144 a_1 a_2^2 a_3 (m_1 m_2 - \\ & 
                  2 a_2 m_4) - 8 (m_1^3 m_2^3 + a_2^3 m_4^3) + 18 a_2 (m_1 m_2 - 2 a_2 m_4) (2 a_3 m_1^2 + 2 a_1 m_2^2 - a_2 m_3 m_5) - \\ & 
                  108 a_2^3 (a_1 m_3^2 + a_3 m_5^2) + 27 a_2^2 (m_1^2 m_3^2 + m_2^2 m_5^2)) \tilde u^6\ .
\end{array}
\end{equation}
}

\subsection{Additional Fricke data for mass deformed $E_8$ del Pezzo} 
\label{app:FrickeE8}

Fricke's theory allows to obtain from the Weierstrass data the Picard-Fuchs equations 
and the periods. For example the coefficients of the Picard--Fuchs equation for mass deformed $E_8$ del Pezzo are (see \eqref{PFequationE8})
 {\footnotesize 
\begin{equation}
\begin{array}{rl} 
f_{9,8}= & 
6 - 12m_1m_2u + 10m_2^2u^2 + 7m_1^2m_2^2u^2 - 240m_3u^2 + 10m_1^2m_3u^2 + 1116m_1u^3 - 6m_1^3u^3 \\ &
- 14m_1m_2^3u^3 + 468m_1m_2m_3u^3 - 14m_1^3m_2m_3u^3 - 3312m_2u^4 - 2304m_1^2m_2u^4 + 12m_1^4m_2u^4 \\ &
+ 8m_2^4u^4 - 456m_2^2m_3u^4 - 212m_1^2m_2^2m_3u^4 + 2400m_3^2u^4 - 456m_1^2m_3^2u^4 + 8m_1^4m_3^2u^4 \\ &
+ 9180m_1m_2^2u^5 + 1202m_1^3m_2^2u^5 - 9936m_1m_3u^5 + 2688m_1^3m_3u^5 - 16m_1^5m_3u^5 + 468m_1m_2^3m_3u^5\\ & 
- 3360m_1m_2m_3^2u^5 + 468m_1^3m_2m_3^2u^5 - 31104u^6 + 16200m_1^2u^6 - 2142m_1^4u^6 + 9m_1^6u^6 \\ &
- 6912m_2^3u^6 - 6216m_1^2m_2^3u^6 + 27648m_2m_3u^6 + 3312m_1^2m_2m_3u^6 - 2952m_1^4m_2m_3u^6 \\ &
- 256m_2^4m_3u^6 + 3456m_2^2m_3^2u^6 + 352m_1^2m_2^2m_3^2u^6 - 9216m_3^3u^6 + 3456m_1^2m_3^3u^6 \\ &
- 256m_1^4m_3^3u^6 + 24624m_1m_2u^7 - 16020m_1^3m_2u^7 + 2466m_1^5m_2u^7 + 9024m_1m_2^4u^7 \\ &
- 40608m_1m_2^2m_3u^7 + 11700m_1^3m_2^2m_3u^7 + 24192m_1m_3^2u^7 - 13488m_1^3m_3^2u^7 \\ &
+ 1860m_1^5m_3^2u^7 - 1808m_1m_2^3m_3^2u^7 + 6336m_1m_2m_3^3u^7 - 1808m_1^3m_2m_3^3u^7 \\ &
- 27648m_2^2u^8 + 27648m_1^2m_2^2u^8 - 5184m_1^4m_2^2u^8 - 4096m_2^5u^8 + 82944m_3u^8 \\ &
- 89856m_1^2m_3u^8 + 29376m_1^4m_3u^8 - 3024m_1^6m_3u^8 + 30720m_2^3m_3u^8 - 10240m_1^2m_2^3m_3u^8 \\ &
- 55296m_2m_3^2u^8 + 32256m_1^2m_2m_3^2u^8 - 4608m_1^4m_2m_3^2u^8 + 1024m_2^4m_3^2u^8 - 7168m_2^2m_3^3u^8\\ & 
+ 2304m_1^2m_2^2m_3^3u^8 + 12288m_3^4u^8 - 7168m_1^2m_3^4u^8 + 1024m_1^4m_3^4u^8 - 93312m_1u^9 \\ &
+ 69984m_1^3u^9 - 17496m_1^5u^9 + 1458m_1^7u^9 - 13824m_1m_2^3u^9 + 3456m_1^3m_2^3u^9 + 72576m_1m_2m_3u^9 \\ &
- 36288m_1^3m_2m_3u^9 + 4536m_1^5m_2m_3u^9 + 1536m_1m_2^4m_3u^9 - 3456m_1m_2^2m_3^2u^9 + 864m_1^3m_2^2m_3^2u^9 \\ &
- 13824m_1m_3^3u^9 + 6912m_1^3m_3^3u^9 - 864m_1^5m_3^3u^9 - 384m_1m_2^3m_3^3u^9 + 1536m_1m_2m_3^4u^9 \\ &
- 384m_1^3m_2m_3^4u^9
 \end{array} 
\end{equation} 
\begin{equation}
\begin{array}{rl} 
g_{9,8}= &
18 - 38m_1m_2u + 38m_2^2u^2 + 21m_1^2m_2^2u^2 - 384m_3u^2 + 38m_1^2m_3u^2 + 1296m_1u^3  \\ &
- 36m_1^3u^3 - 44m_1m_2^3u^3 + 672m_1m_2m_3u^3 - 44m_1^3m_2m_3u^3 - 3024m_2u^4  \\ &
- 2556m_1^2m_2u^4 + 44m_1^4m_2u^4 + 24m_2^4u^4 - 696m_2^2m_3u^4 - 228m_1^2m_2^2m_3u^4  \\ &
+ 2880m_3^2u^4 - 696m_1^2m_3^2u^4 + 24m_1^4m_3^2u^4 + 8856m_1m_2^2u^5 + 1210m_1^3m_2^2u^5  \\ &
- 11232m_1m_3u^5 + 3264m_1^3m_3u^5 - 50m_1^5m_3u^5 + 584m_1m_2^3m_3u^5 - 3456m_1m_2m_3^2u^5  \\ &
+ 584m_1^3m_2m_3^2u^5 - 23328u^6 + 15552m_1^2u^6 - 2538m_1^4u^6 + 27m_1^6u^6 - 6336m_2^3u^6  \\ &
- 5904m_1^2m_2^3u^6 + 24192m_2m_3u^6 + 4320m_1^2m_2m_3u^6 - 3168m_1^4m_2m_3u^6  \\ &
- 320m_2^4m_3u^6 + 3744m_2^2m_3^2u^6 + 48m_1^2m_2^2m_3^2u^6 - 9216m_3^3u^6 + 3744m_1^2m_3^3u^6  \\ &
- 320m_1^4m_3^3u^6 + 13824m_1m_2u^7 - 13824m_1^3m_2u^7 + 2592m_1^5m_2u^7 + 8192m_1m_2^4u^7  \\ &
- 36864m_1m_2^2m_3u^7 + 11136m_1^3m_2^2m_3u^7 + 27648m_1m_3^2u^7 - 14976m_1^3m_3^2u^7  \\ &
+ 2016m_1^5m_3^2u^7 - 1536m_1m_2^3m_3^2u^7 + 5120m_1m_2m_3^3u^7 - 1536m_1^3m_2m_3^3u^7  \\ &
- 24192m_2^2u^8 + 29376m_1^2m_2^2u^8 - 5832m_1^4m_2^2u^8 - 3584m_2^5u^8 + 72576m_3u^8  \\ &
- 86400m_1^2m_3u^8 + 29592m_1^4m_3u^8 - 3132m_1^6m_3u^8 + 26880m_2^3m_3u^8 - 9536m_1^2m_2^3m_3u^8  \\ &
- 48384m_2m_3^2u^8 + 28224m_1^2m_2m_3^2u^8 - 4032m_1^4m_2m_3^2u^8 + 896m_2^4m_3^2u^8  \\ &
- 6272m_2^2m_3^3u^8 + 2112m_1^2m_2^2m_3^3u^8 + 10752m_3^4u^8 - 6272m_1^2m_3^4u^8 + 896m_1^4m_3^4u^8  \\ &
- 93312m_1u^9 + 69984m_1^3u^9 - 17496m_1^5u^9 + 1458m_1^7u^9 - 13824m_1m_2^3u^9 + 3456m_1^3m_2^3u^9 \\ & 
+ 72576m_1m_2m_3u^9 - 36288m_1^3m_2m_3u^9 + 4536m_1^5m_2m_3u^9 + 1536m_1m_2^4m_3u^9  \\ &
- 3456m_1m_2^2m_3^2u^9 + 864m_1^3m_2^2m_3^2u^9 - 13824m_1m_3^3u^9 + 6912m_1^3m_3^3u^9  \\ &
- 864m_1^5m_3^3u^9 - 384m_1m_2^3m_3^3u^9 + 1536m_1m_2m_3^4u^9 - 384m_1^3m_2m_3^4u^9 \\ &
 \end{array} 
\end{equation}      }

\subsection{Relation between local $\IF_2$ and local $\IF_0$}
\label{rel-ff}
It turns out that the local $\IF_0$ and the local $\IF_2$ geometries are very 
closely related. 

\begin{center}
\resizebox{3cm}{!}
{\begin{tikzpicture}{xscale=1cm,yscale=1cm}
\coordinate[label=below:${2}$,label=above:$m$](A) at (-2,0);
\coordinate[label={[label distance=.4mm]225:${0}$},label={[label distance=.4mm]45:$\tilde u $}](B) at (0,0);
\coordinate[label=below:${1}$](C) at (2,0);
\coordinate[label=above:${3}$](D) at (0,2);
\coordinate[label=below:${4}$](E) at (0,-2);
\draw (A)--(B)--(C);
\draw (D)--(B)--(E);
\draw (A)--(E)--(C)--(D)--(A);
\fill (A) circle (.1);
\fill (B) circle (.1);
\fill (C) circle (.1);
\fill (D) circle (.1);
\fill (E) circle (.1);
\end{tikzpicture}} 
\end{center}

Let us denote the symmetric classes of the two $\mathbb{P}^1$  inside 
of $\IF_0=\IP^1\times \IP^1$ as $[F_1]$ and 
$[F_2]$. The Mori vectors for local $\IF_0$ are 
\be
l^{(F_1)}=l^{(1)}=(-2;1,1,0,0),\quad l^{(F_2)}=l^{(2)}=(-2;0,0,1,1). 
\ee
The K\"ahler parameters are denoted by $t_k$, $k=1,2$, while the corresponding complex parameters 
\be
t_k=\frac{1}{2 \pi i} \log(z_k)+{\cal O}({\underline z})
\ee
are denoted by $z_k$. The mirror curve can be written in the form
\be 
H(x,y)=1+e^x+ z_1 e^{-x} + e^y+ z_2 e^{-y}
\ee
or equivalently 
\be
 \hat H(\hat x, \hat y)= e^{\hat x}+ m_{\IF_0} e^{-\hat x} + e^{\hat y}+ e^{-\hat y}-\tilde u \ . 
\label{curvef0} 
\ee
Here we used the reparametrization
\be
z_{F_1}=z_1=\frac{m_{\IF_0}}{\tilde u^2}, \quad z_{F_1}=z_2=\frac{1}{\tilde u^2}
\ee
and rescaled  $e^x\mapsto \frac{e^{\hat x}}{\tilde u}$, $e^y\mapsto \frac{e^{\hat y}}{\tilde u}$ 
and $\tilde u\mapsto -\tilde u$. Note that  $\tilde u$ is always naturally associated 
to the unique inner point in the reflexive 2d polyhedra that represent
$\IF_k$, $k=0,1,2$ and the other $13$ toric almost del Pezzo surfaces 
$S$. This parameter corresponds to the canonical class $K_{S}$. One has 
$[F_1][F_2]=1$ and $[F_i]^2=0$, $i=1,2$.   

The mirror curve for local $\IF_2$ is written down in (\ref{curvef2}). The equivalence between the two geometries can be seen in various ways. 
First of all, the BPS numbers are equal albeit in shifted classes. 
\be
n^{(g)\ \IF_0}_{i,j}=n^{(g)\ \IF_2}_{i,i+j} \ . 
\ee
More importantly, we have the following relation. The $J$ invariant for the elliptic curve (\ref{curvef0}) reads   
\be 
J_{\IF_0}(u_{\IF_0},m_{\IF_0})=\frac{\left(16 \left(m_{\IF_0}^2-m_{\IF_0}+1\right) u_{\IF_0}^2-8 (m_{\IF_0}+1) u_{\IF_0}+1\right)^3}{m_{\IF_0}^2 u_{\IF_0}^4 \left(16 (m_{\IF_0}-1)^2 u_{\IF_0}^2-8 (m_{\IF_0}+1) u_{\IF_0}+1\right)},
\label{JF0}
\ee
where we introduced 
\be
u_{\IF_0}=\frac{1}{\tilde u^2_{\IF_0}}.
\ee
For the elliptic curve (\ref{curvef2}) the $J$ invariant $J_{\IF_2} (u,m)$ is 
\begin{equation}\label{equ:F0F2Dict}
J_{\IF_2} (u,m)=\frac{\left(16 m^2 u^2-8 mu-48 u^2+1\right)^3}{16 m^2 u^6-8 m u^5-64 u^6+u^4}. 
\end{equation}
Now it is easy to see that 
\be 
\label{relff}
J_{\IF_2} \left(u=\sqrt{m_{\IF_0}} u_{\IF_0} ,m= \frac{1+m_{\IF_0}}{\sqrt{m_{\IF_0}}}\right)=J_{\IF_0}(u_{\IF_0},m_{\IF_0}).
\ee
%


\end{document}